\documentclass[a4paper, 12pt]{article}
\usepackage[utf8]{inputenc}
\usepackage[english]{babel}
\usepackage{natbib}
\usepackage[margin=1in]{geometry}
\usepackage{graphicx}
\usepackage{subcaption}
\usepackage{rotating}
\usepackage{wrapfig}
\usepackage{lscape}
\usepackage{setspace}
\usepackage{caption}
\usepackage[pagebackref=true]{hyperref}
\hypersetup{colorlinks=true, citecolor=blue}

\graphicspath{ {Figure_for_Paper/} }

\title{Understanding Vaccine Hesitancy: Empirical Evidence from India\thanks{ I thank Esther Duflo for lucidly explaining the puzzles associated with vaccination in India, which became the primary motivation for this research. Additionally, I thank Nathan Nunn, Kensuke Teshima, Lucy Xiaolu Wang, Chikako Yamauchi, and numerous seminar and conference participants for their excellent feedback. All remaining errors are my own. I am grateful for financial support from the Japan Society for the Promotion of Science (No. 21K13307). This paper is revised extensively and supersedes earlier versions circulated under the title “Why is the Vaccination Rate Low in India?”}}

\author{Pramod Kumar Sur\thanks{Asian Growth Research Institute (AGI) and Osaka University. Email: pramodsur@gmail.com}}

\date{First Version: January 2021
\vskip 0.1cm
This Version: February 2023}

\begin{document}
\maketitle

\begin{abstract}
\noindent
\linespread{1.0}\selectfont
Why do vaccination rates remain low even in countries where long-established immunization programs exist and vaccines are provided for free? We study this paradox in the context of India, which contributes to the world's largest pool of under-vaccinated children and about one-third of all vaccine-preventable deaths globally. Combining historical records with survey datasets, we examine the Indian government’s forced sterilization policy, a short-term aggressive family planning program implemented between 1976 and 1977. Using multiple estimation methods, including an instrumental variable (IV) and a geographic regression discontinuity design (RDD) approach, we document that the current vaccination completion rate is low in places where forced sterilization was high. We also explore the heterogeneous effects, mechanisms, and reasons for the mechanism. Finally, we examine the enduring consequence and present evidence that places more exposed to forced sterilization have an average 60 percent higher child mortality rate today. Together, these findings suggest that government policies implemented in the past can have persistent adverse impacts on demand for health-seeking behavior, even if the burden is exceedingly \textit{high}.

\end{abstract}
\textbf{Keywords:} Vaccination, immunization, family planning, forced sterilization, institutional delivery, antenatal care, child mortality

\newpage

\section{Introduction}
Vaccines are among the greatest advances in global health and development, saving millions of lives every year (UNICEF).\footnote{\url{https://www.unicef.org/immunization}} It is also one of the safest methods to protect children from life-threatening diseases. However, despite having access to vaccines for more than 20 life-threatening diseases and almost all of them available for free, about 20 million children still do not receive vaccines each year \citep{world2020immunization}. As a response, in 2019, the World Health Organization declared vaccine hesitancy to be one of the top ten threats to public health.\footnote{\url{https://www.who.int/news-room/spotlight/ten-threats-to-global-health-in-2019}}

India contributes to the world's largest pool of under-vaccinated children in the world \citep{centers2013global}. Additionally, about one in three child deaths due to vaccine-preventable diseases globally occur in India alone \citep{black2010global}. Moreover, India has one of the lowest vaccination rates in the world.\footnote{According to India’s National Family and Health Survey 2015–16 (NFHS-4) only about 43\% of children between 12 and 23 months of age were fully immunized in 2015–16 (see Figure A1 for details).} For instance, India’s vaccination rate is even lower than that of its nearest neighbors, Bangladesh, Bhutan, and Nepal, all of which have a lower GDP than India.\footnote{UNICEF estimates. See \url{https://www.unicef.org/rosa/media/6901/file/South_Asia_Immunization_Regional_Snapshot_2018.pdf}}
The statistics on India’s lower vaccination rate are particularly puzzling because India already has had a well-established immunization program since 1978, and vaccination services are provided for free. Besides, it is home to the largest vaccine maker in the world.\footnote{Serum Institute of India, based in Pune, India, is the largest vaccine manufacturer in the world. See \url{https://www.seruminstitute.com/about_us.php}} So why is the vaccination rate low in India?

The main objective of this paper is to 1) understand the paradox of lower vaccination in India, 2) examine plausible mechanisms for this paradox, 3) explore the \textit{reasons} for the mechanisms, and finally, 4) assess the present-day consequences.\footnote{India did not have any adult vaccination program until the recent COVID-19 vaccination. Therefore, this paper primarily focuses on the lower vaccination paradox among children.} Addressing these issues is not only fundamental from a scientific and academic standpoint but also essential in terms of ethical reasons and policy aspects for the following reasons. First, the cost of poor vaccination levels in India is exceedingly high. For example, according to India’s most recent estimates, in 2015, more than 300,000 children died due to vaccine-preventable diseases---constituting about \textit{two-thirds} of all types of deaths in children \citep{liu2016global}. Second, considering that most vaccines are administered for infectious and communicable diseases, there is a greater need to improve vaccination uptake to achieve the level of herd immunity and limit the spread of diseases. Thus, increasing efforts have been undertaken recently to improve India’s vaccination rate through various government programs.\footnote{Such programs include Mission Indradhanush (in 2014), Intensified Mission Indradhanush (IMI) (in 2017), IMI 2.0 (in 2019), and IMI 3.0 (in 2021).} However, little systematic evidence exists on the causal pathways through which individual, social, and historical characteristics influence decision-making for vaccinations \citep{francis2018factors}. Finally, considering the recent experience of the coronavirus disease 2019 (COVID-19) pandemic and the need for (near) universal vaccination, including children, policymakers and practitioners need to understand the factors affecting India’s lower vaccination paradox to carve out a pragmatic policy and maximize the uptake of new vaccines.

In this paper, we provide the first empirical investigation of the importance of a domestic policy implemented by the government in the past in shaping current vaccination practices in India. In particular, we examine whether the aggressive family planning program, under which the forced sterilization policy was implemented during the state of emergency rule in India in the 1970s, could \textit{partly} explain the lower vaccination paradox today.

Between June 1975 and March 1977, India went through a brief period of authoritarian rule under Prime Minister Indira Gandhi.\footnote{This period is popularly known as “The Emergency.”} During this period, the prime minister proclaimed a national emergency, under which the Indian constitution was suspended for a wide range of civil liberties. A distinctive feature synonymous with this period that affected the general population was an aggressive family planning policy through forced sterilization (hereafter, forced sterilization policy) in the latter part of the emergency period. After about a year of emergency rule, in April 1976, the Ministry of Health and Family Planning introduced a National Population Policy (NPP) under which a family planning program was aggressively undertaken, mostly through sterilizing individuals. Between April 1976 and March 1977, over 8.2 million sterilizations were performed, more than three times the number in the previous year (see Figure 1). The aggressive nature of the program led to serious consequences, including medical complications, deaths, and sterilization of ineligible individuals. Additionally, archival records suggest that disincentives were provided, coercion was enforced, and public officials delivered disinformation to motivate individuals to undergo sterilization during this period \citep{shah_commission_of_inquiry_third_1978, Panandiker1978Family}.

Our main hypothesis is that the forced sterilization policy undertaken during the emergency rule period may have had an unintended effect on India’s vaccination practice. There are reasons to expect that the policy could have had unintended consequences, primarily on vaccination. First, the same government department (Ministry of Health and Family Planning) that implemented the highly controversial forced sterilization policy introduced the first immunization program a year later, in 1978. Second, the health care staff (e.g., community health workers, auxiliary nurse midwives) who coerced and disinformed individuals to get sterilized during the emergency period are the ones who also engage in advising and motivating parents to vaccinate their children. Finally and most importantly, anecdotal evidence suggests that one of the main challenges for the recent vaccination campaigns—Intensified Mission Indradhanush (IMI) in 2017 and the current COVID-19 vaccination—is the concern about the circulation of disinformation about vaccines, rumors about adverse effects, and conspiracy theories, including vaccines causing sterilization, impotency, and infertility \citep{Gurnani2018Improving, BBC2021India, HindustanTimes2021, IndiaToday2021}.\footnote{Also, see \citet{Vardhan2021Dr}—The Health Minister of India—refuting a series of claims regarding the vaccine rumors and disinformation on Twitter in a series of tweets.} Thus, we examine the legacy of the forced sterilization policy on vaccine hesitancy in India.

To measure exposure to the forced sterilization policy, we use the historical records of sterilization from the yearbooks published by the Ministry of Health and Family Planning, Government of India.\footnote{These yearbooks report statistics on family planning programs performed between April and March every year at the state and union territory (UT) levels. Hereafter, we refer to ‘states and union territories’ in India as ‘states’ for simplicity.}
Our primary measure of exposure to the forced sterilization policy is the number of excess sterilizations performed between April 1976 and March 1977 (after the introduction of the NPP), normalized by its performance in the previous year. We also corroborate our primary measure of exposure to the forced sterilization policy with different sterilization measures, including the total number of sterilizations performed in 1976–77, the excess number of sterilizations performed in 1976–77, total and excess sterilizations on a natural logarithm scale, and an alternative measure of exposure to sterilization measured by vasectomies, which constituted the majority of the sterilizations performed during this period.

Our main finding is that higher exposure to the forced sterilization policy is associated with lower vaccination completion rates. We examine vaccination completion rates using data from India’s national representative NFHS-4 survey conducted in 2015–16. We construct a vaccination index that measures the share of vaccination completion rate. We find that higher exposure to the forced sterilization policy is associated with lower vaccination completion rates. Our results are robust to a variety of controls, consideration of specific cohorts of children, and a number of alternative measures of exposure to the forced sterilization policy.

After presenting evidence that the forced sterilization policy has a negative association with India’s current vaccination completion rate, we next turn to the task of addressing concerns over reverse causality and omitted variable bias using an instrumental variable (IV) estimation approach. We construct an instrument based on the unique history of the emergency period. Historical accounts, including India's court judgment and previous studies, suggest that sterilization during this period was aggressively undertaken because of the active role played by Sanjay Gandhi, the younger son of the prime minister \citep{Gwatkin1979Political, Vicziany1982Coercion, Chandra2017In, IndianNationalCongress2011Congress, Nayar2013Emergency, Williams2014Storming}. In fact, family planning was an integral part of his self-declared five-point program implemented during this period. Although Mr. Gandhi did not hold any formal position in the government, he and his close colleagues in New Delhi continuously influenced political leaders, particularly in the states adjacent to the national capital \citep{shah_commission_of_inquiry_third_1978}. As a result, northern parts of India, especially states adjacent to New Delhi, were later popularly known as the “vasectomy belt” because of the large number of male sterilizations performed during this period. \citet{Gwatkin1979Political} observes that distance from New Delhi to state capitals—a proxy measure for Mr. Gandhi’s influence—which was previously not relevant, emerged as an important determinant of excess sterilization during this period. Gwatkin also presents evidence that the “distance is itself capable of explaining two-thirds of the variation in excess sterilization performance among the states.” Considering these historical accounts, we use the distance from New Delhi to state capitals as an instrument to capture the variation in exposure to the forced sterilization policy.\footnote{Distance from New Delhi to state capitals is not the only reason for variation in forced sterilization. For our empirical approach (i.e., IV) to work, all we need is that they are a source of exogenous variation, primarily driven by the personal effort of Sanjay Gandhi.}

The IV estimates also suggest that the forced sterilization policy has had a significant negative effect on the vaccination completion rate in India. Not only are the negative coefficient estimates statistically significant, but they are also economically meaningful. Our IV estimates indicate that an average increase in excess sterilization—from zero to about 3.5 times—decreases the completion of all vaccinations by about 8.1 percentage points. This is relative to a sample mean of 32.1\% for our sample as a whole. It suggests that the forced sterilization policy has a sizable effect (about 25\% on average) on explaining India’s lower vaccination paradox.

To establish our IV results, we also need to ascertain the validity of our identification strategy hold. In particular, a potential concern in IV strategy is that our instrument is not exogenous, which is impossible to test directly. To address this concern, we perform a battery of counterfactual or falsification exercises. We primarily group them into five. First, we formally test Gwatkin’s insight. Because Sanjay Gandhi had no personal influence over sterilization before 1976 (because he did not implement his five-point program before), our instrumental variable—if exogenous—should have no predictive power on sterilizations that were performed before 1976. We perform a battery of placebo tests and confirm that the distance from New Delhi to state capitals has no predictive power for sterilizations performed in previous years. Second, we examine whether or not the sterilization policy’s forceful nature is only associated with our instrument. To explore this, we run a counterfactual exercise considering female sterilization, or tubectomy, which was not the main focus during India’s forced sterilization period. We find that our instrument has no predictive power for the increase in female sterilizations performed during 1976–77. Third, we present evidence that the source of variation we exploit is less likely to be confounded by demographic differences across states. We show that our instrument is not correlated with population across states from the 1971 census, rural population in 1971, the share of the Muslim population in 1971, or the population growth rate between 1961 and 1971. Fourth, we also document that the states closer to New Delhi are not systematically different based on development characteristics measured by net domestic product per capita, labor force participation rate, and the share of the population working in the organized sector across states, thus bolstering our overall causal mechanism. Fifth and finally, we present evidence that our instrument is not confounded by the political differences across states as well. Precisely, we show that distance from the national capital (New Delhi) does not predict voting behavior toward PM Indira Gandhi’s Indian National Congress (INC) party before the forced sterilization period. Overall, this extensive set of falsification exercises increases our confidence both in the validity of our identification strategy and, more importantly, in the specific channel via which our instrument is hypothesized to impact the excess sterilization.

As an alternative and complementary strategy, we use two proxy measures of forced sterilization available at a \textit{granular level}. In particular, we focus on the constituency-level variation in the vote share of the INC party in the 1977 election and the change in its vote share between 1971 and 1977. As numerous scholars have argued, and we will explain in detail later, the debate over forced sterilization policy became the focal point during the 1977 election campaign (which happened in March just after the authoritarian rule was relaxed), and the INC's vote share declined substantially, particularly in places that had been deeply affected by the forced sterilization drives.\footnote{Indeed, the 1977 election was the first time since India’s independence in which INC party was defeated in the parliament election. See, for example, \citet{Banerjee2011Poor, Gwatkin1979Political, weiner1978india, Williams2014Storming}.} In both cases, we find results consistent with our interpretation. We find that vaccination rates are low in places where the INC vote share was lower in 1977. Additionally, vaccination rates are also low in places where INC's vote share declined substantially in the 1977 election (in comparison with the immediate general election in 1971).

As a final step in our empirical estimation strategy, we examine the effect of forced sterilization on vaccine hesitancy through a geographic regression discontinuity design (RDD) approach at a local level. We focus our attention on the current adjacent states of Odisha and Chhattisgarh. As we explain in detail later, these states had a similar GDP per capita and performed an equal number of sterilizations before 1976-77. However, they had a stark difference in the number of sterilizations performed during the forced sterilization period.\footnote{For example, Madhya Pradesh (now split into Madhya Pradesh and Chhattisgarh since 2000) performed the highest number of sterilizations (more than one million) among all Indian states. It also reported over 40 times higher numbers of complaints related to compulsion or use of force in family planning than the neighboring state of Odisha (only 36 cases were reported in Odisha).} No other adjacent states with similar development characteristics faced such a stark difference in forced sterilization and complaints related to compulsion or use of force in family planning.

Before presenting our geographic RDD results, we first establish the validity of our design hold. Precisely, we present evidence to convince us that—except for the treatment (vaccination rate in our case), which varies discontinuously at the border—the other covariates change continuously. We conduct an extensive set of exercises to establish the validity of our design. First, we show that the distributions of observed covariates are continuous across the border. Second, we present evidence of no manipulations or bunching at the cutoff and no effects at artificial cutoff points. Third and finally, we document that there is no evidence of a jump at the counterfactual border between Madhya Pradesh and Chhattisgarh that belonged to the same state during the forced sterilization period.

After establishing the validity of our geographic RD design hold, we document a sharp discontinuity in the vaccination rate at the border between Odisha and Chhattisgarh. Precisely, we find that the vaccination rate is, on average, about 24-33 percentage points lower in Chhattisgarh, where forced sterilization was high. These results, \textit{although at a local level}, further bolster our interpretation that places, where forced sterilization was severe have a lower vaccination rate today.

We then turn to examine the heterogeneous effects. First, we explore each vaccine separately to obtain some insights into whether the lower vaccination rate we observe differs for some specific vaccines. We document that higher exposure to the forced sterilization policy has the largest effect on vaccines given at birth. Additionally, we find evidence suggesting that the vaccination rate declines with higher doses for vaccines administered multiple times. Second, exploring earlier rounds of NFHS surveys conducted in 1992–93, 1998–99, and 2005–06, we present evidence that the effect of forced sterilization on vaccination is not unique to the current period but rather persists over time.

Next, we explore plausible mechanisms. First, considering the results from our heterogeneous analysis, we explore whether the place of delivery is a possible channel. Place of delivery—at home or at an institution such as a healthcare facility—is an important determinant for vaccination because some vaccines are given immediately after birth. We document that exposure to the forced sterilization policy has a large, positive, and significant effect on non-institutional delivery.

Digging a little further, we also check the reasons for non-institutional delivery. The NFHS-4 survey asks mothers to report reasons for not delivering their child in a healthcare facility. We find that exposures to forced sterilization on supply-side constraints are negative. In contrast, the effects on demand-side factors are positive. These results suggest that, despite having access to medical facilities and the ability to afford medical expenses to deliver at the hospital, mothers are less likely to seek out institutional delivery primarily due to demand-side reasons.

We also test the effect of a plausible indirect channel of information provision.\footnote{Several studies have shown that information provision is essential to generate take-up rates in health-seeking behavior (for a review, see \citet{Dupas2017Impacts}.} As we noted earlier, one of the main challenges of India’s recent vaccination campaign was the concern about the circulation of disinformation, including vaccines causing sterilization, impotency, and infertility. In such an environment, the provision of reliable and accurate information can help increase the vaccination rate. We study one such platform, i.e., antenatal care (ANC). ANC is not only essential to reduce the health risks for mothers and their unborn babies during pregnancy but can also be an important source of reliable and accurate information regarding a child’s future health-seeking behavior, such as vaccination practice. We find that exposure to the forced sterilization policy has a large, negative, and statistically significant effect on visiting a healthcare center for ANC and the number of additional visits conditional on receiving any ANC. These results suggest that a lack of reliable and accurate information provision may also be an important mechanism.

Finally, we examine the relevance of the forced sterilization policy on present-day consequences measured by child mortality. We find that child mortality is currently high in states with greater sterilization exposure. The effect size is quite large. An average increase in excess sterilization increases the probability of death of a child under the age of five by about 2.7 percentage points, relative to a sample mean of 4.5\%. These results highlight the importance of history for understanding present-day health outcomes and, more broadly, how historical policies affect the demand for health-seeking behavior, even if the burden is exceedingly high.

In addition to the historical literature discussed previously, our paper is related to several diverse literature streams in economics. First, we directly contribute to the literature on understanding the factors associated with the vaccination take-up rate in general and in India in particular. Recent studies suggest that the lower vaccination rate in India is associated with a child’s individual-level characteristics and other household factors \citep{francis2018factors, Ghosh2017Demand, Shrivastwa2015Predictors, Srivastava2020Explaining}. Although we do not dispute such findings, these characteristics alone, however, cannot explain all the differences.\footnote{For example, according to the NFHS-4 estimates, the vaccination completion rate is still low among male children, Hindu households, urban residents, mothers with 12 or more years of education, forward castes, and households in the highest wealth quintile (see Figure A1 in the Appendix for a detailed description).} Moreover, recent studies have also highlighted that the absolute demand for vaccination in India is low even when there is a reliable supply of free immunization services with incentives \citep{Banerjee2010Improving, Chernozhukov2020Generic}, and sometimes these incentives backfire \citep{Chernozhukov2020Generic}. We build on this literature in three ways. First, we compile novel historical data and provide the first empirical investigation of the importance of historical policies in shaping India’s lower vaccination paradox. Second, we provide a plausible causal pathway, the mechanisms, and the reasons for the mechanisms through which historical characteristics influence decision-making about childhood vaccinations. Third, we explore the present-day consequence of lower vaccination measured by child mortality.

This paper is also related to the broader literature on understanding the barriers associated with health-seeking behavior in developing countries \citep{Dupas2011Health}. Recently, experimental methods such as randomized controlled trials have been extensively used to examine both demand- and supply-side barriers to healthcare utilization (for a review, see \citet{Dupas2017Impacts}). We contribute to this literature on understanding the demand-side barriers to health-seeking behavior by considering historical policy intervention as a natural experiment. As we demonstrated earlier, India is a typical case where the demand for health-seeking behavior, such as vaccination, is low, even if the burden is exceedingly high. We present evidence suggesting that government policies implemented in the past could have a long-term and persistent effect on adverse demand for health-seeking behavior.

This work is also related to the literature on the unintended consequences of health interventions. Recent studies have found that the disclosure of information related to unethical medical intervention has had adverse effects on medical trust and health-seeking behavior \citep{Alsan2018Tuskegee, martinez2022vaccines}. Relatedly, \citet{Lowes2021Legacy} provide evidence that historical medical campaigns during the colonization period in Africa have had a long-term impact on health outcomes and trust in medicine. We build on this work by presenting evidence on how a domestic policy (not externally imposed) implemented by the government in the past could have a long-term spillover effect on vaccination, institutional delivery, ANC, and child mortality in India. Additionally, we provide evidence of the evolution of vaccination rates over time.

Finally, this paper also contributes to the literature on understanding the importance of history on current development \citep{Nunn2009importance}. This field has been studied extensively, beginning with the seminal work by \citet{Acemoglu2001colonial} (for a review, see \citet{Nunn2014Historical}). We build on this literature in two ways. First, we demonstrate that a \textit{short-term policy}—that lasted for less than a year—has had a large, negative, and significant long-term impact on later development outcomes measured by healthcare indicators.\footnote{In India’s case, the government implemented the forced sterilization policy in April 1976 and ended it less than a year later, in January 1977.} Second, we present evidence showing that historical events can affect subsequent policies implemented by the same organization or government agencies, even when the policies were well-intentioned. This has important implications, particularly for public policy, because several countries currently pursue evidence-based policy-making through experimentation. We show empirically that a policy failure could have spillover effects and affect subsequent policies in the long run.

The remainder of the paper is structured as follows. Section 2 provides a brief background to the authoritarian rule, the forced sterilization policy, and the immunization program in India. Section 3 describes the historical and contemporary data used in the empirical analysis. Section 4 presents the main results. Section 5 examines the heterogeneous effect of the forced sterilization policy on vaccination. Section 6 presents a direct and an indirect mechanism. Section 7 explores the enduring consequences, and Section 8 concludes. The online appendix provides additional robustness checks and results.

\section{Background}
\subsection{The Emergency and Forced Sterilization in India}

On June 25, 1975, Prime Minister Indira Gandhi proclaimed a national emergency in India. The exact reason for the proclamation of emergency is controversial to this day. However, historians, sociologists, and political scientists agree that a combination of political and economic problems facing her and India could be the most likely factor.\footnote{For a detailed overview of this period, see \citet{Dhar2000Indira, Nayar2013Emergency, jaffrelot2021india}.}

In 1971, Mrs. Gandhi won the national election under the radical slogan of ending poverty (\textit{garibi hatao}). However, food production decreased in the succeeding years because of poor rainfall. Furthermore, the balance of payments was in turmoil because of a sharp rise in oil prices and the subsequent slump in export demand. Things became more complicated in June 1975, when Allahabad High Court found Mrs. Gandhi guilty of various corrupt election practices in the 1971 national election, jeopardizing her continuation as prime minister. The court decision led to opposition protests and demanded that Mrs. Gandhi resign. Instead of resigning, she seized the moment and proclaimed a national emergency, justifying the situation as a threat to India’s internal stability \citep{Hewitt2007Political}.

The emergency rule allowed Mrs. Gandhi to suspend a wide range of civil liberties under the Indian constitution. Her government used this period to repress the opposition and institute censorship in the name of law and order. Thousands, including leading opposition leaders, were arrested, the press was censored, and public gatherings and strikes were declared illegal. With all the power in her hands, she undertook a series of new legislative and constitutional amendments to govern the country and extend the emergency period. Furthermore, she delayed parliamentary elections several times, indicating her intent to remain in power, an impression strengthened by (unofficially) elevating her younger son Sanjay to the position of heir apparent \citep{Gwatkin1979Political}. However, in January 1977, Mrs. Gandhi unexpectedly called an election and released opposition leaders from jail, lifted press censorship, and permitted public meetings once again. The emergency period officially ended in March after the Indian National Congress party’s defeat in the lower house of the Indian parliament election (Lok Sabha).

A distinguishing characteristic synonymous with this brief period of authoritarian rule was the aggressive family planning drive through forced sterilization.\footnote{For a detailed overview of the family planning program during the emergency rule period, the complete reliance on sterilization only, and the forceful nature of the program, see \citet{Panandiker1978Family, shah_commission_of_inquiry_third_1978, Gwatkin1979Political}.} It started in April 1976, about a year after the start of the emergency rule. It began with the National Population Policy (NPP) for India, introduced by the Ministry of Health and Family Planning to the parliament on April 17. The NPP’s principal aim was to reduce population growth by boosting the family planning program. The new policy incorporated a series of fundamental changes to reduce population growth. The legislation primarily included a substantial increase in monetary compensation for sterilization acceptors, encouragement for state-level incentives and disincentives for family planning, disenfranchisement of states that failed to control fertility rates, allocation of central assistance to states according to family planning performance, and, most controversially, the provisions for state governments to pass compulsory sterilization legislation \citep{Singh1976National}.

With the NPP’s introduction, the central government authorized and endorsed various coercive measures for sterilization and, in extreme cases, the provision for compulsory sterilization. The central and state governments substantially increased the financial rewards for sterilization acceptors. Through a range of incentives and disincentives, they pressured their employees to get sterilized and motivated others to do so. In some cases, compulsory quotas were imposed on government employees to produce people for sterilization. In other cases, citizens were required to produce sterilization certificates to access basic facilities, such as housing, irrigation, ration cards, and public healthcare facilities. Several extreme measures were also undertaken in some states. For example, the state government in Maharashtra passed a bill allowing compulsory sterilization of couples with three or more children \citep{shah_commission_of_inquiry_third_1978, Panandiker1978Family}.\footnote{This was not approved by the central government and eventually returned to the state for revision.} In the words of the New York Times, Maharashtra had \textit{“become the first political entity in the world to legislate population control by forced sterilization”} \citep{NewYorkTimes1976India}.

Historical records, court rulings, and anecdotal evidence from the field suggest that quotas were imposed, incentives and disincentives were provided, coercion was enforced, and disinformation was delivered to motivate individuals to undergo sterilization during this period.\footnote{For a detailed discussion on quota enforcement, incentives and disincentives, coercion, disinformation, and fear of sterilization during the emergency, see \citet{Panandiker1978Family, shah_commission_of_inquiry_third_1978}.} For example, in Uttar Pradesh, over 24,000 employees of the Department of Health and Family planning were not paid their salary in June 1976 for failing to complete their sterilization quota for the April–June quarter \citep{Panandiker1978Family}. Anecdotal evidence of some of the extreme coercive measures can be seen from the following incident in Uttawar, a village in the state of Haryana, as reported in \citep{Mehta2015Sanjay}:\footnote{The same incident is also reported in \citet{shah_commission_of_inquiry_third_1978, Gwatkin1979Political}.}

\textit{Uttawar is a village 80 kms south of Delhi [..]. At 3 a.m. one morning, the inhabitants of the village were awakened by police loudspeakers which ordered all the men to assemble at the bus-stop [..]. Frightened, unsure, the villagers did as they were ordered, and it was only when they arrived at the bus stop that they discovered that their village, like in some crazy western movie, had been surrounded. 400 men assembled at the bus-stop, but the police suspected that some were still hiding. In the process of unearthing more volunteers [for sterilization], the police pillaged, broke and looted. [..] These ‘find and operate’ activities continued for three weeks in which a total of 800 sterilizations were notched up; and Uttawar ‘had the dubious distinction of probably having every eligible male sterilized’.}

A unique feature of the family planning program during this period was that almost all government departments were involved in family planning, and it was organized and administered locally. Additionally, the nature of the emergency rule and the executive power allowed the central government to give directions to states as to how the policies were to be exercised. The central government encouraged the states to decide on and implement incentives and disincentives for sterilization.\footnote{For a detailed description of the incentives and disincentives implemented at the state level, see \citet{shah_commission_of_inquiry_third_1978}.} All government departments were engaged in the family planning program, and specific targets were allotted to each of them. Coordination and supervision were delegated by the Chief Secretary (the top-ranking civil servant in the state) to the collectors or magistrates—the highest-ranking administrative civil servants at the district level. Under their guidance, most sterilizations were performed in temporary camps organized by the health departments.\footnote{There were exceptions as well. For example, the military and the railway department were given special sterilization targets, which were not a part of the state administrative unit.}

The aggressive family planning drive led to over 8 million sterilizations in 1976–77, more than three times the number in the previous year. During the peak, over 1.7 million sterilizations were performed in September 1976 alone, a figure that equaled the annual average for the ten preceding years. The majority of the sterilizations performed during this period involved men undergoing vasectomy. Out of about 8.3 million sterilizations performed in 1976–77, about 6.2 million (about 75\%) were achieved through vasectomy.

The main reason for the heavy reliance on vasectomy was the simplicity of the procedure. Tubectomy (female sterilization) constitutes major abdominal surgery, whereas vasectomies are relatively quick to perform, and patients can be discharged on the same day of the operation. During the period of emergency rule, the authorities relied on vasectomy, as sterilization was mostly performed in temporary camps. The existing infrastructure also struggled to cope with a large number of operations because of increased pressure and intentions to meet the target \citep{Gwatkin1979Political}. Therefore, although vasectomy was not a part of the constructed family planning scheme during this period, it was necessary to reach the required target owing to the authoritarian rule, political pressure from the center, and relatively lower time window \citep{scott2017my}.

The aggressive nature of the program led to serious consequences, including medical complications, deaths, and sterilization of ineligible individuals. For example, according to the report published by the \citet{shah_commission_of_inquiry_third_1978}, 1,778 complaints of deaths related to sterilization and 548 reports regarding sterilizations of unmarried persons had been registered.

Anecdotal evidence also suggests that the forced sterilization policy’s legacy remained in peoples’ minds and could be felt even after the emergency rule came to an end. For example, to repair family planning’s legacy, the Indian government changed the name of the \textit{Department of Family Planning} to the \textit{Department of Family Welfare}. \citet{Basu1985Family} found that the family planning program shifted from vasectomy to tubectomy during the post-emergency period when women emerged as the primary target. Additionally, \citet{Tarlo2000Body} notes that the word “emergency” itself became synonymous with “sterilization.” The emergency period remains controversial today and is still regarded as one of the darkest periods in the history of Indian democracy.

\subsection{Immunization Programs in India}

The immunization program in India was introduced in 1978 as the Expanded Program of Immunization (EPI) by the Ministry of Health and Family Welfare. In 1985, it was renamed the Universal Immunization Program (UIP) when its reach was expanded beyond urban areas. The UIP was implemented in a phased manner to cover all districts by the year 1989–90. In 1990, the IUP became universalized in geographical coverage to cover all infants in India. Since 2005, UIP has been under the National Health Mission—an initiative strengthening the health system in rural and urban areas—and serves as a key area of health intervention in India \citep{Lahariya2014brief}.

UIP is one of the largest public health programs in the world, targeting about 27 million newborns annually. Under UIP, immunization is currently provided against 12 vaccine-preventable diseases.\footnote{According to the National Health Mission, India, immunization is provided nationally against nine diseases (diphtheria, pertussis, tetanus, polio, measles, rubella, severe form of childhood tuberculosis, hepatitis B, and meningitis and pneumonia caused by Haemophilus influenza type B) and sub-nationally against three diseases (rotavirus diarrhea, pneumococcal pneumonia, and Japanese encephalitis).} Immunization services are primarily administered through Integrated Child Development Services (ICDS)—a publicly funded program through which the Government of India promotes early-childhood health and education services. ICDS provides immunization services through Anganwadi centers—a type of childcare and preschool education center in India. According to the NFHS-4 survey in 2015–16, about half of the children received most vaccinations from Anganwadi centers. Additionally, both public and private healthcare facilities also provide immunization services.

Despite having a longstanding history of immunization programs and the free availability of vaccines, India continues to have one of the lowest vaccination take-up rates globally and contributes to the largest pool of under-vaccinated children in the world. According to the most recent estimates, more than 300,000 children aged 1–59 months died from vaccine-preventable diseases in 2015, contributing to about one-third of total deaths globally. Vaccination coverage in India also varies considerably within states. The highest numbers of under-vaccinated children are found in central and northern states such as Madhya Pradesh, Uttar Pradesh, Bihar, and Rajasthan.

\section{A Brief Description of Data and their Sources}
\subsection{Historical Data}

The historical data on sterilization for this paper comes from the historical yearbooks published by the Ministry of Health and Family Planning, Department of Family Planning, Government of India. Along with various demographic and health statistics, the yearbooks report yearly statistics on family planning programs performed between April and March every year. Notably, the historical yearbooks include the numbers and types of sterilization performed at the state level.

We collected sterilization data from the Ministry of Health and Family Welfare archives. In Figure A2 in the Appendix, we present examples of the archival data used in this paper. Figure 1 presents the total number of sterilizations along with the types of sterilization performed in India every year since the beginning of the program in 1956. As we can see, there is a sharp increase in the total number of sterilizations performed in 1976–77. We also see that most sterilizations performed during this period were vasectomies.

In Figure A3 in the Appendix, we present the total number of sterilizations performed at the state level in 1975–76 (the year before the announcement of NPP in April 1976). Additionally, in panel A of Figure 2, we present the total number of sterilizations performed at the state level in 1976–77 (the year of the implementation of NPP). To provide a visual representation, we group the total number of sterilizations performed each year into several broad categories and denote a greater number of sterilizations performed by darker shades. As we can see, the number of sterilizations had become distributed across all of India in 1975–76. However, there was a shift in sterilization performance towards the central and northern parts of India after the NPP’s announcement in 1976–77. Panel B of Figure 2 presents a better measure of state-level variation in exposure to the forced sterilization policy as measured by the number of excess sterilizations performed in 1976–77 normalized by performance in 1975–76. As we can see, exposure to the forced sterilization policy was particularly high in central and northern parts of India, especially in states adjacent to New Delhi. This is because a large number of sterilizations performed during this period were the result of the personal influence of Mr. Sanjay Gandhi (we will describe the role of Sanjay Gandhi in forced sterilization in detail in Section 4.2).

The historical data on the Indian National Congress party's vote share is taken from the general election results of the Lok Sabha—the lower house of the Indian parliament. The data come from the statistical reports published by the Election Commission of India.\footnote{See \url{https://eci.gov.in/statistical-report/statistical-reports/}} Recently, they have been digitized by the Trivedi Center for Political Data at Ashoka University.\footnote{See \url{http://lokdhaba.ashoka.edu.in/}} We use the geocoded location of NFHS-4 clusters and match them with the parliamentary constituency-level map of India (before the delimitation of boundaries in 2008).\footnote{The constituency level shape file is geocoded through the assembly-level map provided by Sandip Sukhtankar (See \url{https://uva.theopenscholar.com/sandip-sukhtankar/data)}. We acknowledge Manasa Patnam and Sandip Sukhtankar for their generous effort in creating and providing the shape file before delimitation.}

Finally, we collected historical data on several demographic and development indicators to examine the validity of our instrument. We collected state-level demographic data on the total population, rural population, the share of the Muslim population, and the population growth rate from the 1971 census (before the implementation of the forced sterilization policy). Additionally, we assembled historical data on domestic per capita, labor force participation rate, and the share of workers in the organized sector as proxy measures for development across states.

\subsection{Modern-Day Data}

We combine the historical data with India’s national representative NFHS-4 data collected in 2015-16 \citep{InternationalInstituteforPopulationSciences(IIPS)2017National}. The NFHS-4 is a stratified two-stage sample survey designed to produce indicators at the district, state, UT, and national levels, with separate estimates for urban and rural areas. The primary sampling units (PSUs) in the NFHS-4 are villages in rural areas and Census Enumeration Blocks (based on the 2011 Census) in urban areas. The dataset in our main analysis includes NFHS-4 data on children.\footnote{The sample from Sikkim and Nagaland are excluded from our main analysis as we have incomplete information on sterilization in these two states.} In the heterogeneous analysis, we also use earlier rounds of NFHS surveys conducted in 1992-93, 1998-99, and 2005-06.

Additionally, we combine data on population and health care to control for potential covariates that could affect exposure to both forced sterilization and the vaccination rate. We collect population data from the 2011 population census to construct state-level population density. Additionally, we collect healthcare facility and healthcare personnel data from Rural Health Statistics to construct hospitals per 1000 population and doctors per 1000 population at the state level.\footnote{We construct the healthcare facility data combining the number of subcenters, primary health centers, and community health centers from the Rural Health Statistics 2014–15. The number of health care personnel includes doctors, specialists (e.g., surgeons, obstetricians \& gynecologists, physicians \& pediatricians), and female health workers/auxiliary nurse midwives (ANMs). Using data on the number of doctors and hospitals from other sources, such as the Directorate General of State Health Services, produces identical results. We use the data from Rural Health Statistics because it includes compressive data on healthcare personnel and healthcare facility in more detail.}

Our primary outcome variable is the vaccination rate. The NFHS-4 data report a total of 13 vaccination details for children under the age of 5 years.\footnote{We exclude vitamin A supplements reported in the survey because supplements are not a vaccine.} The reported vaccines are against polio (Polio 0–3), tuberculosis (BCG), hepatitis B (Hepatitis-B 0–3), diphtheria, pertussis, tetanus (DPT 1–3), and measles. For our main analysis, we construct a vaccination index that measures the share of vaccination completion rate. The key benefit of considering a vaccination index measure instead of individual vaccines is that each vaccine or combination of doses is generally effective for preventing certain illnesses. Therefore, an index of vaccination completion can be considered an important health indicator. In the heterogeneous analysis, we also explore each vaccine separately as our outcome variable. Figure A4 presents the state-level variation in the percentage of children who have received all the vaccines.

We also use additional outcome variables to examine the mechanism through which the forced sterilization policy has influenced decision-making for childhood vaccinations. Our first additional outcome variable, from the NFHS-4 data, is the non-institutional delivery of a child. We consider this variable because the place of delivery—at home or at an institution such as a healthcare facility—is an important determinant of vaccination because some vaccines are given immediately after birth. In the NFHS-4, about 20\% of children are born at home (non-institutional delivery). We test whether exposure to the forced sterilization policy has had any effect on the place of delivery of a child.

Our second additional outcome variable, from the NFHS-4 data, is the reasons for non-institutional delivery among women. We use this variable to understand whether demand- or supply-side factors affect a mother’s intention to deliver her child at home. The NFHS-4 asks mothers the reasons for the non-institutional delivery of their last childbirth and reports a total of nine factors.\footnote{The reasons include: costly, facility not open, too far/no transportation, no female provider, no trust in a health care facility/poor service quality, not allowed by the husband or family, not necessary, not customary, and others.} First, we consider each possible reason separately as our outcome of interest. Second, we combine the information on reasons reported and construct two indexes, demand-side and supply-side, and examine whether demand- or supply-side factors affect the mother’s intention to deliver the child at home.

Our third additional outcome variable, also from the NFHS-4 data, is the mother’s data on ANC visits during pregnancy. We consider this variable to test the channel of information provision because an antenatal visit to health care centers can also be an essential source of receiving reliable and accurate information regarding a child’s future health-seeking behavior, such as vaccination practice. The NFHS-4 also provides information on the mother’s ANC records for her most recent pregnancy. We construct two outcome variables from these data: 1) whether the mother received ANC and 2) the number of visits conditional on receiving ANC. In our sample, about 87\% of mothers received ANC, and conditional on receiving ANC, the average number of visits was about 5.8 times. We test whether exposure to the forced sterilization policy has any effect on ANC.

Our fourth and final additional outcome variable is child mortality. We consider this variable to test the consequences of childhood vaccination. The NFHS-4 has information about the mortality record of children under five years old in the household. In our sample, about 12,000 (about 4.5\%) children under the age of five have died. Using the mortality records, we test whether exposure to the forced sterilization policy has had any persistent effect on child mortality.

\section{Understanding Vaccine Hesitancy}
\subsection{Correlation and OLS Estimates}

We begin by examining the relationship between historical exposure to the forced sterilization policy and India’s current vaccination completion rate. In Figure 3, we present a simple correlation plot between exposure to the forced sterilization policy and the vaccination index in 2015–16. In panel A, we present the correlation between the state-level \textit{total} number of sterilizations performed in 1976–77 and the vaccination index in 2015–16. In panel B, we present the correlation considering a better measure of exposure to the forced sterilization policy, as measured by state-level \textit{excess} sterilizations performed in 1976–77 normalized by performance in 1975–76, the year before (we discuss this variable in detail below). As we can see, a strong negative association is apparent in the raw data. Additionally, we observe that a specific outlier state does not drive this simple negative association.

We next examine this relationship by controlling for individual, household, geographic, and healthcare characteristics that are also potentially important for India’s current vaccination rate. Our baseline estimating equation is:

\begin{equation}
Y_{ihcs} = \alpha + \beta Forced\ Sterilization_{s} + \gamma_1 X^{'}_{ihcs} + \gamma_2 X^{'}_{hcs} + \gamma_3 X^{'}_{cs} + \gamma_4 X^{'}_{s} + \epsilon_{ihcs}
\end{equation}

where $Y_{ihcs}$ denotes one of our vaccination measures for child i living in household h in NFHS-4 cluster c of Indian state s. The variable $Forced\ Sterilization_s$ denotes one of our measures of exposure to the forced sterilization policy in state s (we discuss this variable in more detail below).

We include $X^{'}_{ihcs}$, a vector of child-level covariates, which includes an indicator variable for the child’s gender, month by year of birth fixed effects, an indicator for whether the child is a twin, and birth order of the child. The vector $X^{'}_{hcs}$ consists of household-level covariates, including the age and sex of the household head, household size, number of household members below the age of 5 years, seven religion fixed effects, four caste fixed effects, 20 education of the mother fixed effects, four household wealth index fixed effects, and an indicator for whether any household member is covered by health insurance. These child-level and household characteristics that we control have been shown to be correlated with the vaccination rate in India. $X^{'}_{cs}$ is a vector of NFHS-4 cluster-level covariates that captures the characteristics of the place where the child lives, such as altitude in meters, altitude squared, and an indicator of whether the cluster is urban. $X^{'}_s$ is a vector of covariates meant to capture state-level characteristics that are likely to be correlated with vaccination, including population density per square kilometer (in log), hospitals per 1000 population, and doctors per 1000 population. Finally, $\epsilon_{ihcs}$  is a random, idiosyncratic error term, capturing all omitted factors, which we allow to be heteroscedastic and correlated across children; in practice, the standard errors we report in our baseline estimates are clustered at the NFHS-4 defined cluster level.\footnote{As we mentioned before, NFHS-4 is a stratified two-stage sample designed to produce indicators at the district, state, and national levels and separate estimates for urban and rural areas. Therefore, undersampling and oversampling are observed in many places. To account for this issue, we conduct all the regression analyses using weights defined in the NFHS-4.}

We present the OLS estimates for the impact of $Forced\ Sterilization_s$ on the vaccination in Table 1. In column 1, we use the total number of sterilizations performed in a state in 1976–77 (expressed in 100,000 individuals) as our measure of exposure to the forced sterilization policy. The estimated coefficient for $Forced\ Sterilization_s$, $\beta$, is negative, which is similar to what we found in Figure 3. This suggests that higher exposure to the forced sterilization policy has an adverse effect on the vaccination completion rate. Because the distribution of the number of sterilizations performed in 1976–77 is skewed, with a small number of observations taking on large values, we report estimates using the natural log of the number of sterilizations performed in 1976–77 in column 2. The results are similar, as we find a significant negative correlation between forced sterilizations and the vaccination rate.

In columns 1 and 2 of Table 1, we use the total number of sterilizations performed in 1976–77 to measure exposure to the forced sterilization policy. One limitation of this measure is that it does not account for the number of sterilizations that would have happened anyway in the \textit{absence} of the NPP. Accounting for this difference is important because sterilization, as a family planning method, has been performed in India since the 1950s, as shown in Figure 1. In column 3 of Table 1, we account for this issue and use an alternative measure of the forced sterilization policy based on the absolute number of excess sterilizations performed in 1976–77 over and above the 1975–76 numbers.\footnote{Alternative measures of excess sterilization performed in 1976–77, such as deducting the average of the last 2 or 3 years, are also possible. Using such alternative measures also produces identical results.} Additionally, in column 4, we report estimates using the natural log of the absolute number of excess sterilizations performed in 1976–77. As we see, the results are similar using these alternative forced sterilization measures.

The estimates we report in columns 3 and 4 of Table 1 use the absolute number of sterilizations to measure the forced sterilization policy. Some shortcomings of these measures are that they 1) do not account for the difference in the size of states and 2) do not account for any state-wide historical factors associated with the level of sterilization performance that we do not capture in our estimation. To account for these issues, column 5 reports the estimates normalizing the excess sterilizations performed using sterilization figures for the previous year (1975–76). Specifically, we define $Forced\ Sterilization_s$ as,

{\footnotesize $${Excess \ Sterilization}_s = \frac{\# \ of\ sterilizations {(1976~1977)}_s - \# \ of\ sterilizations {(1975~1976)}_s} {\# \ of\ sterilizations {(1975~1976)}_s}$$}

We normalize the previous years’ figures to account for the effect of emergency rule in India and isolate the impact of forced sterilization policy from India’s emergency rule.\footnote{Using alternative measures such as normalizing by the average of the last 2 or 3 years produce nearly identical results.} This is because India’s emergency rule could itself affect our outcome in several ways, as this period was largely governed by autocratic rule and involved numerous policy changes. As we see, the results we obtain in column 5 remain robust to this alternative specification.

For the remainder of our analysis, we use state-level excess sterilizations performed in 1976–77 normalized by the sterilization figure in 1975–76 (the specification from column 5). This provides a better measure that accounts for India’s emergency rule and is normalized by both size and state-level historical characteristics associated with sterilization performance. However, as illustrated in Table 1, our results do not rest on this choice only.

In Section B of the Appendix, we present a series of robustness tests. We only briefly discuss them here. We present the results of Table 1, adding each set of controls sequentially (Tables B1-B5), an analysis with children aged between 12 and 23 months to capture the Indian government’s official vaccination estimate (Table B6), and considering excess vasectomy (Table B7), which constituted the majority of sterilization operations, as an alternative measure of $Forced\ Sterilization_s$. The findings are overall robust to these alternative specifications, specific cohorts, and different measures of the forced sterilization policy.

\subsection{Instrumental Variable Estimates}

In the previous section, we presented results suggesting a negative association between historical exposure to the forced sterilization policy and vaccination. We also showed several alternative estimations to provide robust evidence. However, the correlation we found may not necessarily identify the causal effect of forced sterilization on vaccination. For example, the correlation could also be explained by some omitted variables that determine both exposures to forced sterilization and the vaccination rate.

To address this concern, in this section, we present results by pursuing an instrumental variable (IV) approach. We need an instrument that is correlated with sterilization performance during the forced sterilization period but does not affect vaccination through any channels other than forced sterilization. We use distance from New Delhi to state capitals as an instrument to capture the variation in exposure to the forced sterilization policy.

The history of forced sterilization policy in the later part of the emergency rule leaves little doubt that our instrument is relevant. Historical accounts, including those of \citet{Gwatkin1979Political, Vicziany1982Coercion, Chandra2017In, IndianNationalCongress2011Congress, Nayar2013Emergency, Williams2014Storming}, describe the forced sterilization policy as aggressively undertaken owing to the active role played by Sanjay Gandhi— the younger son of the Prime Minister. It is a well-known fact that family planning was a key element of Mr. Gandhi’s self-declared five-point program and became his central theme of public addresses during the latter part of the emergency period \citep{shah_commission_of_inquiry_third_1978}. Mr. Gandhi and his close colleagues in New Delhi were at the center of the action and continuously influenced central government bureaucrats and regional political leaders, particularly those in the states adjacent to the national capital. \citet{Gwatkin1979Political} describes that \textit{the distance to state capitals from New Delhi (as a proxy of Mr. Gandhi’s influence), which was previously irrelevant, emerged as an important determinant of excess sterilization performance and was itself capable of explaining two-thirds of the variation in performance among states.} Consequently, northern parts of India were later popularly known as the “vasectomy belt” because a large number of (male) sterilizations were performed during this period.

To provide a visual understanding, we present Gwatkin’s insight on distance from New Delhi as an important determinant of excess sterilization performance in Figure 4. In panel A, we plot the association between the distance to state capitals from New Delhi and $Forced\ Sterilization_s$ as measured by excess sterilizations in 1976–77.\footnote{For newly formed states after 1977, the distance from New Delhi to state capitals is based on old states.} In panel B, we present the exact correlation but instead consider excess sterilizations in 1975–76, the immediately previous year. As we can see, the association is negative in panel A; however, we do not see any correlation in panel B. Based on these insights, we use the distance to state capitals from New Delhi as an instrument to capture the exogenous variation in exposure to the forced sterilization policy (we will return to this explanation and analyze it in detail in sub-section 4.4).

We present the IV estimates in Table 2. Panel A presents the first-stage estimates for the instrument we considered in our analysis. As we expect, the instrument is a strong predictor of the forced sterilization policy as measured by excess sterilizations. In panel B and panel C, we present the second-stage and reduced form estimates, respectively. The second stage estimates in Column 5 indicate that an average increase in excess sterilizations (from zero to about 3.5 times) decreases the completion of all vaccinations by about 8.1 percentage points. This is relative to a baseline completion of 32.1\% for the sample as a whole, which suggests a large effect (about 25\%) of exposure to the forced sterilization policy on the current vaccination rate in India.\footnote{The difference between OLS and IV estimates is more likely due to the difference in Local Average Treatment Effect (LATE) and average effect (ATE).}

\subsection{Adjusting Standard Errors for Alternative Clustering}

Thus far, we have shown all our baseline estimates by clustering our standard errors at the NFHS-4 primary sampling unit (PSU) level. We adjust our standard error clustering at the NFHS-4 PSU level primarily because of the design of the NFHS-4 survey \citep{abadie2022should}. Our primary assumptions for this way of clustering rely on the fact that children in the same village (or Census Enumeration Block in urban areas) are more likely to have been subject to common unobserved forces that may affect their current vaccination behavior.

However, likely, the within-group correlation of the residuals could also exist at different levels. For example, the standard errors may be correlated at a higher level of administrative boundaries than the PSU as our primary explanatory variable, and the instrument in our estimation does not vary across these clusters. Additionally, we also need to account for the issue of spatial correlation in our standard errors. As \citet{Kelly2019Standard} argues, persistence regressions are spatial regressions. Places in the real world are not scattered randomly across the landscape but instead clump together. Therefore, spatial data tend to be autocorrelated. In our case, for example, Indian villages might be clustered according to geography rather than administrative boundaries.\footnote{The same can also be true for Indian districts and states.} Thus, simply adjusting standard errors according to administrative boundaries alone—such as village, district, or state—may produce biased results.

We present the estimates with four different types of clustering choices in Table 2. First, we report our baseline standard errors based on NFHS-4 PSU in parentheses. Second, we report the standard errors adjusted for clustering at the current district levels—the second highest level of administrative units available to us. Third, we report the standard errors adjusted at the current state levels—the third highest level of administrative units. Fourth and finally, to account for the spatial autocorrelation in our estimation, we adjust the standard errors using the spatial correction proposed by  \citet{Conley1999GMM}.\footnote{We use the procedures proposed by  \citep{Colella2019Inference} to calculate Conley standard errors (acreg command in STATA) with cutoffs at a distance of 100 km beyond the observations belonging to the same cluster. We use the bartlett option that allows for weights in the matrix linearly decreasing as the distance increases with values very close to one for near observations and almost zero for those close to the distance cutoff. Different choices of distances (such as 25, 50, and 200 kilometers) and the use of binary weights (no bartlett option) produce nearly identical results.} As we can observe, our results are overall robust to adjusting standard errors for these alternative levels of clustering.

In Section C of the Appendix, we present a series of alternative analyses showing that our results are also robust to consideration of specific cohorts and an alternative measure of $Forced\ Sterilization_s$. For the remainder of our analysis, we report standard errors adjusted for clustering at the current state level. We opt for the state-clustered standard errors because they tend to be most conservative for the IV estimates and very similar for the first stage—as we can see from Table 2 and Section C of the Appendix.\footnote{Some potential econometric concerns about clustering standard errors at the state level may be that we have few clusters and these clusters are of unequal size. For example, we have only 34 clusters based on the number of current states in India. Similarly, the size of the largest cluster (Uttar Pradesh) and the smallest cluster (Andaman and Nicobar Islands) varies a lot.}

\subsection{Is Distance from New Delhi a Valid Instrument?}

We focus on distance from New Delhi to the state capitals as our instrument because it was uniquely relevant for forced sterilization, as suggested by historians and previous studies—generating a strong first stage. However, we need to ascertain that our instrument is exogenous and satisfies the exclusion restriction. In particular, our identification would be threatened if distance affected vaccination through channels other than the forced sterilization policy. For example, a primary concern could be that states near New Delhi have experienced a differential trend in sterilization historically. An additional concern is that the demographic, development, or political characteristics may be different. In this section, we explore this hypothesis more rigorously in a set of empirical tests that can shed light on the validity of the instrument.

\subsubsection{Historical Trends in Sterilizations are Not Associated with our Instrument}

Our first exercise consists of examining sterilizations performed before 1976. Because Sanjay Gandhi did not implement his self-declared five-point program before 1976, our instrumental variable—if exogenous—should have no predictive power on sterilization performance before 1976. Additionally, note that family planning was an integral part of the health interventions undertaken by the Indian government in the 1960s and early 1970s. In fact, the Indian health ministry was named the “Ministry of Health and Family Planning” during this period. Furthermore, the yearbooks—published by the ministry—contain specific sections dedicated to family planning (particularly sterilization), and no other health interventions have a dedicated section in the yearbooks published during this period. Therefore, this falsification exercise can also be considered a proxy measure to test whether state-level existing health interventions (such as healthcare spending) or health infrastructure (such as hospitals and doctors) correlate with our instrument.

In sub-section 4.2 above, we presented some evidence through a correlation plot suggesting that our instrument was not associated with sterilization performance in the immediately previous year. We formally test this by estimating several placebo exercises considering excess sterilizations performed between 1972 and 1975. We also present the associations disaggregated by excess vasectomies (male sterilization) and tubectomies (female sterilization). We present the results in the top panel of Figure 5. As we can see, the distance to state capitals from New Delhi has no predictive power for excess sterilizations performed in the previous four years. Additionally, the comparison with the effects of excess sterilization and excess vasectomy in 1976–77, shown in Figure 5 (in red color), indicates that the quantitative magnitude of this impact is also small.

\subsubsection{Female Sterilization is Not Associated with the Instrument}

Our second falsification exercise consists of female sterilization or tubectomy. During the period of emergency rule, the forced sterilization program primarily focused on men undergoing vasectomy, as seen in Figure 1. As we explained in detail in Section 2, tubectomy was not the focus during India’s forced sterilization period, primarily due to the complication of the operation and the longer period of hospitalization for recovery. We formally test whether distance to state capitals from New Delhi predicts the variation in excess tubectomy performed during the forced sterilization period. Row 11 of Figure 5 presents the results. As we can see, our instrumental variable is completely unrelated to excess tubectomy performed during the forced sterilization period.

\subsubsection{The Instrument is Orthogonal to Historical Differences in Demographic Characteristics}

Our third falsification exercise consists of testing whether historical differences in demographic characteristics across states are associated with our instrument. Historical demographic characteristics may affect sterilization because the \textit{primary} reason for implementing the coercive family planning program was to check the increasing population growth in India. We use several demographic characteristics from the historical censuses to test this hypothesis. We present the results in rows 16-20 of Figure 5. As we can see, our instrument is not confounded by the historical demographic structure across states. In particular, we find that the state-level population, absolute rural population, share of the rural population, the share of the Muslim population in 1971, and population growth rate between 1961 and 1971 are not systematically associated with our instrument.

\subsubsection{The Instrument is not Related to Historical Difference in Development Characteristics}

Our fourth falsification exercise tests whether historical differences in development characteristics are associated with our instrument as well. Using historical development indicators from archival sources, we present the results in rows 21-23 of Figure 5. The results suggest that more developed states—as expressed by higher net domestic product per capita, labor force participation rate, and the share of the population working in the formal sector—are not correlated with distance from New Delhi.\footnote{We use the share of employment in the organized sector in 1977 because the dataset is not available before that period.}

\subsubsection{Distance from New Delhi is not Related to Historical Differences in Political Behavior}

A final concern in our IV strategy is that distance from New Delhi to state capitals, even if orthogonal to pre-1976 sterilization, demographic, and development characteristics may be working through other channels. The most important alternative that could be confounding our results is that our instrument may be correlated with voting behavior in earlier elections—in particular, the 1971 election in which Indira Gandhi became the Prime Minister and later instituted authoritarian rule in India. We investigate whether this is the case. We look at the vote share of the INC party received in the 1971 election, the immediate general election before the emergency rule. We present the results in the last row of Figure 5. As we can observe, there is no evidence of a statistical association between our instrument and the INC’s vote share in the 1971 election.

Overall, we find no evidence of a higher level of sterilizations before 1976-77 or excess female sterilizations during the forced sterilization period in states closer to New Delhi. Our instrument is also orthogonal to pre-forced sterilization level demographic, development, and political differences across states. These extensive sets of falsification exercises strengthen our interpretation that the instrument that we employ in our IV estimation is plausibly exogenous—primarily driven by the personal efforts of Sanjay Gandhi, the younger son of the prime minister.

\subsection{Alternative Measures of Forced Sterilization at a Granular Level}

Until now, we have examined the impact of the forced sterilization policy on the lower vaccination paradox using variation in sterilizations performed in 1976-77. A major limitation of this measure could be that our identification is at the state level, as the sterilization records are not available at a more granular level, such as the district or village. It may not be an ideal level of variation as we have at most 34 states in our analysis, which is quite small. To account for this issue, we use two alternative proxy measures of forced sterilization that are available at a more granular level. In particular, we first use the constituency level variation in vote share of Indira Gandhi's Indian National Congress (INC) party in the 1977 parliament election (lower house or Lok Sabha), the immediate election after the implementation of the autocratic rule. Second, we use the change in INC's vote share in 1977 compared to the 1971 election (the immediate last general election to 1977) as an alternative proxy measure of forced sterilization.

Several scholars, including  \citet{Banerjee2011Poor, Gwatkin1979Political, weiner1978india, Williams2014Storming}, have argued that the draconian sterilization policy played an important role in Indira Gandhi and her INC party's defeat in the 1977 parliament election. Indeed, nasbandi (the term used for “sterilization” in India) became the focal point of the 1977 election campaign, and the INC's vote share declined substantially in places that had been deeply affected by forced sterilization drives  \citep{weiner1978india}. As \citet{Banerjee2011Poor} note,

	\textit{…when in 1977 India finally held elections, discussions of sterilization policy were a key part of the debate, as captured most memorably by the slogan “Indira hatao, indiri bachao (get rid of Indira and save your penis).” It is widely believed that Indira Gandhi's defeat in the 1977 elections was in part driven by popular hatred for this program. The new government immediately reversed the policy. (p. 105)}

We visually present the relationship between INC's vote share and excess sterilization in Figure 6. Panel A shows the association between state-level variation in excess sterilization and constituency-level variation in INC's vote share in the 1977 election. As we can see, a negative association is evident. To test whether such an association exists in earlier elections, we present the association between excess sterilization and INC's vote share in 1971, the immediate parliament election before 1977, in Panel B. The relationship in Panel B is flat, suggesting that a limited association exists between excess sterilization and INC's vote share in 1971. Overall, Figure 6 indicates that forced sterilization was not strategically targeted towards places where INC's vote share earlier was high or low.\footnote{For a detailed econometric analysis of the legacy of forced sterilization on voting behavior in India, see \citet{sur2022legacy}.}

We present the regression results in Table 3. Columns 1 and 3 present the OLS and IV estimates of the variation in the INC party's vote share in the 1977 election on vaccination, respectively. Additionally, columns 2 and 4 report the OLS and IV estimates of change in INC's vote share in 1977 compared with the 1971 election on vaccination, respectively.\footnote{As we can see, the number of observations drops in columns 2 and 4 in Table 5. This is primarily because the INC party did not contest its candidates from every constituency in the 1971 and 1977 elections.} As we can see, the constituency-level variation in INC's vote share has a positive and significant association with vaccination. In particular, we find that the places where the INC's vote share in the 1977 election was high, suggesting less exposure to forced sterilization, have a higher vaccination completion rate today. Additionally, we find that places where INC's vote share declined sharply in the 1977 election (in comparison with the 1971 election), have lower vaccination completion rates.

\subsection{Regression Discontinuity Design Approach}

As a final step in our empirical estimation strategy, we examine the effect of forced sterilization on vaccine hesitancy through a regression discontinuity design (RDD) approach at a local level. We focus our attention on the current states of Odisha and Chhattisgarh (Chhattisgarh was a part of Madhya Pradesh state during the emergency period). We focus on these two states because they are adjacent to each other. More importantly, they had a similar GDP per capita and performed a similar number of sterilizations in the previous year.\footnote{Comparing other adjacent states such as Madhya Pradesh and Maharashtra is possible. However, as we explain in detail below, these states are not historically comparable because they had substantially different GDP per capita and the number of sterilizations performed in the previous years. Additionally, the observed covariates are not well balanced at the border, which is the primary concern while conducting an RDD analysis.} However, they had a starkly different trajectory in terms of the number of sterilizations performed during the forced sterilization period (See Figure A5 for a descriptive comparison).

During the forced sterilization period, Madhya Pradesh (now split into Madhya Pradesh and Chhattisgarh since 2000) performed the highest number of sterilizations among all Indian states. Out of about 8 million sterilizations performed in India in 1976-77, over 1 million sterilizations took place in Madhya Pradesh alone. In fact, the absolute number of excess sterilizations was about \textit{eight} times higher than the previous year. In contrast, the sterilization performance was modest in the adjacent state of Orissa (present-day Odisha).\footnote{Odisha performed about 300,000 sterilizations which were about one and half times higher than the previous year.} The number of complaints related to the forceful nature of the sterilizations between these two states was even more striking. According to the report produced by the \citet{shah_commission_of_inquiry_third_1978}, Madhya Pradesh reported 1477 complaints related to compulsion and use of force in family planning which was over 40 times higher than the neighboring state of Odisha (only 36 cases were reported in Odisha). No other adjacent states with similar development characteristics faced such a stark difference in forced sterilization and reported complaints related to compulsion and use of force in family planning during this period. Thus, considering this phenomenon as a natural experiment in history, we conduct a geographic RDD design at the border between Odisha and Chhattisgarh and test our hypothesis at a local level.

Before presenting the main results, we first need to establish the validity of our geographic RDD design hold. Precisely, we want to present sufficient evidence to convince us that— except for the treatment, which varies discontinuously at the border—the other covariates change continuously. First, we check whether the distributions of observed covariates (that we included in our OLS and IV regression) are balanced at the border. We present the covariate balance results in Figure 7.\footnote{All the results presented in this section (except mentioned) use optimal bandwidths based on the data-driven method proposed in \citet{Calonico2015Optimal}. We set the local polynomial of order 2, which has a minimum asymptotic mean squared error (MSE) of the RD point estimator in our sample, as proposed by \citet{pei2021local}. Finally, we use a uniform kernel, as suggested by \citet{IMBENS2008615, lee2010regression}, which is simple, transparent, and easy to interpret. (Use of triangular kernel also produces identical results.)} As we can see, all the baseline covariates are balanced at the cutoff and statistically indistinguishable from zero.\footnote{Among 15 baseline characteristics, we only find that the share of children living in urban areas is slightly higher in Odisha and is marginally statistically significant at the 10 percent level.} Second, we look at the density of the forcing variable (distance to the border between Odisha and Chhattisgarh) to see whether there is any manipulation or bunching at the cutoff point. We present the results in Figure A6 in the Appendix. As we can see, there is no evidence of manipulation at the border. Third, we run the RDD specification by relocating the cutoff points to artificial boundaries on either side of the border. The primary intuition behind this test is that because the true cutoff point is the only score value at which the probability of receiving the treatment changes discontinuously, it should also be the only score value at which the outcome changes discontinuously. In other words, this is an exercise to test for a zero effect in a setting where we know that the effect should be 0, thereby increasing our confidence that we have chosen the correct specification. We chose the median (or close to the median) threshold point in each direction, as proposed by \citet{IMBENS2008615}, which increases the statistical power and makes it more likely to find an effect. We present the results in Figure A7 in the Appendix. We find no effect at these artificial threshold points both in the treatment as well as in the control groups.

Finally, we conducted a counterfactual or placebo exercise between Madhya Pradesh and Chhattisgarh border, which belonged to the same state during the forced sterilization period.  Utilizing the same geographic RDD design approach, we check whether the vaccination rates between Madhya Pradesh and Chhattisgarh are different. We essentially run this exercise to test that our outcome, apart from the real treatment status, should not have an abrupt discontinuity around the counterfactual border. In addition, we can also test whether the demarcation of state borders (due to geographic characteristics such as mountains and rivers) is affecting our results. We present the results in Figure A8 in the Appendix. As we can see, the vaccination rates in the state of Chhattisgarh and Madhya Pradesh are statistically indifferent at the border.

After establishing the validity of our design hold, we present our main RDD result in Panel A of Figure 8. As we can see, there is a sharp discontinuity in the vaccination rate at the cutoff point. Precisely, we find that the vaccination rate is, on average, about 24 percentage points lower in Chhattisgarh. The results are similar if we present the donut hole RDD design approach in Panel B of Figure 8 to account for the random allocation of the GPS coordinates.\footnote{To keep the anonymity of the clusters, the NFHS randomly allocate the latitude and longitude of the clusters up to 5 kilometers within a district.} Additionally, in Figure A9 in the Appendix, we also find similar results if we include the covariates that are not somehow balanced at the cutoff. Finally, we examine whether the data-driven bandwidth approach proposed by \citet{Calonico2015Optimal} is affecting our results. We present the evidence with alternative bandwidth choices in Figure A10 in the Appendix. As we can see, the chosen data-driven bandwidth is not affecting our results.\footnote{We only find a negative effect but with huge standard errors at the 20 KM, primarily due to a small sample size.}

Overall, all the results we presented until now come to the same conclusion suggesting that exposure to the forced sterilization policy has a direct impact on the current vaccination rate in India.

\section{Heterogeneous Effects of Forced Sterilization on Vaccination}

In this section, we examine the heterogeneous effects. We begin by exploring each vaccine separately from the NFHS-4 survey. Additionally, we use the previous three rounds of NFHS surveys conducted between 1992 and 2006 to explore the evolution and persistence of the vaccination rate over time. Because the GPS coordinates are not reported in earlier survey rounds, from now on, we primarily rely on our IV strategy to present the results and compare our outcomes of interest.

\subsection{Exploring Each Vaccine from NFHS-4}

We first explore each vaccine separately to understand whether the lower vaccination rate we observe differs for some vaccines or any particular doses. Understanding the heterogeneous effect is important because 1) different vaccines are given to children at different points in time, and 2) multiple doses of the same vaccines are given for full immunization. For example, according to India’s National Immunization Schedule, the first dose of polio and hepatitis B vaccines (Polio 0 and Hepatitis 0) should be given immediately after birth, whereas the measles vaccine is generally given between the age of 9 and 12 months. Similarly, vaccines such as hepatitis, DPT, and polio are given to children multiple times for full immunization.

We plot the IV regression coefficients for each vaccine in Figure 9.\footnote{We present the results in tabular form in Table D1 in the appendix.} For reference, we also include the average effect size of the vaccination index in Figure 9. The figure suggests three interesting findings. First, the effect sizes of all the vaccines, except measles, are negative. Second, we find that higher exposure to the forced sterilization policy has the largest effect on vaccines given at birth (i.e., Hepatitis 0 and Polio 0). Third, although not precisely estimated, we also find some indications that the vaccination rate declines with higher doses for vaccines administered multiple times, such as hepatitis, DPT, and polio. In Section D of the Appendix, we present alternative estimates examining specific cohorts, such as those between the age of 13 and 24 months (Table D2), and an alternative measure of the forced sterilization policy (Table D3). Overall, our analysis provides evidence that the forced sterilization policy has heterogeneous effects on vaccination and, in particular, the largest and statistically significant effects on vaccines given at birth.

\subsection{Evolution Over Time: Evidence from Earlier Survey Rounds}

Until now, we have examined the impact of forced sterilization on vaccination records from the NFHS-4 survey conducted in 2015–16. In this sub-section, we take advantage of earlier survey rounds to explore the evolution and persistence of the vaccination rate over time. In particular, we examine the impact of forced sterilization on vaccination from three previous NFHS survey rounds conducted in 1992–93, 1998–99, and 2005–06.
We should keep in mind that a direct comparison of NFHS-4 estimates with earlier rounds is difficult as the number of reported vaccines has changed over time. For example, NFHS-1 and NFHS-2 report eight and nine vaccines, respectively, whereas NFHS-4 reports 13 vaccines. However, one thing we can do is compare specific vaccines that have been reported in each survey period. Hence, we report the effect size of each vaccine separately, along with the average effect size.

Figure 10 presents the result of three earlier surveys.\footnote{Tables D4–D6 in the Appendix present the results in tabular format.} The results in Figure 10 suggest two important findings. First, the effect size of all the vaccines used to be negative and highly statistically significant in the earlier years, such as in 1992–93 and 1998–99. Second, we also document similar patterns consistent with the findings from the NFHS-4 survey. In particular, we find that exposure to forced sterilization has the largest effect on vaccines given at birth and the vaccination rate generally declines with higher doses for vaccines administered multiple times. Overall, the findings suggest that the effect of forced sterilization on vaccination is not unique to the current period but rather persists over time.

\section{Understanding the Mechanisms: Why are the Effects Still Persistent?}

Up to this point, we have found that the forced sterilization policy has had a significant and sizable effect on India’s vaccination rate. We have also found that the policy has heterogeneous effects on different vaccines. In particular, we found that the policy has significant effects on vaccines given at birth. In this section, we aim to understand better the plausible channels or mechanisms through which the forced sterilization policy has affected India’s current vaccination rate. First, we explore whether the place of delivery of a child is a possible channel considering the results of our heterogeneous analysis. Second, we examine an indirect channel of information provision through ANC.

\subsection{Place of Delivery}

Place of delivery—at home or in a healthcare facility—is an important determinant for vaccinations because some vaccines are given immediately after birth. We test whether exposure to the forced sterilization policy had any effect on the place of delivery. The NFHS-4 includes a question on the place of birth of the child. About 20\% of the children in our sample had non-institutional delivery, such as at the home of the respondents (the mother), their parents, or others.

For a simple visualization, we first present the association between exposure to the forced sterilization policy and the percentage of children who had non-institutional delivery at the state level through a scatter plot in Figure 11. As we can see, a strong positive association is apparent in the raw data. We then present the IV results in Table 4. As we can see, the coefficient of excess sterilization is sizable, positive, and statistically significant. This suggests that exposure to the forced sterilization policy has a large, positive, and significant effect on non-institutional delivery. Table E1 of the Appendix presents the results considering alternative measures of forced sterilization policy. Again, the results are similar, suggesting that higher exposure to the forced sterilization policy positively affects non-institutional delivery.

\subsection{Reasons for non-institutional Delivery}

We dig a little further and also check the reasons for non-institutional delivery. The NFHS-4 also asks mothers an additional question to explain the reasons for not delivering their child in a healthcare facility. The question is only asked of mothers for their most recent delivery. Both demand- and supply-side reasons are reported, such as higher costs, the facility not open, the facility being far or no transportation, no female provider, no trust in the healthcare facility or poor service quality, not being allowed by the husband or family members, not necessary, not customary, and others.\footnote{Respondents were allowed to provide multiple responses.} We present the IV estimates of each answer separately in Figure 12.\footnote{We present the results in tabular form in Table E2 in the Appendix.} Additionally, we present estimates by indexing the supply- and demand-side reasons in Table 5.

Notably, the coefficients of individual answers and average effect size coefficients in Table 5 suggest that the impact of exposure to the forced sterilization policy on supply-side constraints is actually negative. This suggests that mothers are less likely to report supply-side reasons as their primary reason for non-institutional delivery in places with greater exposure to forced sterilization in 1976–77. In contrast, the results from Figure 12 and Table 5 suggest that the effects on demand-side reasons are positive. It implies that demand-side reasons are the main drivers of non-institutional delivery in states where forced sterilization was high.\footnote{The supply-side constraints, in general, are less likely to be a problem, particularly in the Indian context, because the Indian government has been promoting and paying for institutional delivery through Janani Suraksha Yojana (the Maternal Protection Scheme) since 2005.}

\subsection{Information Provision through Antenatal Care (ANC)}

In this section, we examine an additional mechanism of information provision. Several studies have shown that provision of information is important to generate a take-up rate in health-seeking behavior.\footnote{See \citet{Dupas2017Impacts} for a review.} We test this channel in India’s context in general and vaccination in particular because one of the main challenges for India’s vaccination campaign is concern about the circulation of disinformation and a segment of the population linking vaccination initiatives to family planning and sterilization \citep{Banerjee2011Poor, Gurnani2018Improving, Nichter1995Vaccinations}. We hypothesize that providing reliable and accurate information can help increase the vaccination take-up rate in such an environment.

We study the information provision mechanism through ANC. ANC is not only important to reduce the health risks for mothers and their babies during pregnancy, but it also can be an essential source of reliable and accurate information regarding a child’s future health-seeking behavior, such as institutional delivery and vaccination practice. The NFHS-4 asks a question about the mother’s ANC records for her most recent pregnancy. In our sample, about 83\% of mothers received ANC, and conditional on receiving ANC, the average number of visits was about 5.7 times. We test whether exposure to the forced sterilization policy had any effect on receiving ANC and the number of visits.

We first present the associations through scatter plots in Figure 13. As we can see, there is a negative correlation between exposure to sterilization on the probability of receiving ANC in panel A and the number of visits conditional on receiving any ANC in panel B. We then present the IV results in Table 6. Column 1 presents the results of exposure to the forced sterilization policy on the probability of receiving ANC. Column 2 additionally reports the results on exposure to the forced sterilization policy on the number of ANC visits conditional on receiving ANC. We find that exposure to the forced sterilization policy has a large, negative, and significant effect on visiting a health care center for ANC and the number of visits conditional on receiving ANC. These results suggest that a lack of reliable and accurate information provision may also be an important channel.\footnote{In Table E3 of the Appendix, we present estimates showing that the results presented in Table 6 are robust to alternative measures of the forced sterilization policy.}

\section{Enduring Consequence}

Finally, we examine the enduring consequence of lower vaccination caused by the forced sterilization policy. To examine the consequence, we test whether exposure to the forced sterilization policy affects child mortality. We test child mortality as the consequence of lower vaccination because previous studies have shown that about two-thirds of deaths in children in India are due to vaccine-preventable diseases \citep{liu2016global}. The NFHS-4 survey has information about the mortality records of children below the age of five in the household. In our sample, about 4.5\% of children have died.\footnote{Unfortunately, we do not have any detailed information about the cause of death, which would have allowed us to examine this in detail. Also note that this sample does not include the number of children who died before their birth, such as due to abortion.} We test whether exposure to the forced sterilization policy has had any effect on child mortality.

We present the association through a scatter plot in Figure 14 and the IV regression results in Table 7. As we can see, a positive association is apparent in Figure 14. The coefficient estimates of excess sterilization in Table 7 are also positive, statistically significant, and quite large. An average increase in excess sterilization—from zero to about 3.5 times—increases the probability of a child under the age of five not being alive by about 2.7 percentage points, relative to a sample mean of 4.5\%. This suggests that exposure to the forced sterilization policy has a large effect (about 60\%) on child mortality. In Table F1 in the Appendix, we present estimates by examining alternative measures of forced sterilization measured by excess vasectomy. Again, the results are similar, suggesting that higher exposure to the forced sterilization policy positively affects child mortality. Finally, the results remain robust and quantitatively similar if we consider the mortality of male children only—to account for the issue of selective abortion, neglect of girl children, and problems associated with the missing girls’ phenomenon in India (see Table F2 in the Appendix).

\section{Conclusion}

In this paper, we have examined the importance of historical policies explaining vaccine hesitancy in India. Using multiple estimation methods, including an instrumental variable and a geographic regression discontinuity design approach, we examined whether the aggressive family planning program under which a forced sterilization policy was implemented between 1976 and 1977 could partly explain India’s lower vaccination rates today.

Our results provide robust evidence suggesting that historical events have had a strong impact on shaping India’s current vaccination paradox. This has implications for understanding the puzzling factors behind the lower demand for health-seeking behaviors, such as vaccination, even if the potential cost of morbidity and mortality is high and services are available for free. The findings from this paper also highlight the unintended consequences associated with domestic policies implemented in the past and the importance of understanding such contexts for the design and implementation of future interventions. We empirically show that a short-term policy failure (or success) could have spillover effects and could affect subsequent policies in the long run.

A key question is what generalizable lessons we can learn from this historical episode in India. The most direct parallel is to the countries that have implemented coercive domestic policies in the past. Coercive domestic policies, especially in the healthcare sector, are not uncommon. For example, sterilization without consent, primarily through coercive methods, has also been implemented in several countries, including Bangladesh, Brazil, China, Germany, Japan, South Africa, and the United States, to name a few (see \citet{Reilly2015Eugenics} for a detailed review).\footnote{Although anecdotal, Indiana state in the US was the first government body in the world to enact an involuntary sterilization law in 1907 \citep{Reilly1991Surgical}. It is also one of the states with the lowest COVID-19 vaccination rate in the US (Data from Our World in Data, Accessed on January 13, 2022).} Peru’s forced sterilization program during Alberto Fujimori’s regime between 1995 and 2000 is an excellent example that directly fits the Indian context. A number of countries also require compulsory vaccination; most notably, the current debates on COVID-19 vaccine mandates are considered involuntary and coercive \citep{Omer2019Mandate, Ward2019Vaccine, UN2021WHO:}.

What are the lessons for India today? Considering the recent experience of the COVID-19 pandemic, vaccination in India’s context is currently of substantive interest as about one-sixth of the world’s population lives there. Anecdotal evidence from India suggests that COVID-19 vaccine hesitancy has been high even after the surge in cases in early 2021 (due to the Delta variant), particularly in India’s vast rural hinterlands \citep{AssociatedPressNews2021Vaccine, BBC2021Covid, TheGuardian2021India}. Our results provide implications for policymakers and practitioners to understand the factors affecting India’s vaccination practices and carve out a pragmatic policy to maximize the uptake of the new vaccines. Perhaps more importantly, the current COVID-19 vaccination drive can be a perfect juncture for the government to provide accurate and reliable information on the importance and benefits of vaccines, which could eventually help reduce parents' hesitancy to vaccinate their children.

\clearpage
\bibliography{vaccination_ref.bib}
\bibliographystyle{chicago}



\clearpage
\centering
\large {\textbf{FIGURES AND TABLES}}

\vskip 1cm
\begin{figure}[htbp]
\begin{center}
\caption{\label{figure:Figure1}\textbf {Number of Sterilizations Performed in India (1956-82)}}
\includegraphics[width=\textwidth]{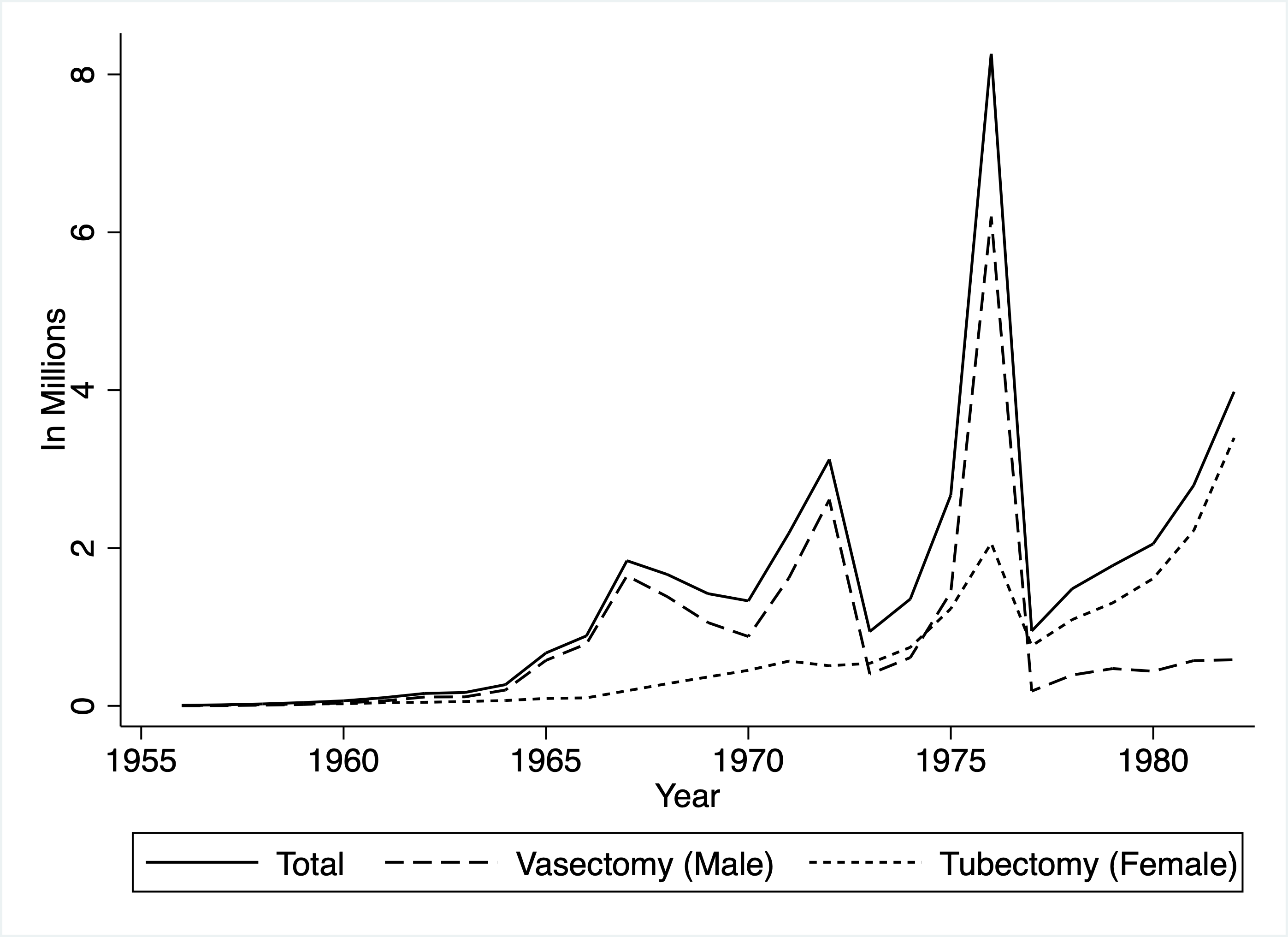}
\end{center}
{\footnotesize Notes: The figure presents the total number of sterilizations along with the types of sterilization performed in India every year since the beginning of the program in 1956. The solid line represents the total number of sterilizations performed every year. The dashed and short dashed lines represent the total number of vasectomies and tubectomies performed every year, respectively.}
\end{figure}


\clearpage
\begin{figure}[htbp]
\begin{center}
\caption{\label{figure:Figure2}\textbf{State-level Variation in Exposure to the Forced Sterilization policy}}
\subcaption{Panel A: Total Number of Sterilizations Performed in 1976-77 (in 100,000)}
\includegraphics[height=10cm]{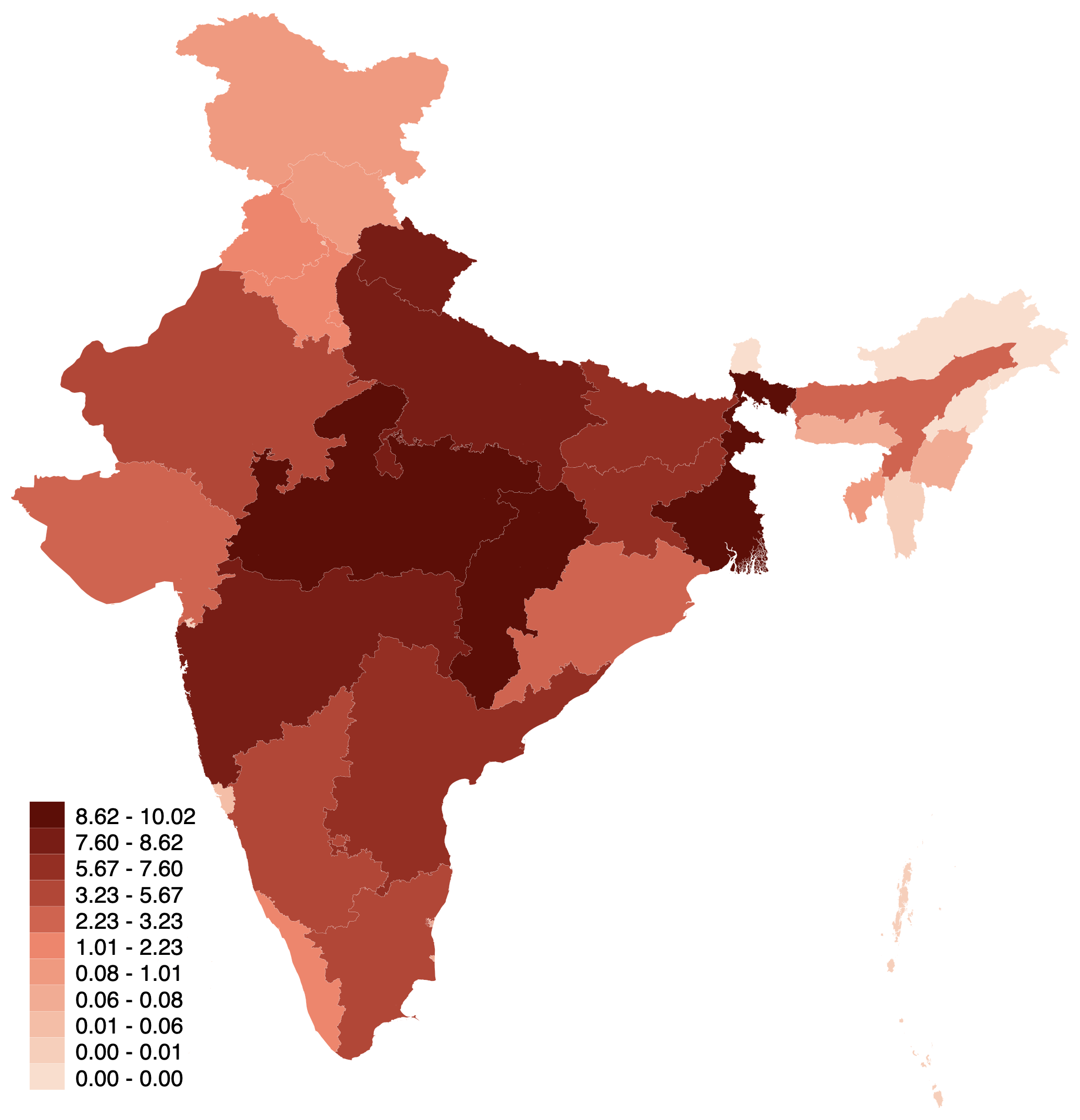}
\subcaption{Panel B: Excess Sterilizations Performed in 1976-77 (Normalized by 1975-76 Numbers)}
\includegraphics[height=10cm]{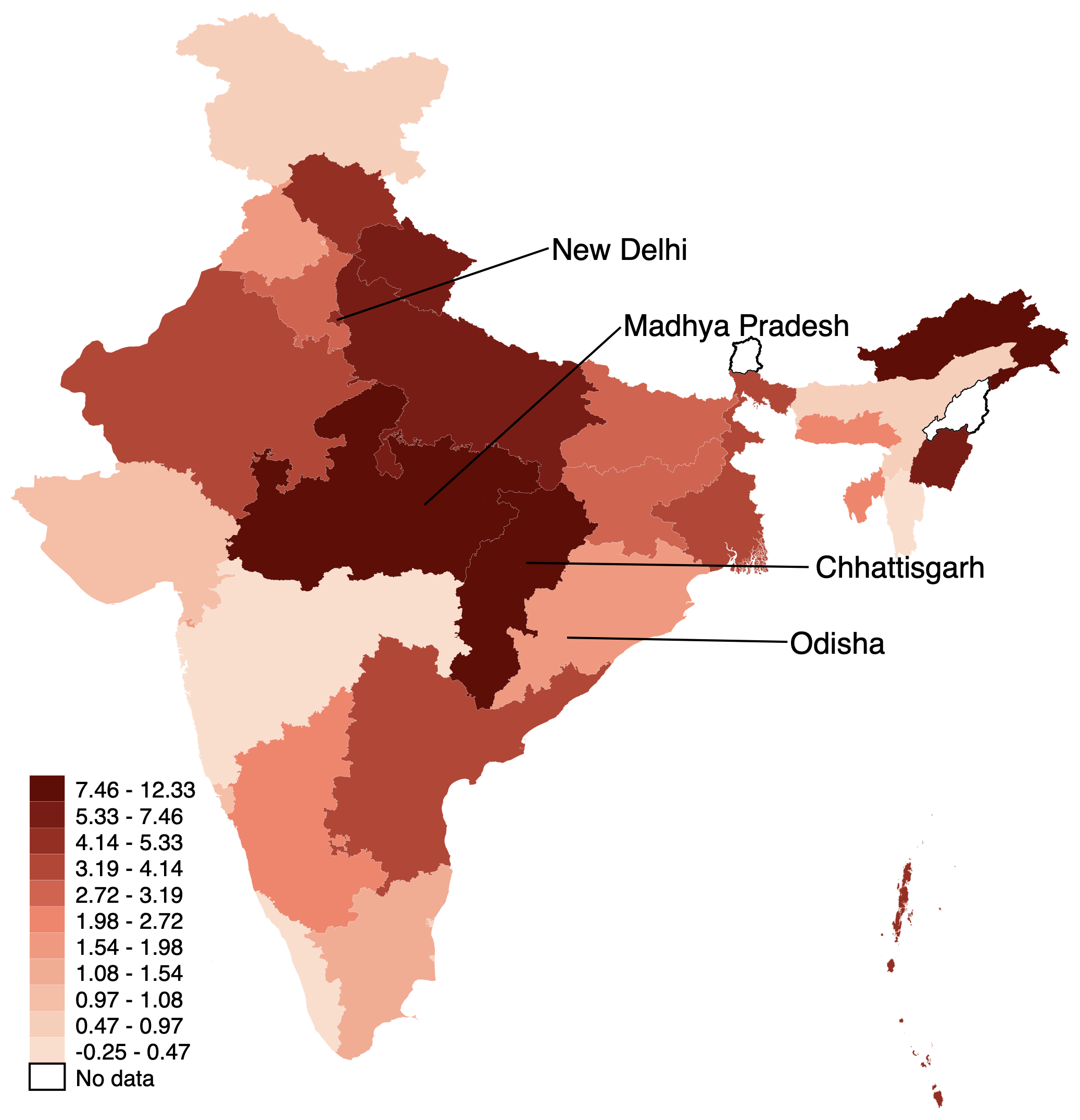}
\end{center}
{\footnotesize Notes: The figure presents the state-level variation in exposure to the forced sterilization policy. Panel A presents the Total Number of Sterilizations Performed in 1976-77 (expressed in 100,000). Panel B presents the excess sterilizations performed in 1976–77 normalized by performance in 1975–76. Darker shades denote a greater number of excess sterilizations performed.}
\end{figure}


\clearpage
\begin{figure}[htbp]
\begin{center}
\caption{\label{figure:Figure3}\textbf{Association Between Forced Sterilization and Vaccination}}
\subcaption{Panel A: Association between the total number of sterilizations performed in 1976-77 and vaccination index}
\includegraphics[height=8cm]{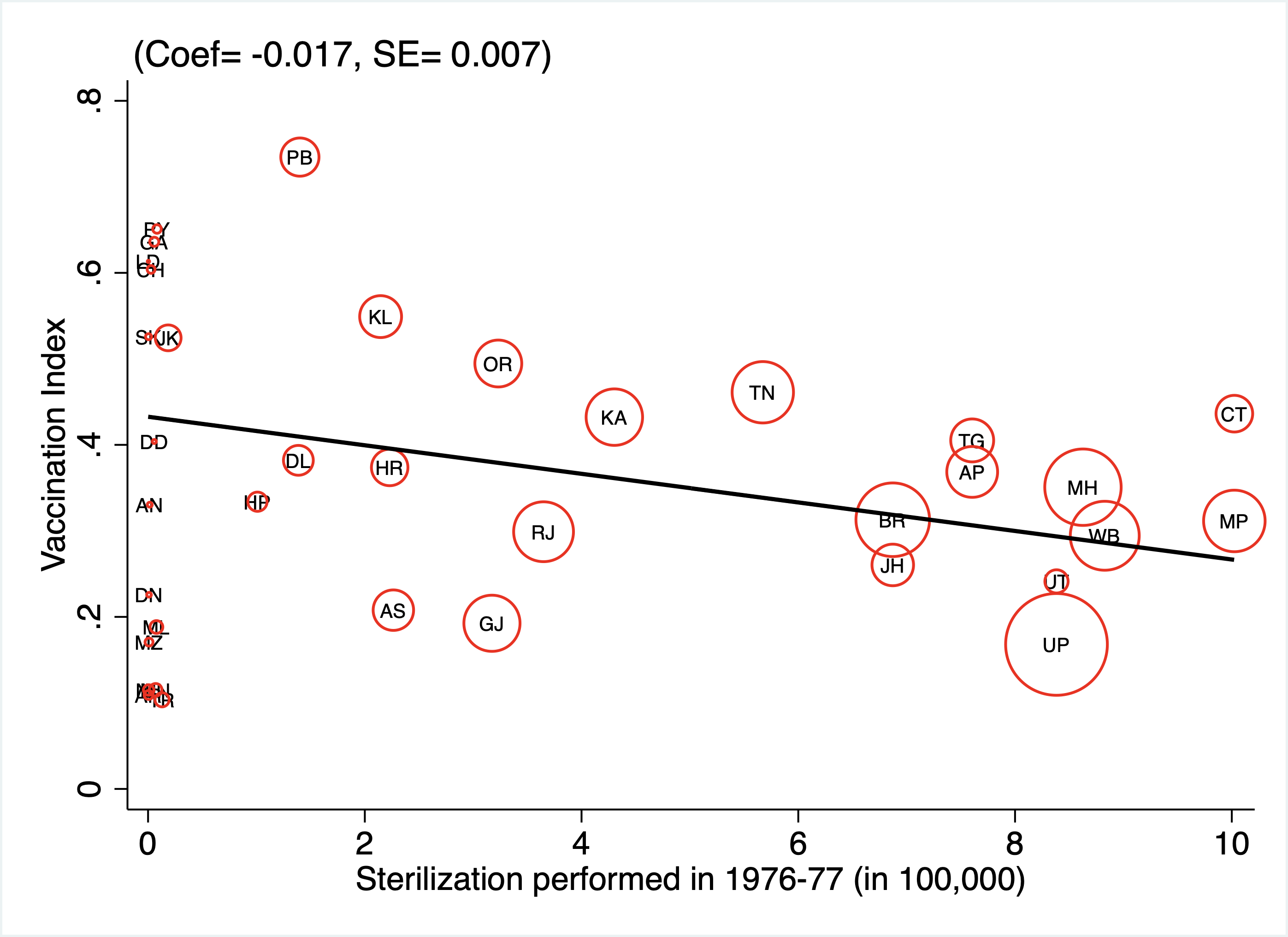}
\subcaption{Panel B: Association between excess sterilizations performed in 1976-77 normalized by 1975-76 figures and vaccination index}
\includegraphics[height=8cm]{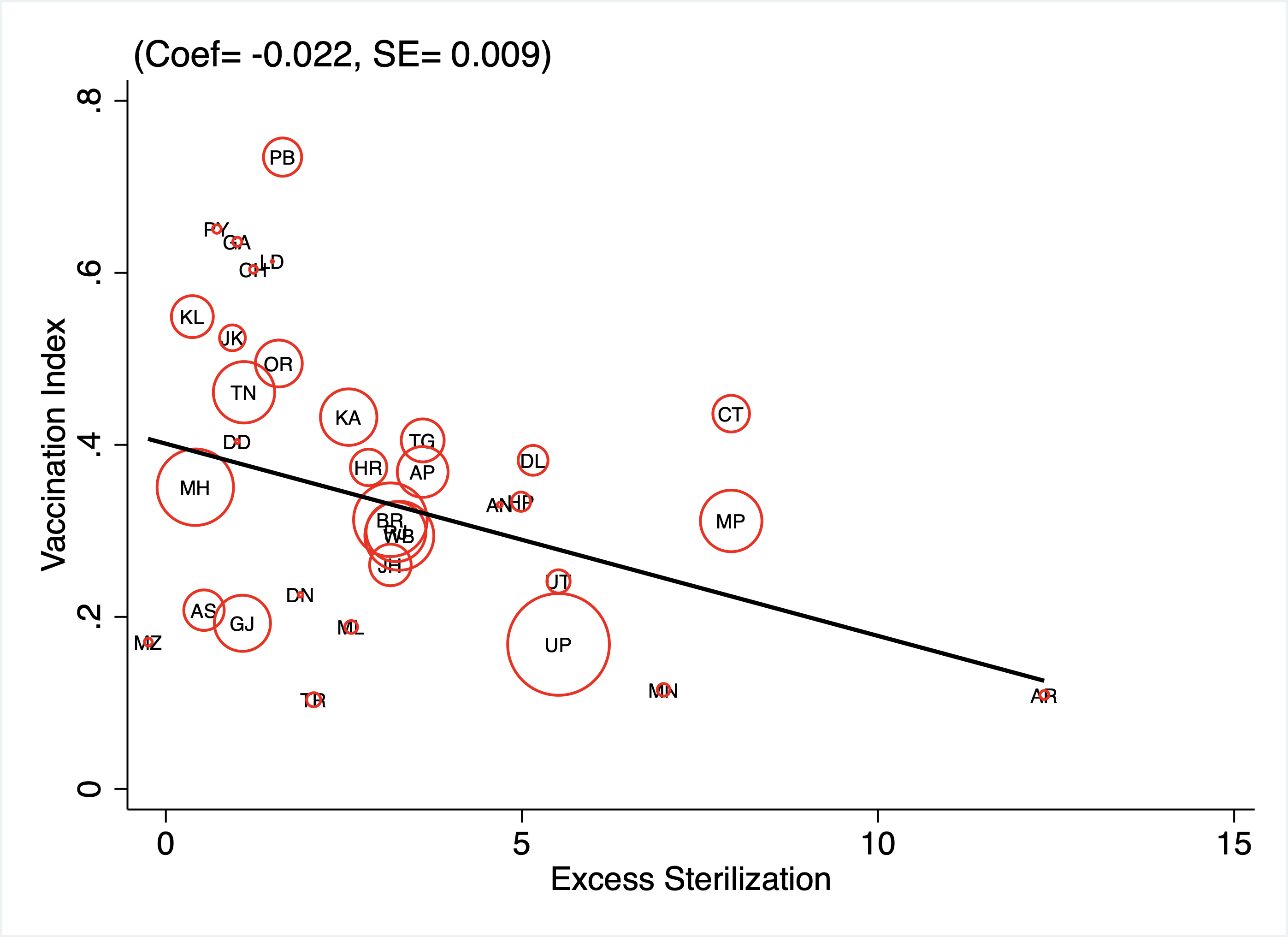}
\end{center}
{\footnotesize Notes: The figure presents correlation plots of exposure to the forced sterilization policy and the vaccination in 2015–16. Panel A presents the correlation between the state-level total number of sterilizations performed in 1976–77 and the vaccination index in 2015–16. Panel B presents the correlation between state-level excess sterilizations performed in 1976–77 and the vaccination index in 2015–16. The size of each circle denotes the population of the state and union territory. The fitted lines are weighted by the population of the state and union territory.}
\end{figure}


\clearpage
\begin{figure}[htbp]
\begin{center}
\caption{\label{figure:Figure4}\textbf{Distance from New Delhi as Instrument }}
\subcaption{Panel A: Association between distance from New Delhi to state capitals (in 100km) and excess sterilization in 1976-77}
\includegraphics[height=8cm]{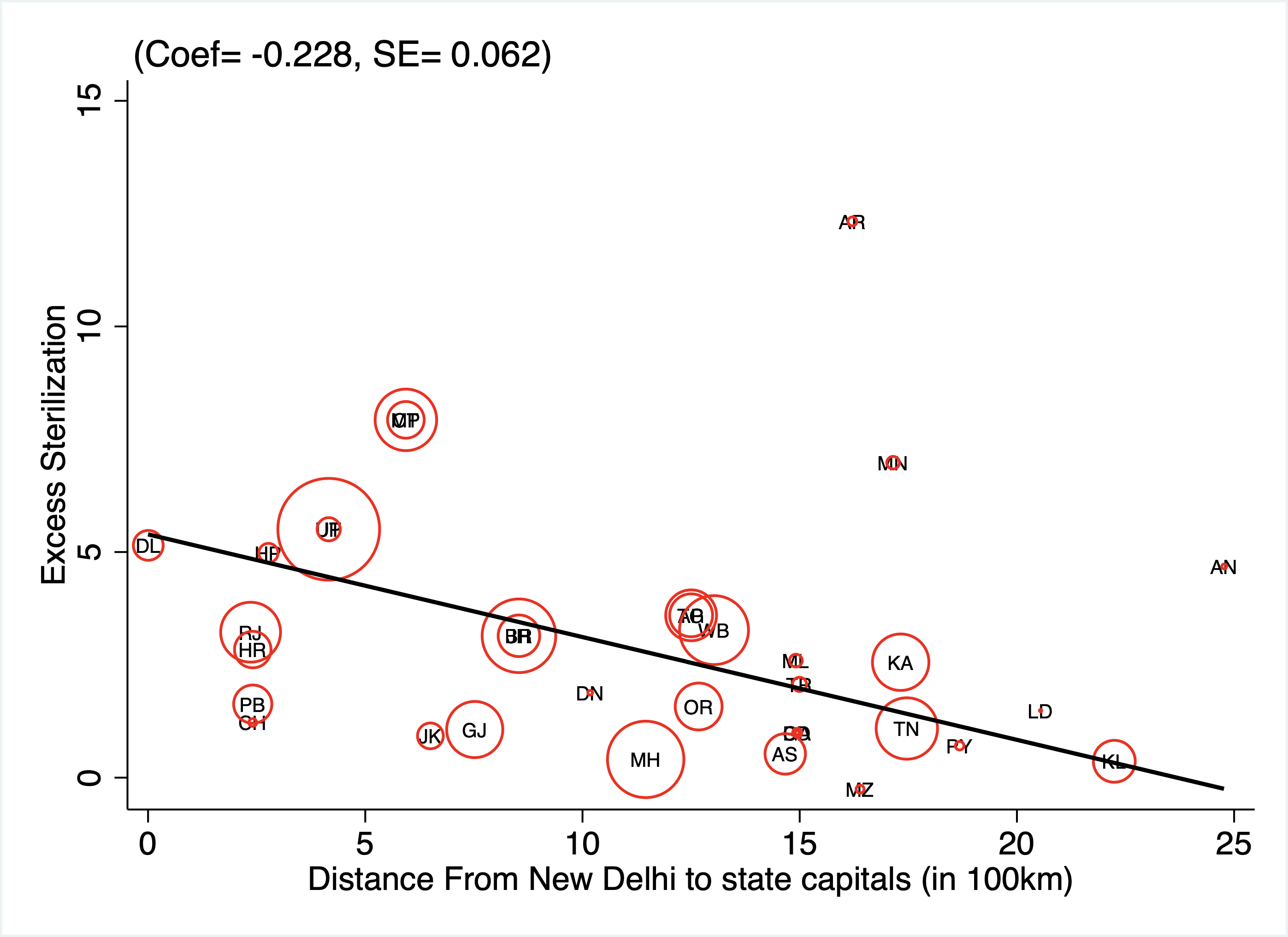}
\subcaption{Panel B: Association between distance from New Delhi to state capitals (in 100km) and excess sterilization in 1975-76 (\textit{previous year}).}
\includegraphics[height=8cm]{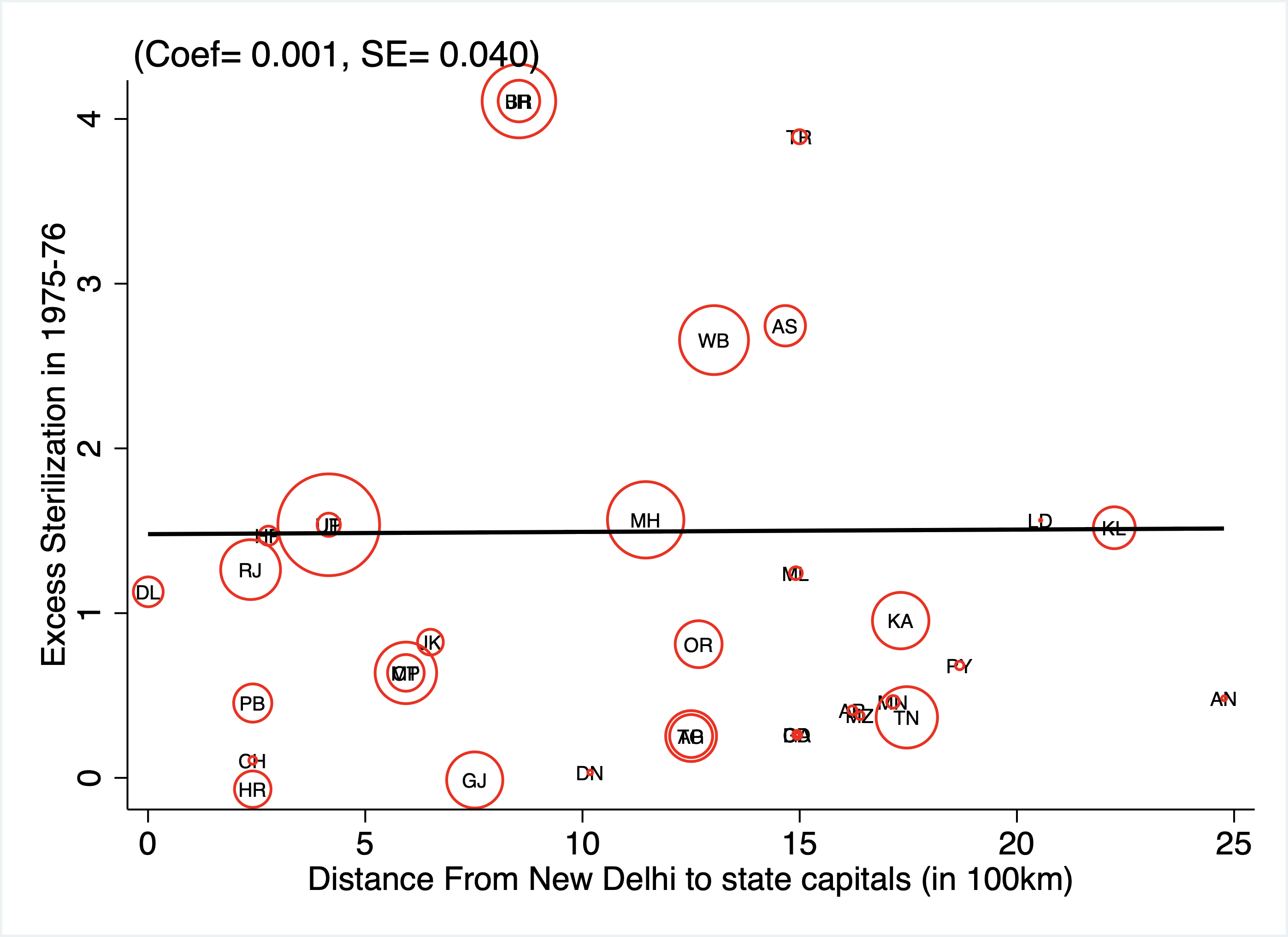}
\end{center}
{\footnotesize Notes: The figure presents the strength and exogeneity of our instrument. Panel A presents the correlation between the distance from New Delhi to state capitals (in 100km) and state-level excess sterilizations performed in 1976–77. Panel B presents the correlation between the distance from New Delhi to state capitals (in 100km) and state-level excess sterilizations performed in 1975–76. The size of each circle denotes the population of the state and union territory. The fitted lines are weighted by the population of the state and union territory.}
\end{figure}


\clearpage
\begin{figure}[htbp]
\begin{center}
\caption{\label{figure:Figure5}\textbf{Test of Validity of the Instrument}}
\includegraphics[width=\textwidth]{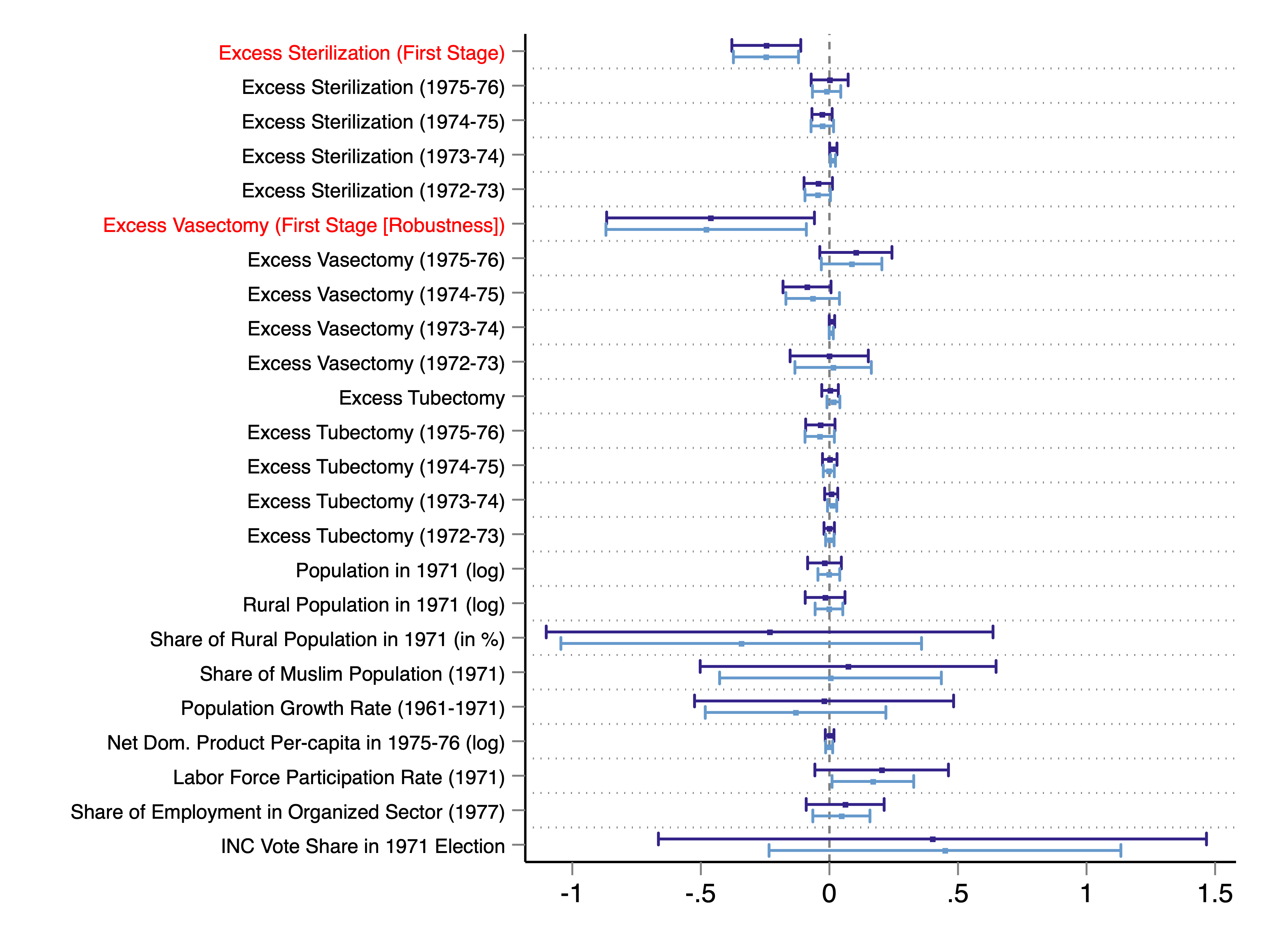}
\end{center}
{\footnotesize Notes: The figure presents several placebo exercises to test the validity of our instrument. The estimates correspond to the specifications from column 1 (top blue) and column 5 (bottom light blue) in Table 2. Please see the notes in Table 2. Each estimate comes from a separate regression. The explanatory variable is the distance from New Delhi to state capitals (expressed in 100 kilometers). The dots are the estimated coefficients, and the horizontal lines represent the 95 percent confidence intervals.}
\end{figure}

\clearpage
\begin{figure}[htbp]
\begin{center}
\caption{\label{figure:Figure6}\textbf{Alternative Measure of Forced Sterilization: INC's vote Share}}
\subcaption{Panel A: Association between excess sterilization in 1976-77 and INC's vote share in 1977 election}
\includegraphics[height=8cm]{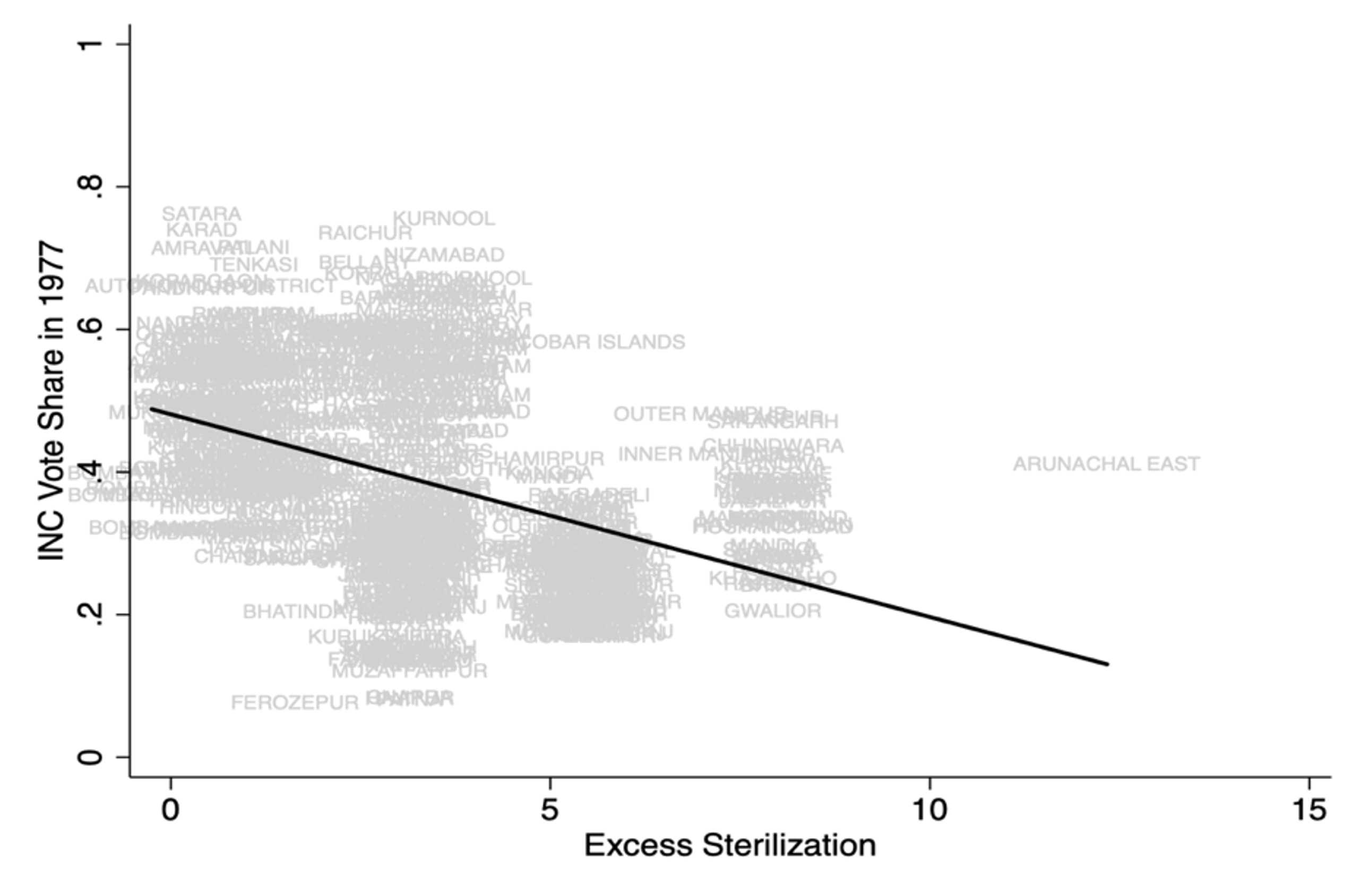}
\subcaption{Panel B: Panel B: Association between excess sterilization in 1976-77 and INC's vote share in 1971 election (\textit{previous election)}}
\includegraphics[height=8cm]{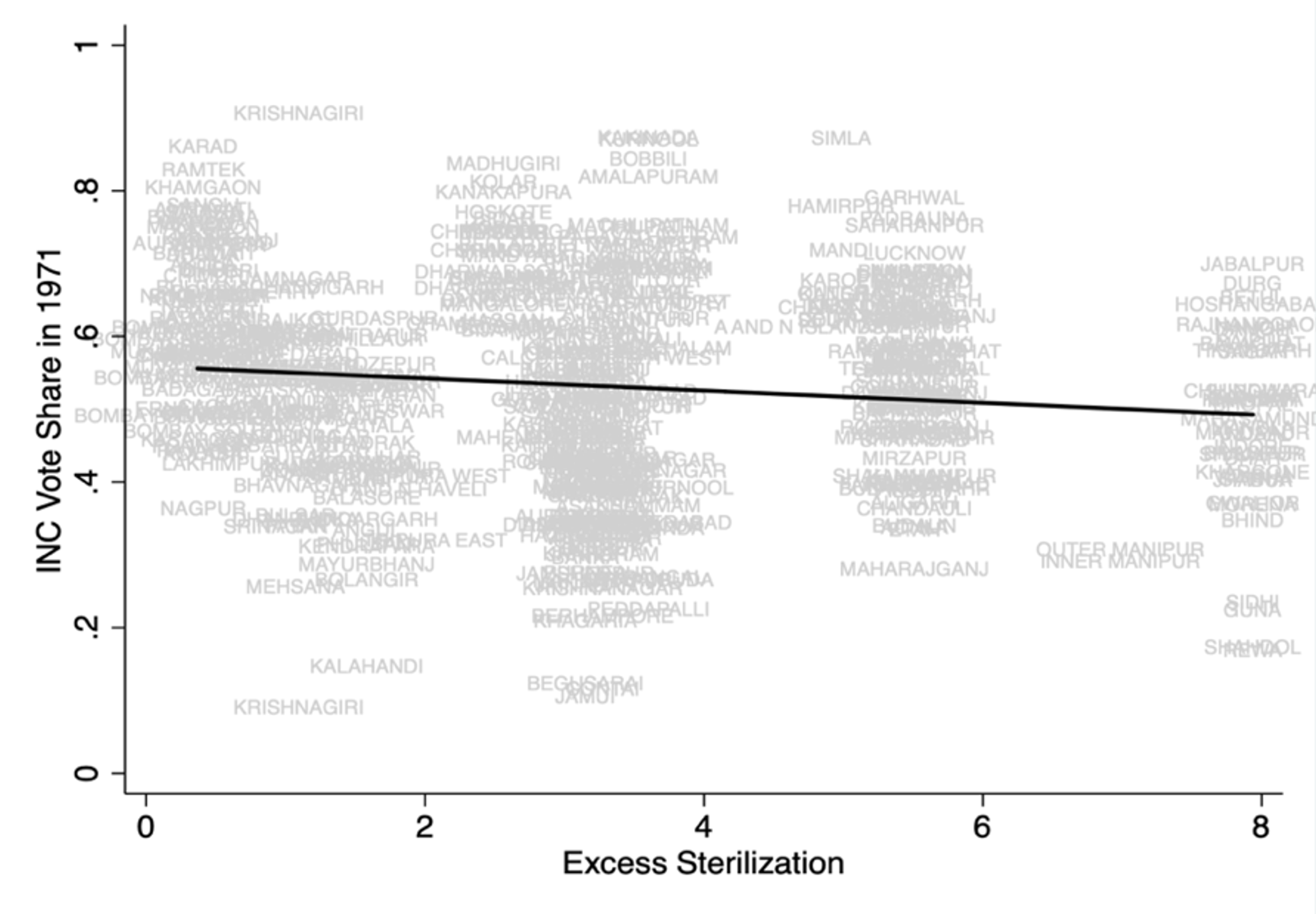}
\end{center}
{\footnotesize Notes: The figure presents an alternative measure of forced sterilization explained by the INC party's vote share. Panel A presents the correlation between state-level excess sterilizations performed in 1976–77 and the constituency-level variation in the INC party's vote share in the 1977 election. Panel B presents the correlation between state-level excess sterilizations performed in 1976–77 and the constituency-level variation in the INC party's vote share in 1971, the immediate election before 1977.}
\end{figure}


\clearpage
\begin{sidewaysfigure}[htbp]

\begin{center}
\caption[width=\textwidth]{\label{figure:Figure7}\textbf{RDD Covariate Balance Tests Between Odisha and Chhattisgarh}}
\includegraphics[height=13cm]{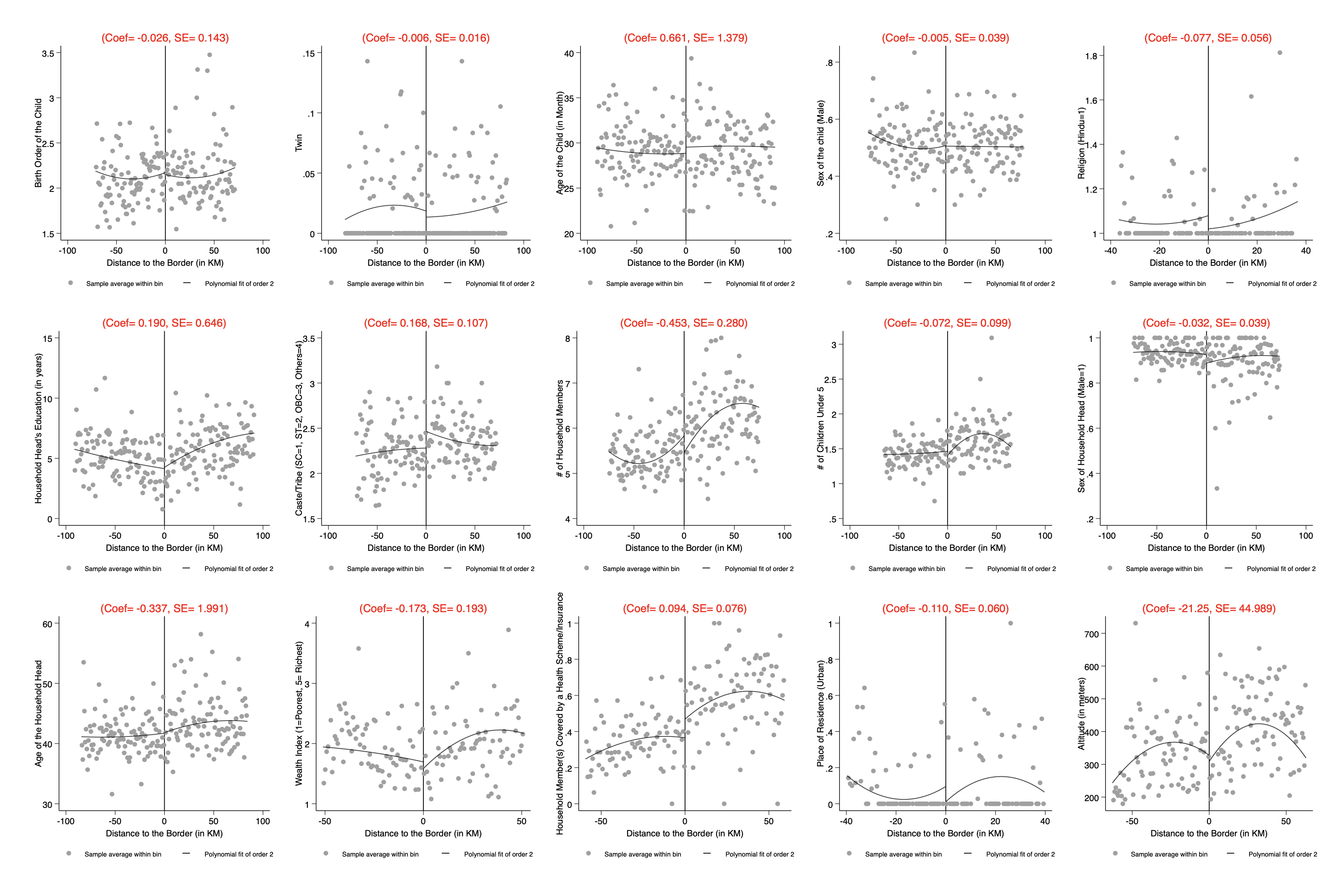}
\end{center}
{\footnotesize Notes: The figure presents the balance in baseline covariates using the RDD approach. All the figures use optimal bandwidths based on the data-driven method proposed by Calonico et al. (2015). We set the local polynomial of order 2, which has the minimum asymptotic mean squared error (MSE) of the RD point estimator in our sample, as proposed by Pei et al. (2021). We use a uniform kernel, as suggested by Imbens and Lemieux (2008) and Lee and Lemieux (2010), which is simple, transparent, and easy to interpret. (Use of triangular kernel also produces similar results.) Left-hand side of each graph presents the samples of Odisha, and the right-hand side presents the samples of Chhattisgarh. }

\end{sidewaysfigure}


\clearpage
\begin{figure}[htbp]
\begin{center}
\caption{\label{figure:Figure8}\textbf{Regression Discontinuity Design (Odisha and Chhattisgarh)}}
\subcaption{Panel A: General RDD Approach}
\includegraphics[height=8cm]{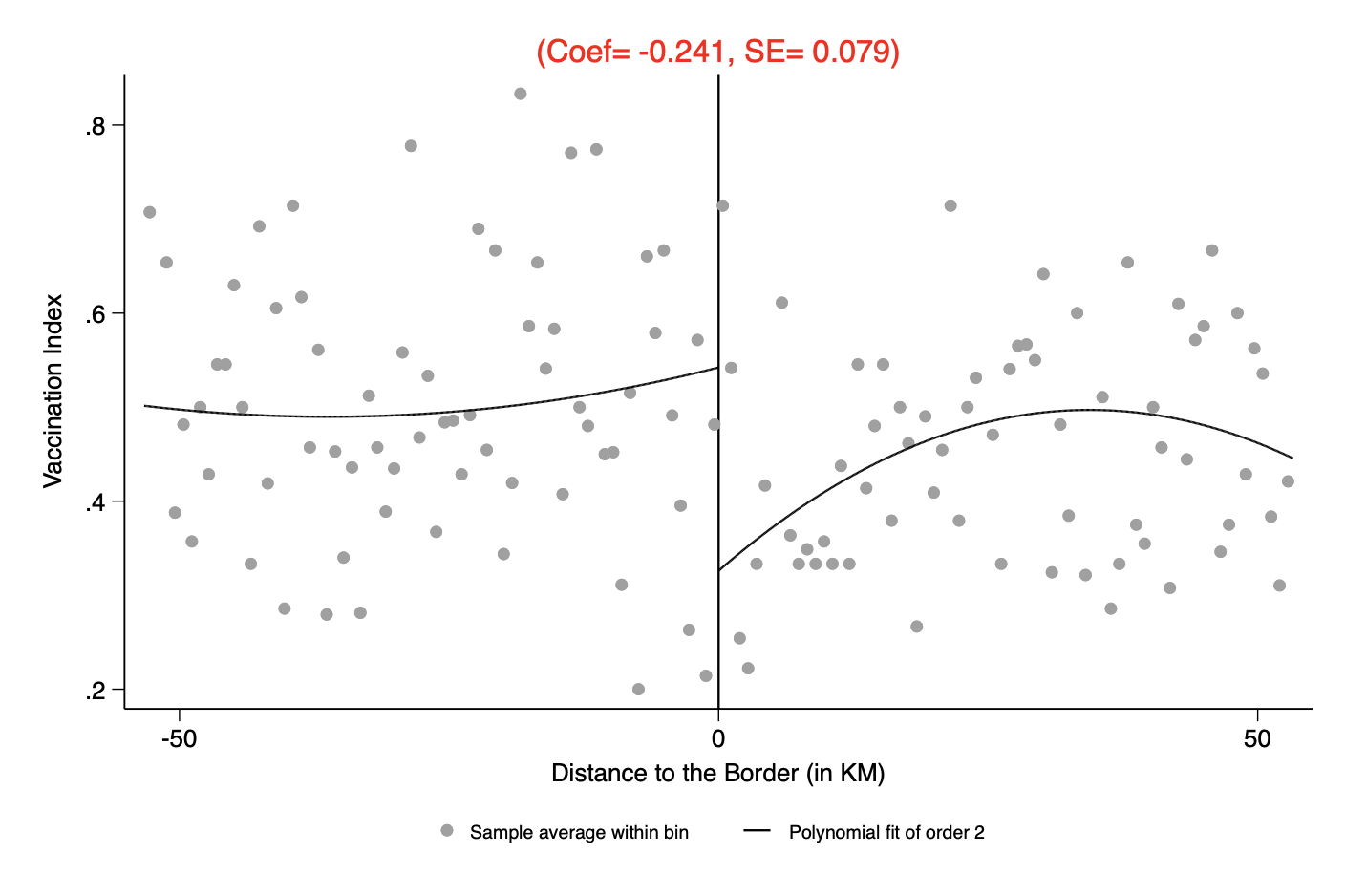}
\subcaption{Panel B: Donut Hole RD Approach (Excluding the observations within 5km from the border)}
\includegraphics[height=8cm]{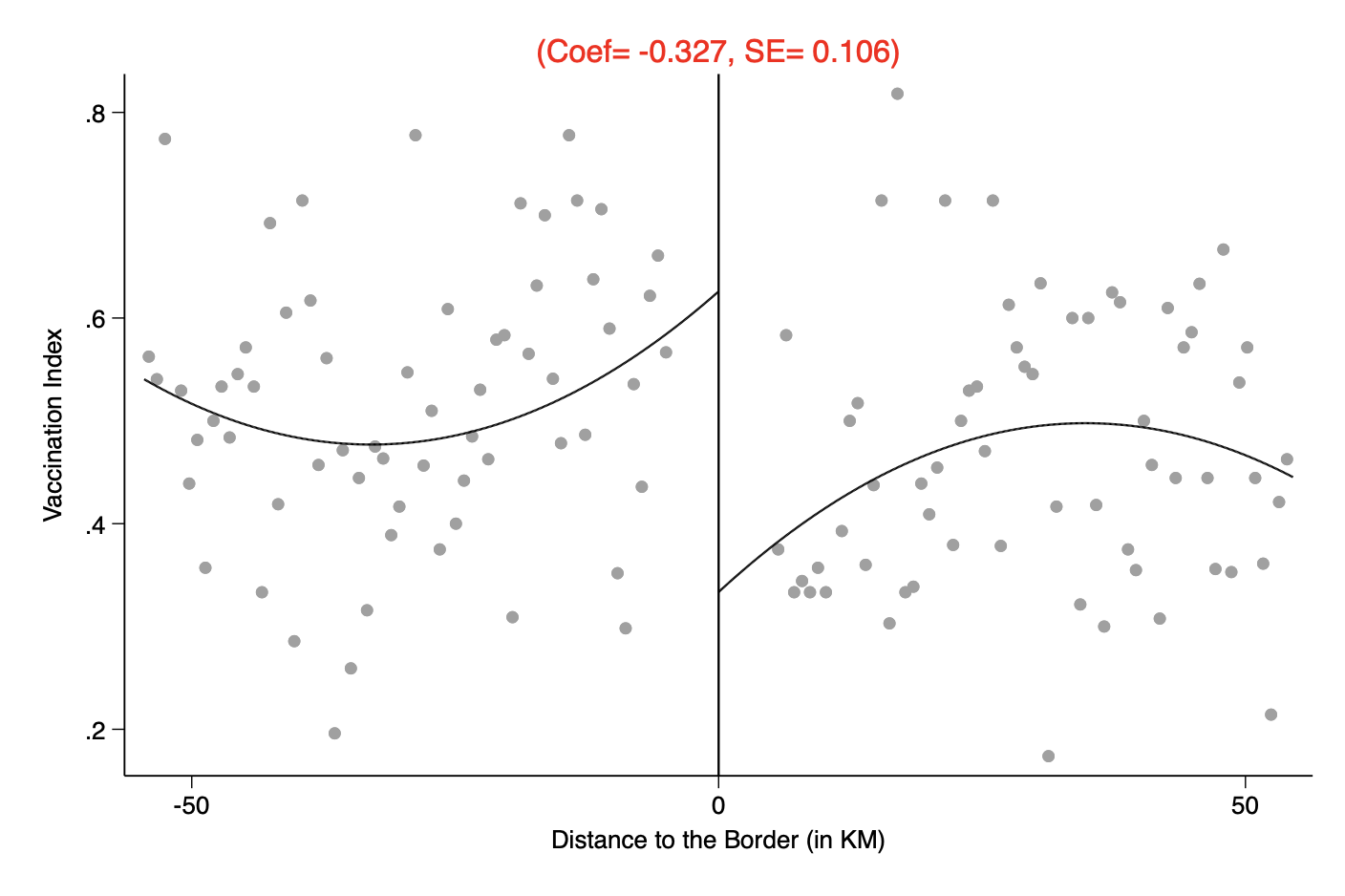}
\end{center}
{\footnotesize Notes: The figure presents the RDD results between Odisha and Chhattisgarh. Panel A presents the result using the general RDD approach. Panel B presents the result using the donut hole RDD approach, excluding the observations within 5 km of the border due to the random allocation of the GPS coordinates up to 5 km in the NFHS-4 survey. The left-hand side of each graph presents the samples of Odisha, and the right-hand side presents the samples of Chhattisgarh. All the figures use optimal bandwidths based on the data-driven method proposed by Calonico et al. (2015). We set the local polynomial of order 2, which has the minimum asymptotic mean squared error (MSE) of the RD point estimator in our sample, as proposed by Pei et al. (2021). We use a uniform kernel, as suggested by Imbens and Lemieux (2008) and Lee and Lemieux (2010), which is simple, transparent, and easy to interpret. (Use of triangular kernel also produces similar results.)}
\end{figure}


\clearpage
\begin{figure}[htbp]
\begin{center}
\caption{\label{figure:Figure9}\textbf{Heterogenous Effects (NFHS-4)}}
\includegraphics[width=\textwidth]{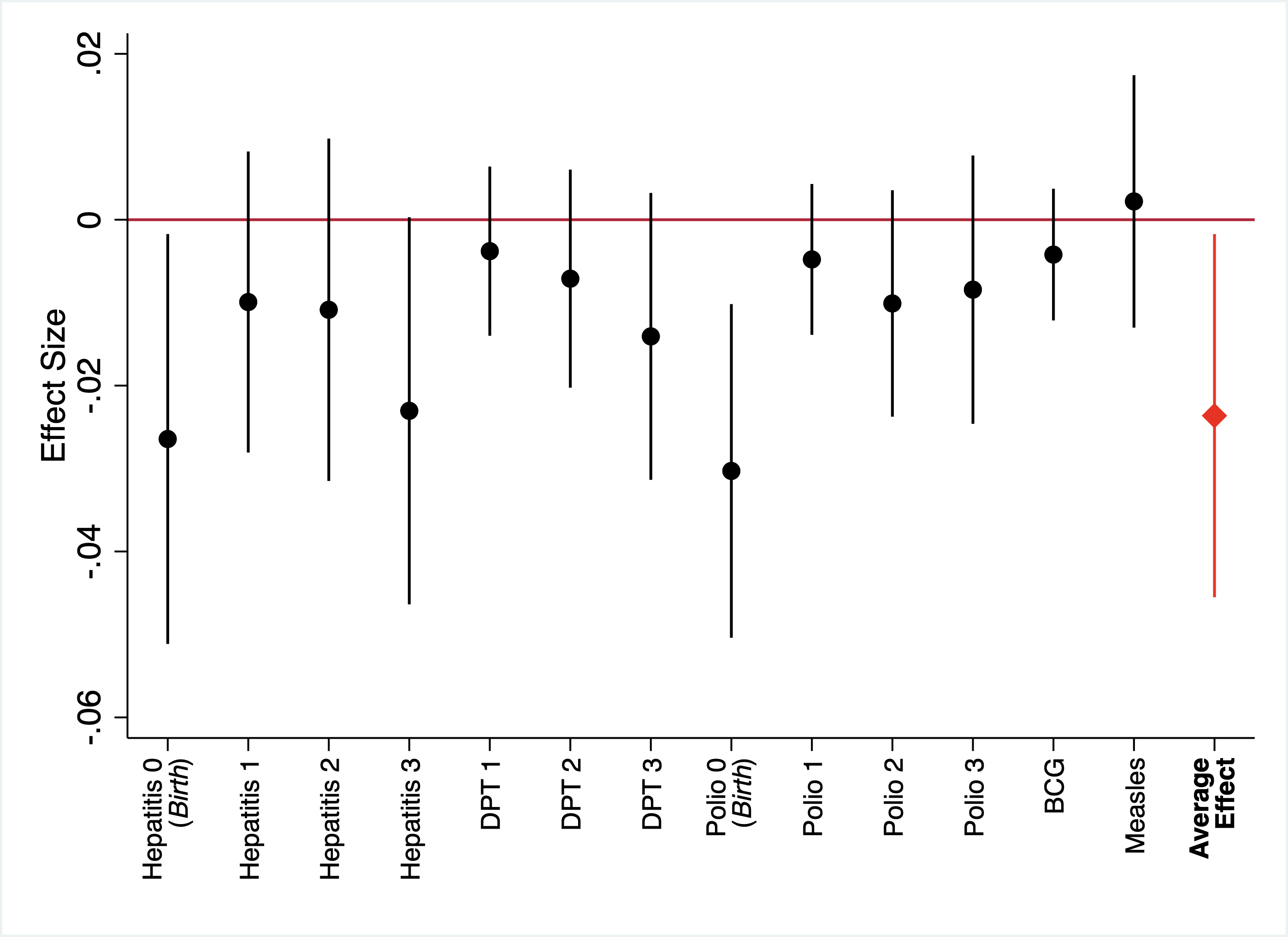}
\end{center}
{\footnotesize Notes: The figure presents the regression coefficients of each vaccine and the average effect size. Each estimate comes from a separate IV regression. The explanatory variable is the excess sterilizations performed in 1976-77 (compared with 1975-76 numbers) normalized by the sterilization performed in 1975-76 at the state level. The dots are the estimated coefficients, and the vertical lines represent the 95\% confidence intervals. See Appendix Table D1 for more information on variable definitions and for the results in table format.}
\end{figure}


\clearpage
\begin{figure}[htbp]
\begin{center}
\caption{\label{figure:Figure10}\textbf{Heterogeneity: Evolution over Time}}
\subcaption{Panel A: Heterogenous Effects from NFHS-1 (1992-93)}
\includegraphics[height=6cm]{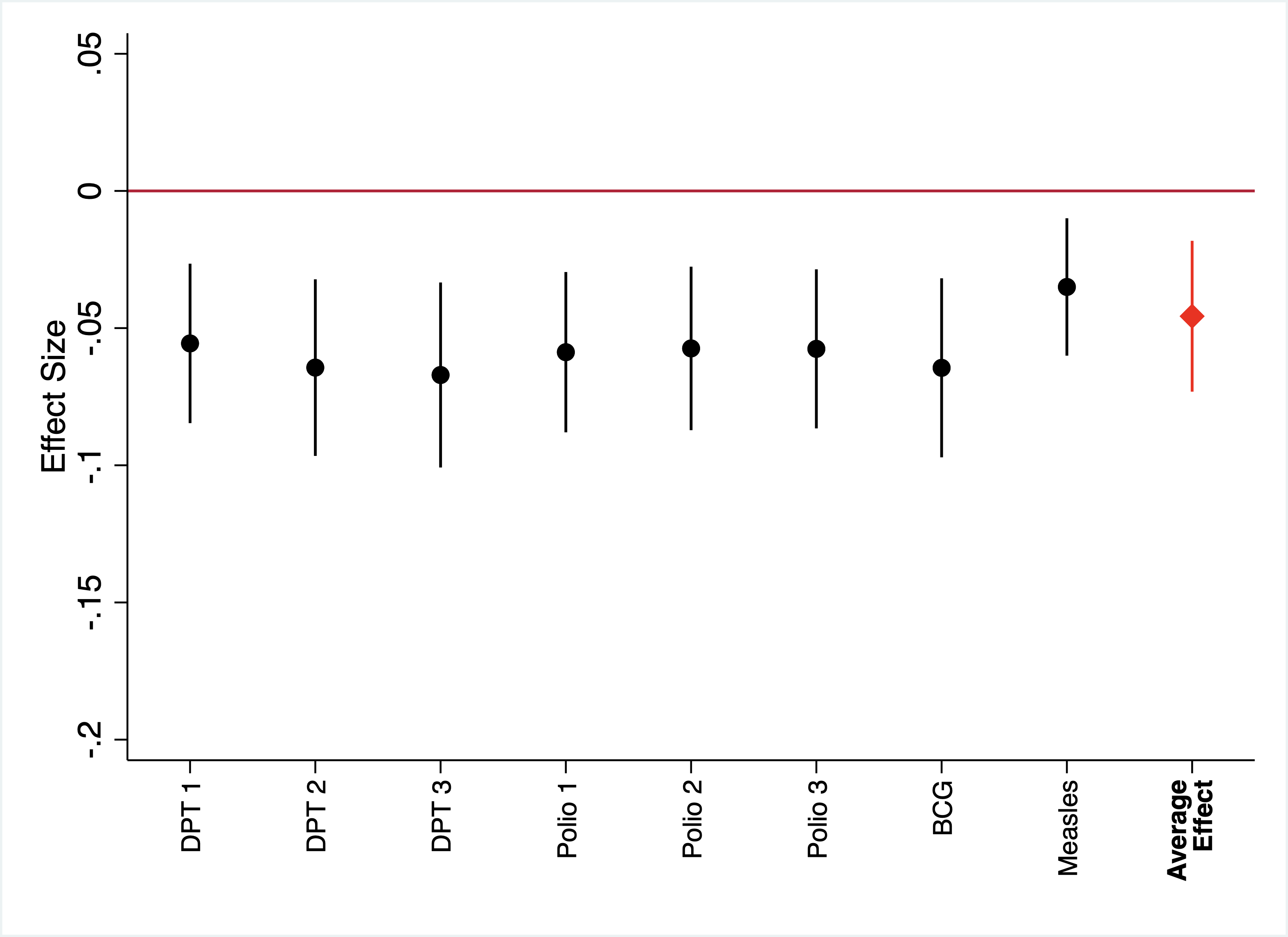}
\subcaption{Panel B: Heterogenous Effects from NFHS-2 (1998-99)}
\includegraphics[height=6cm]{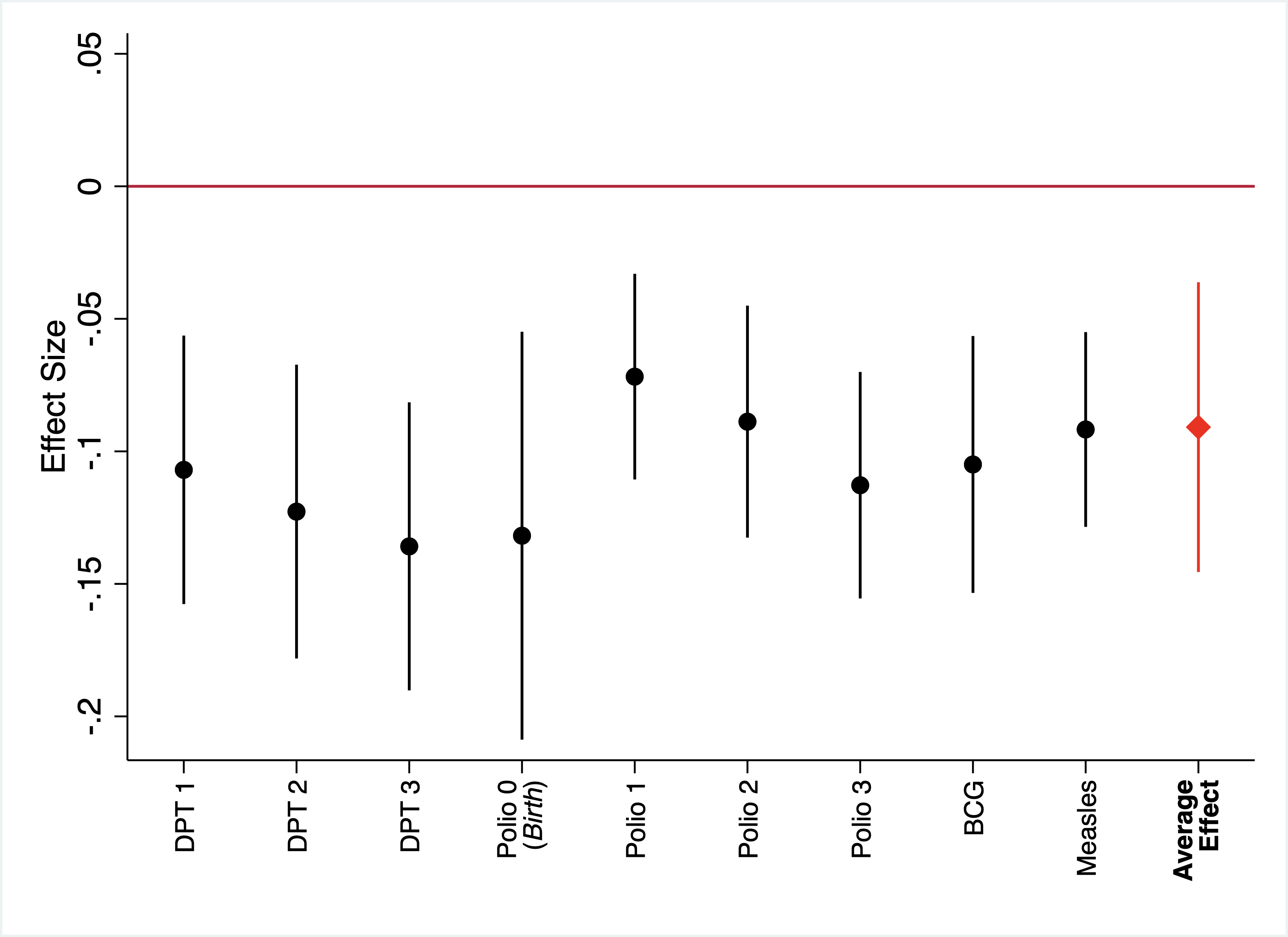}
\subcaption{Panel C: Heterogenous Effects from NFHS-3 (2005-06)}
\includegraphics[height=6cm]{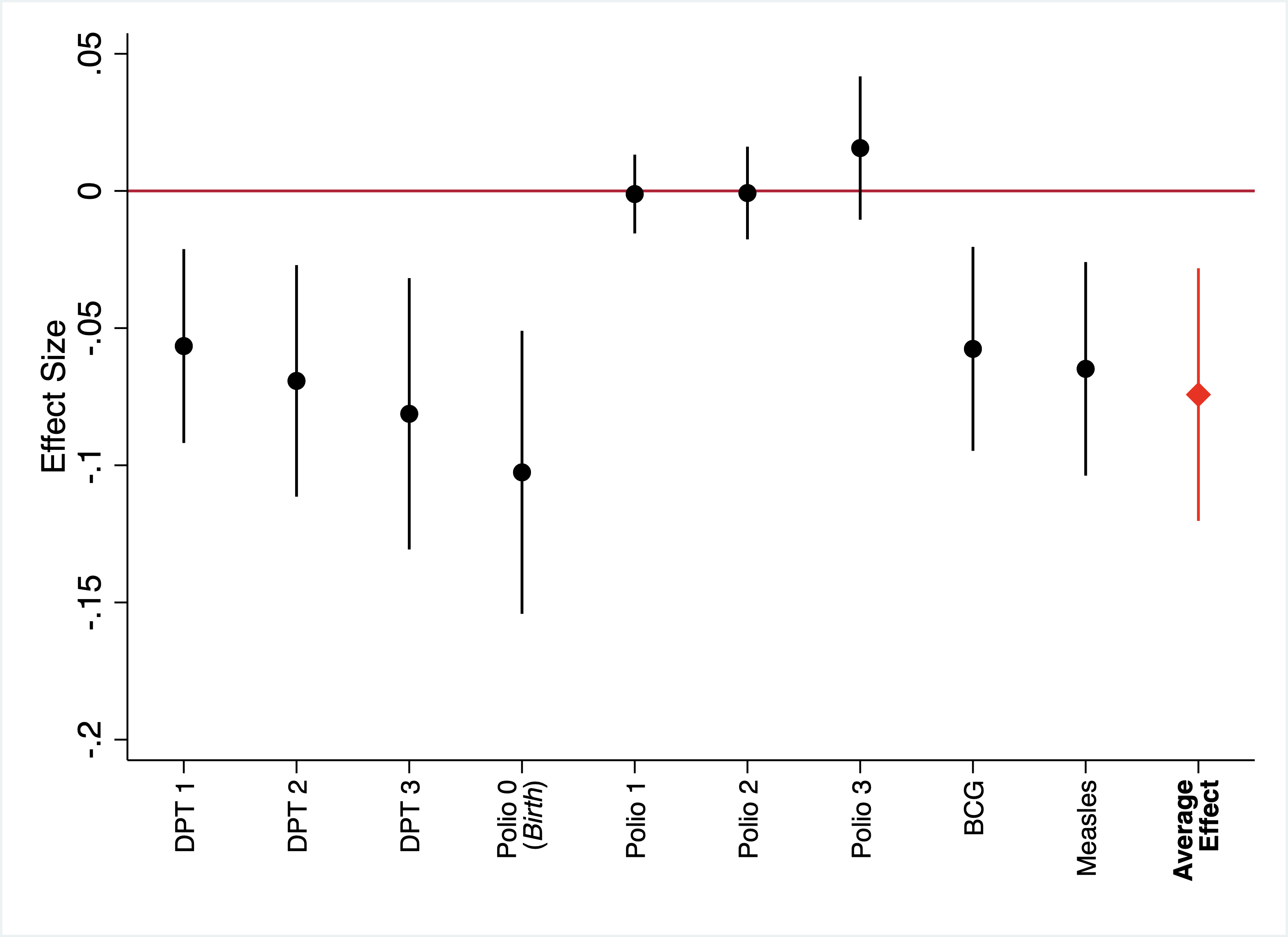}
\end{center}
{\footnotesize Notes: The figure presents the regression coefficients of each vaccine and the average effect size of the NFHS-1, NFHS-2, and NFHS-3 surveys. Each estimate comes from a separate IV regression. The explanatory variable is the excess sterilizations performed in 1976-77 (compared with 1975-76 numbers) normalized by the sterilization performed in 1975-76 at the state level. The dots are the estimated coefficients, and the vertical lines represent the 95\% percent confidence intervals. See Appendix Table D4-D6 for more information on variable definitions and for the results in table format.}
\end{figure}


\clearpage
\begin{figure}[htbp]
\begin{center}
\caption{\label{figure:Figure11}\textbf{Mechanism - Non-institutional Delivery}}
\includegraphics[width=\textwidth]{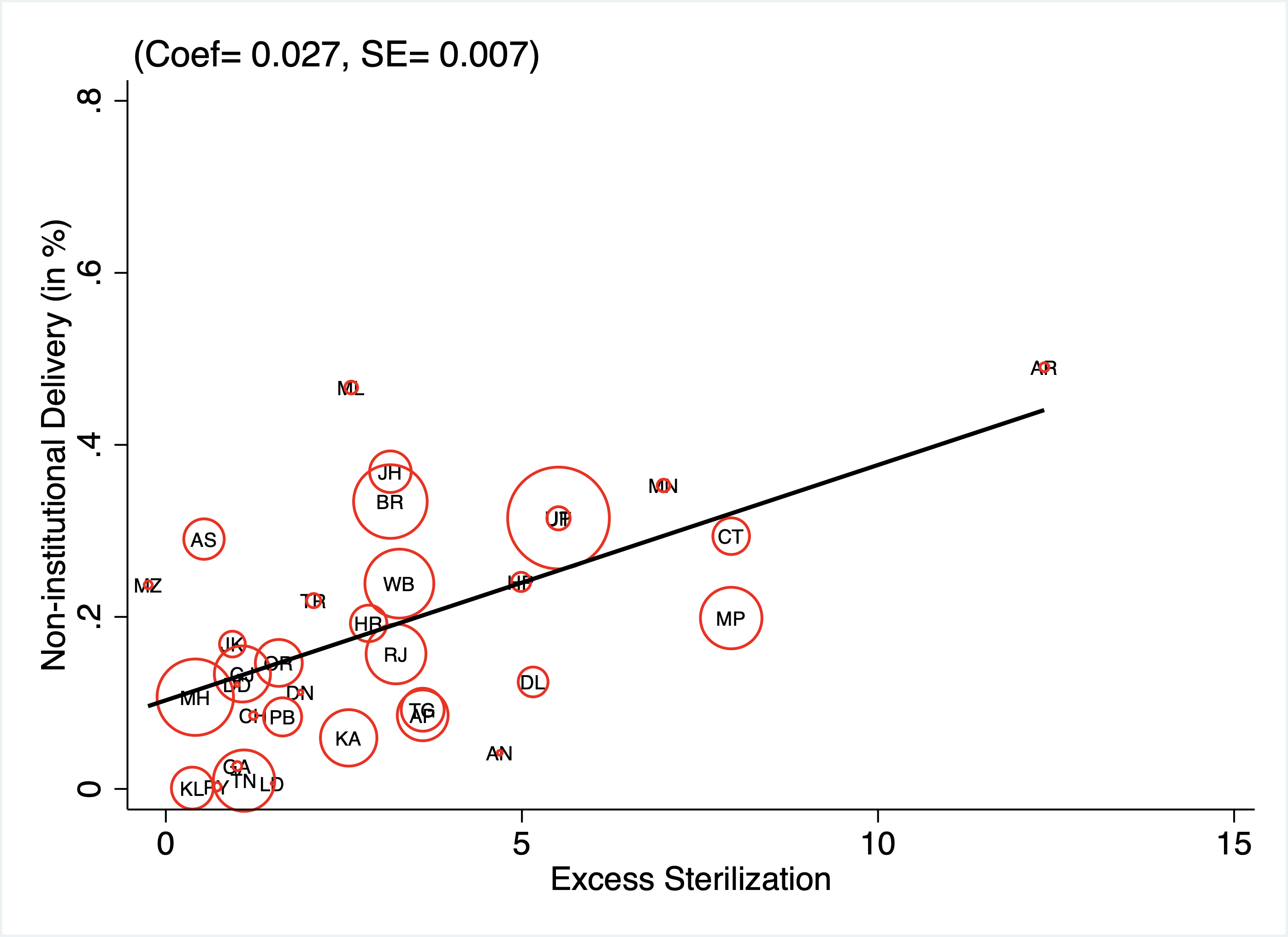}
\end{center}
{\footnotesize Notes: The figure presents the association between state-level exposure to the forced sterilization policy and the percentage of children who have had non-institutional delivery. The size of each circle denotes the population of the state and union territory. The fitted line is weighted by the population of the state and union territory.}
\end{figure}


\clearpage
\begin{figure}[htbp]
\begin{center}
\caption{\label{figure:Figure12}\textbf{Mechanism - Reasons for Non-institutional Delivery}}
\includegraphics[width=\textwidth]{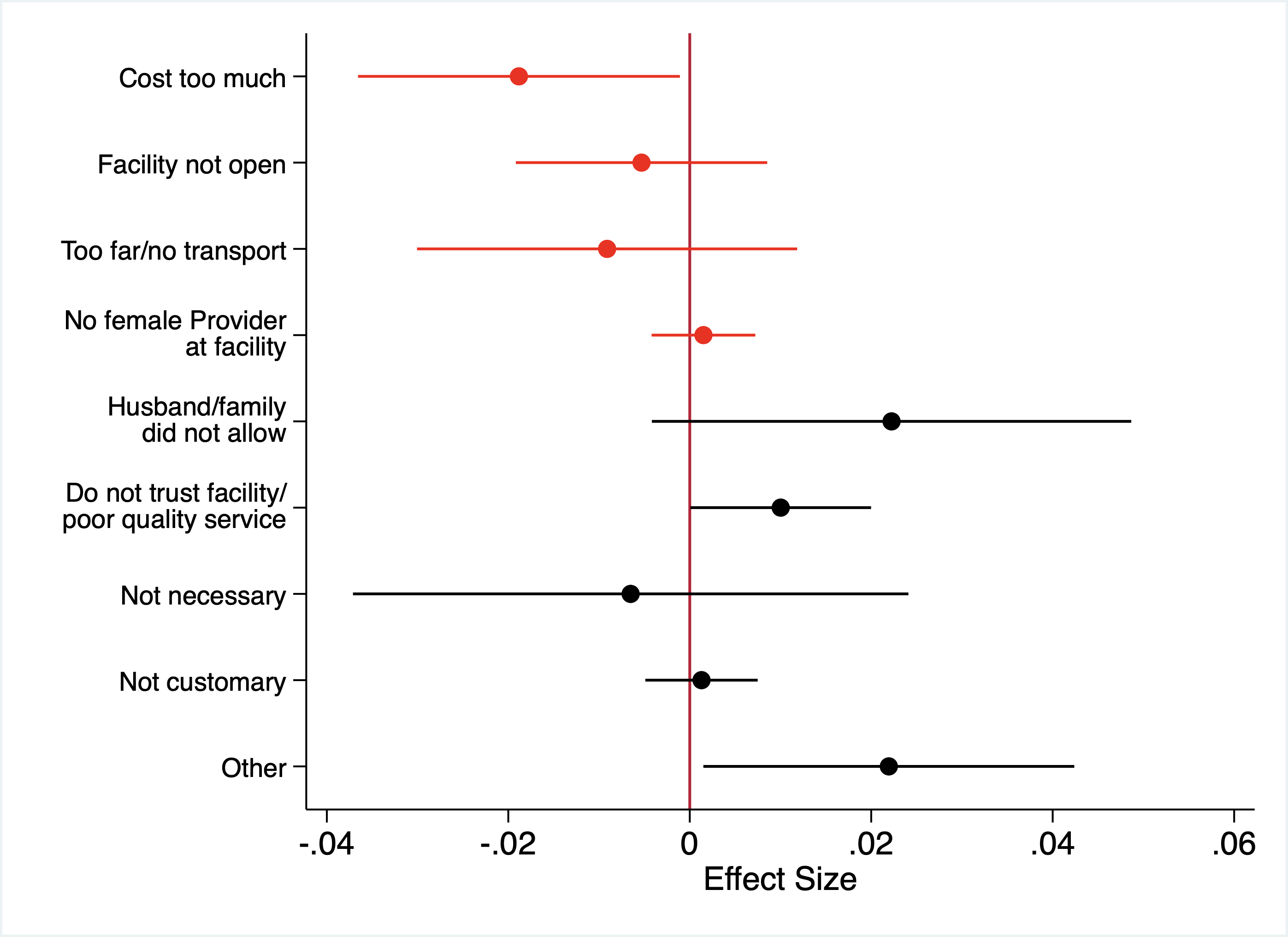}
\end{center}
{\footnotesize Notes: The figure presents the regression coefficients of each reason for non-institutional delivery. Each estimate comes from a separate IV regression. The explanatory variable is the excess sterilizations performed in 1976-77 (compared with 1975-76 numbers) normalized by the sterilization performed in 1975-76 at the state level. The dots are the estimated coefficients, and the horizontal lines represent the 95\% confidence intervals. Labels in red and black are supply-side reasons and demand-side reasons, respectively. See Appendix Table E2 for more information on variable definitions and for the results in table format.}
\end{figure}


\clearpage
\begin{figure}[htbp]
\begin{center}
\caption{\label{figure:Figure13}\textbf{Mechanism - Antenatal Care (ANC)}}
\subcaption{Panel A: Association between excess sterilization and the probability of receiving ANC }
\includegraphics[height=8cm]{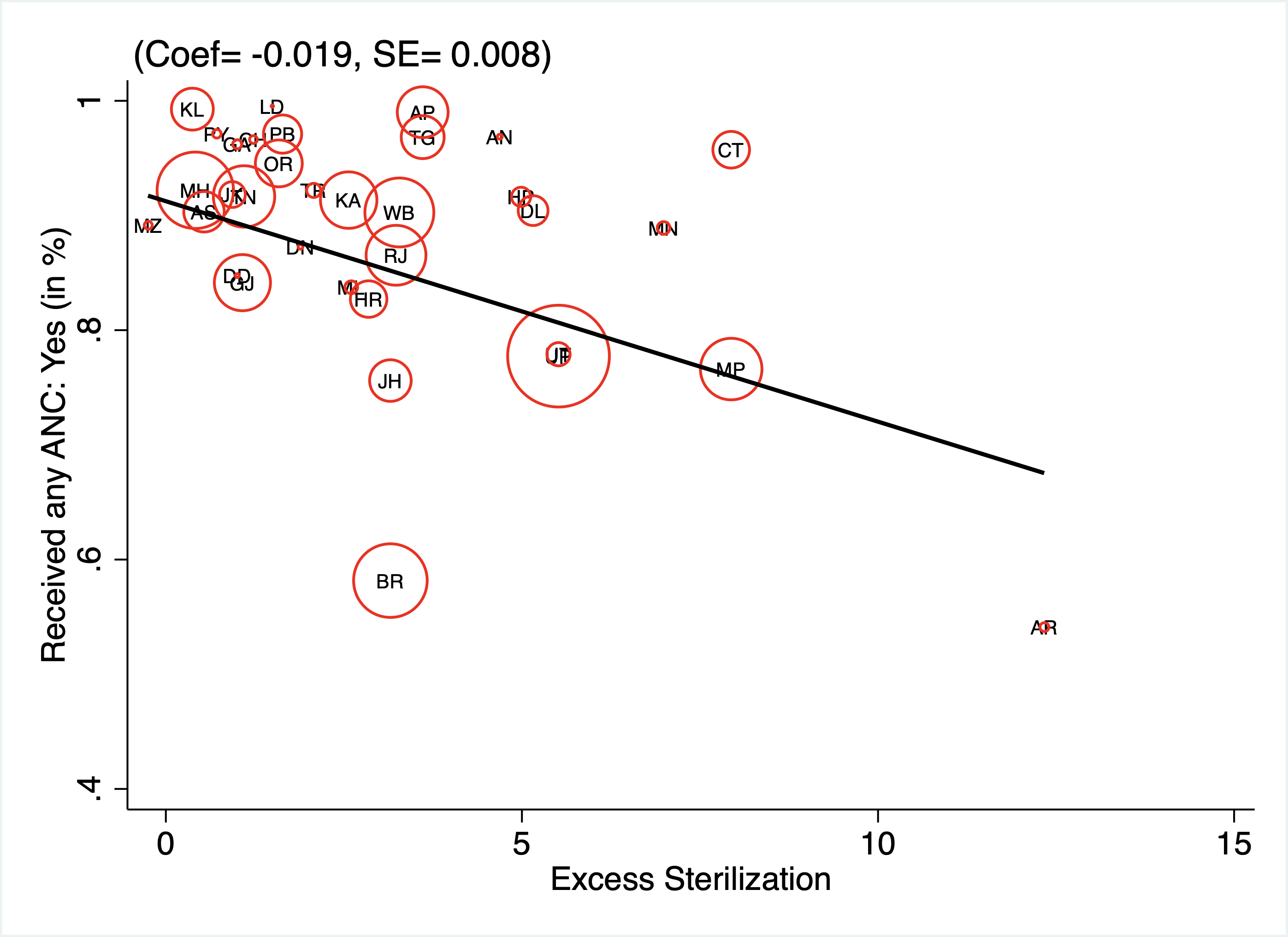}
\subcaption{Panel B: Association between excess sterilization and the number of visits conditional on receiving ANC}
\includegraphics[height=8cm]{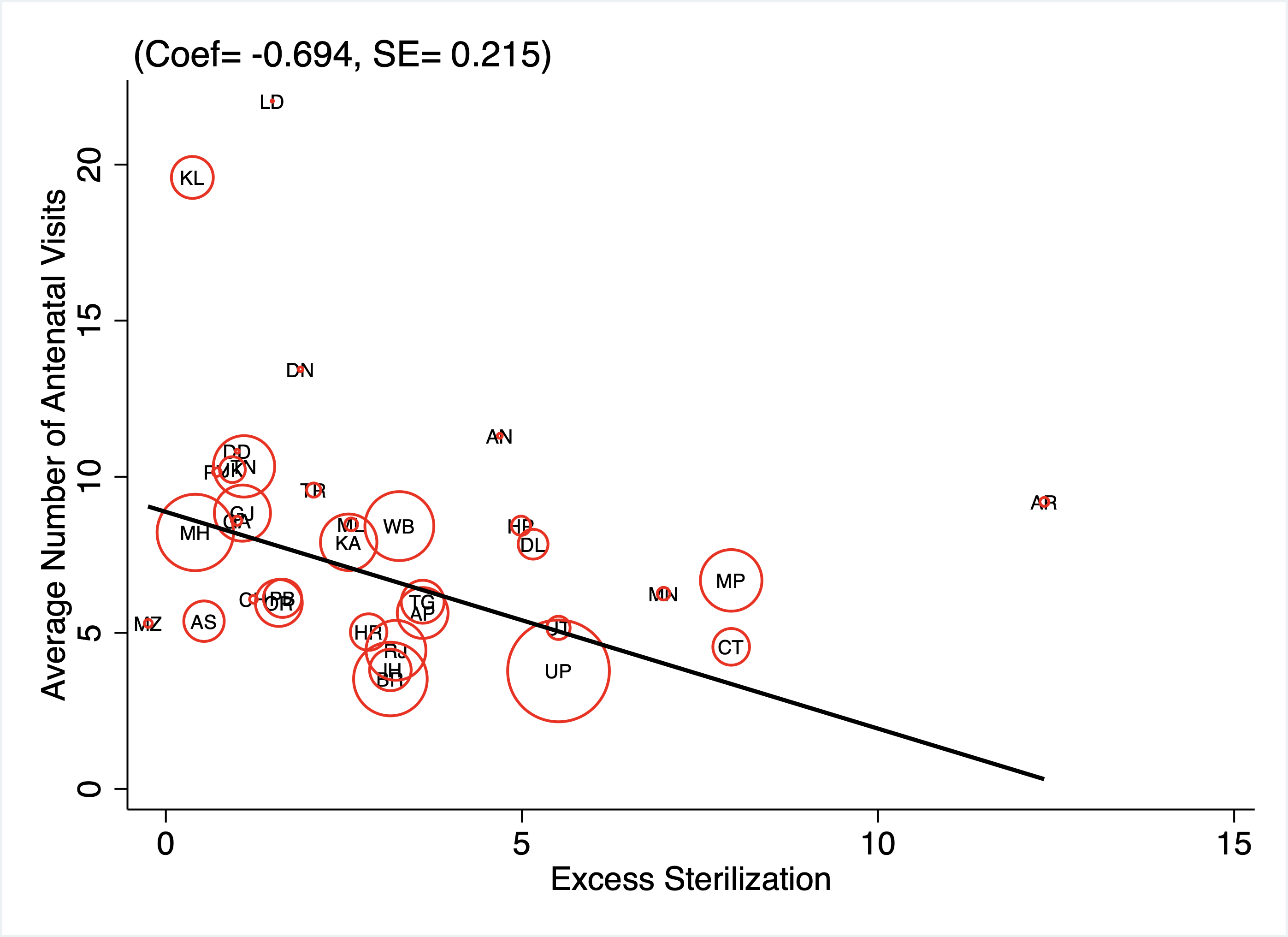}
\end{center}
{\footnotesize Notes: The figure presents the association between exposure to the forced sterilization policy on antenatal care (ANC). Panel A presents the association between state-level excess sterilizations performed in 1976–77 and the probability of receiving ANC. Panel B presents the correlation between state-level excess sterilizations performed in 1975–76 and the number of visits conditional on receiving any ANC. The size of each circle denotes the population of the state and union territory. The fitted lines are weighted by the population of the state and union territory.}
\end{figure}


\clearpage
\begin{figure}[htbp]
\begin{center}
\caption{\label{figure:Figure14}\textbf{Consequence – Child Mortality}}
\includegraphics[width=\textwidth]{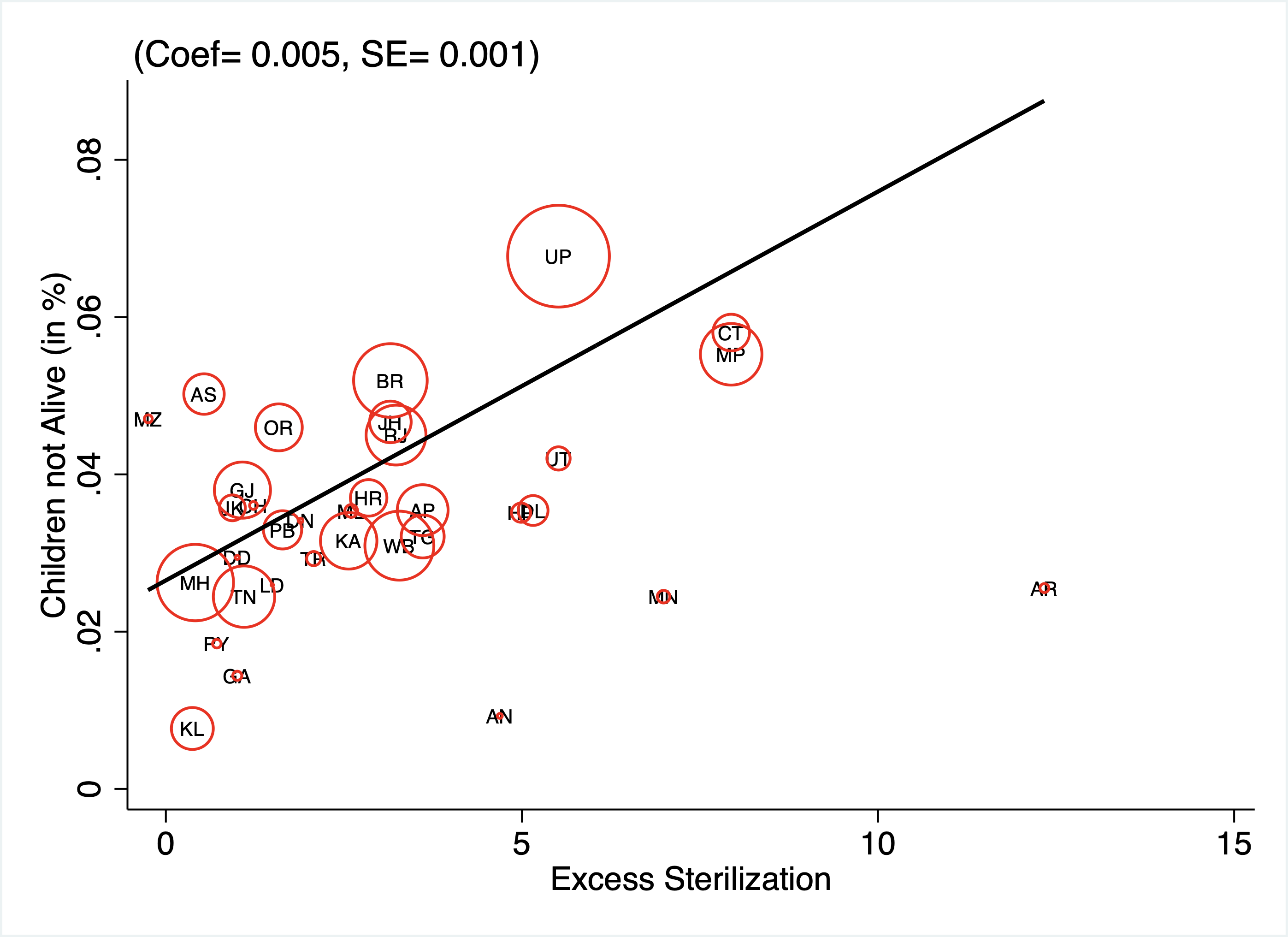}
\end{center}
{\footnotesize Notes: The figure presents the association between the state-level exposure to the forced sterilization policy and the percentage of children who are not alive. The size of each circle denotes the population of the state and union territory. The fitted line is weighted by the population of the state and union territory.}
\end{figure}


\clearpage
\begin{table}[htbp]
\begin{center}
\caption{\label{figure:Table1}\textbf{OLS Estimates - Different Measures of Sterilization}}
\includegraphics[width=\textwidth]{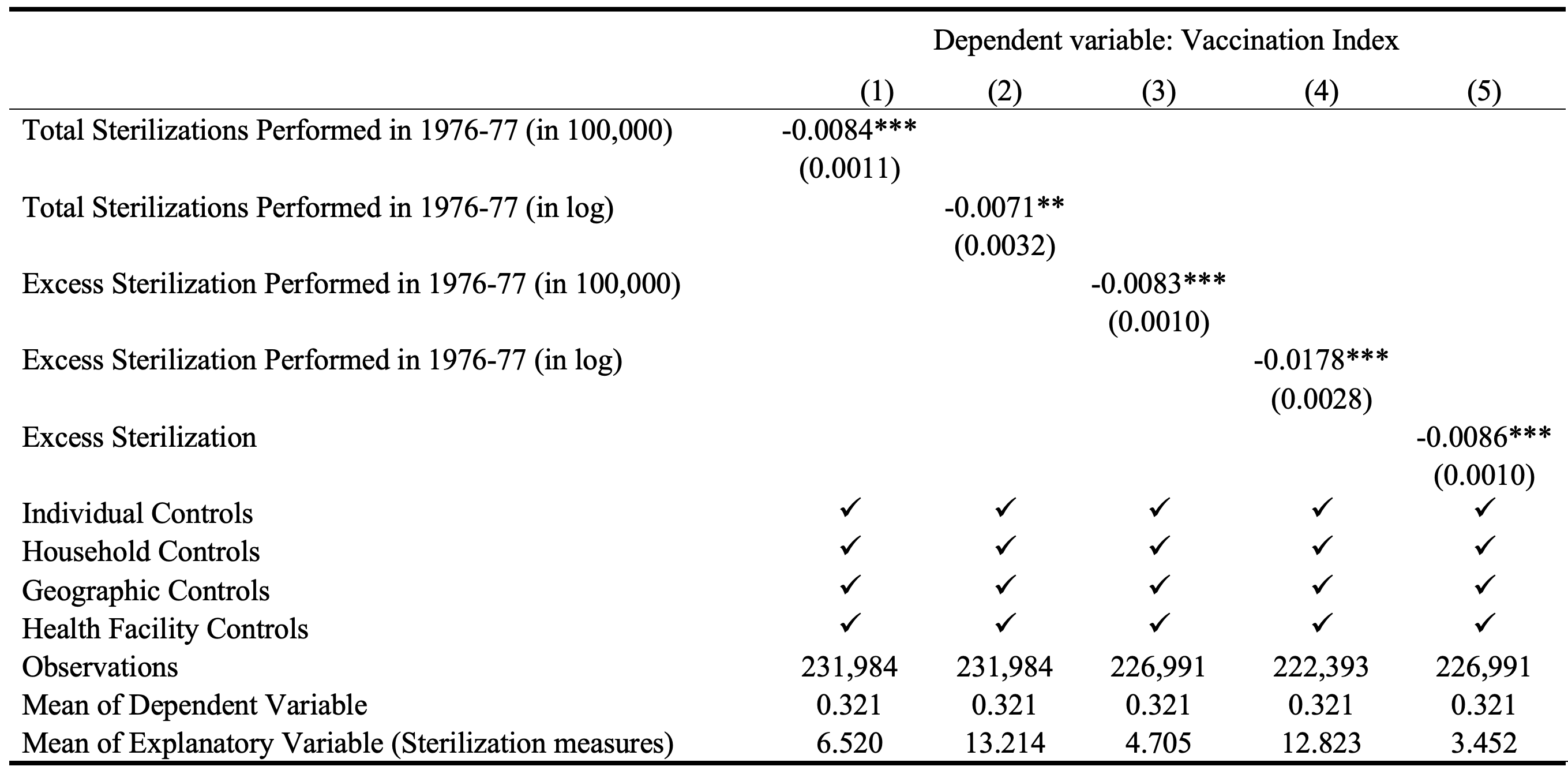}
\end{center}
{\footnotesize Notes: The table presents OLS estimates. Data are from India's National Family and Health Survey 2015-16 (NFHS-4). The unit of observation is a child below the age of 5. Vaccination Index is an index that includes BCG, measles, and three doses each of DPT, four doses of polio, and four doses of hepatitis B. Total Sterilizations Performed in 1976-77 (in 100,000) measures the total number of sterilizations performed in a state in 1976-77 expressed in 100,000 individuals.  Total Sterilizations Performed in 1976-77 (in log) measures the natural log of the number of sterilizations performed in 1976-77. Excess Sterilization Performed in 1976-77 (in 100,000) measures the number of excess sterilizations performed in 1976-77 over and above the 1975-76 numbers expressed in 100,000 individuals. Excess Sterilization Performed in 1976-77 (in log) measures the natural log of the excess number of sterilizations performed in 1976-77 over and above the 1975-76 numbers. Excess Sterilization measures the number of excess sterilizations performed in 1976-77 (compared with 1975-76 numbers) normalized by the sterilization performed in 1975-76 at the state level. Individual controls are for a gender indicator variable of the child, month by year of birth fixed effects, an indicator for whether the child is twin, and the birth order of the child. Household controls include age and sex of the household head, household size, number of household members below the age of 5, seven religion fixed effects, four caste fixed effects, 20 education of the mother fixed effects, four household wealth index fixed effects, and an indicator for whether any household member is covered by health insurance. Geographic controls include the altitude of the cluster in meters, altitude squared, state-level population density per square kilometers (in log), and an indicator of whether the place of residence is urban. Health facility controls include hospitals per 1000 population and doctors per 1000 population at the state level. Robust standard errors in parentheses clustered at the NFHS-4 cluster (PSU) level.  *** p$<0.01$, ** p$<0.05$, * p$<0.1$
}
\end{table}


\clearpage
\begin{table}[htbp]
\begin{center}
\caption{\label{figure:Table2}\textbf{Instrumental Variable Estimates}}
\includegraphics[width=\textwidth]{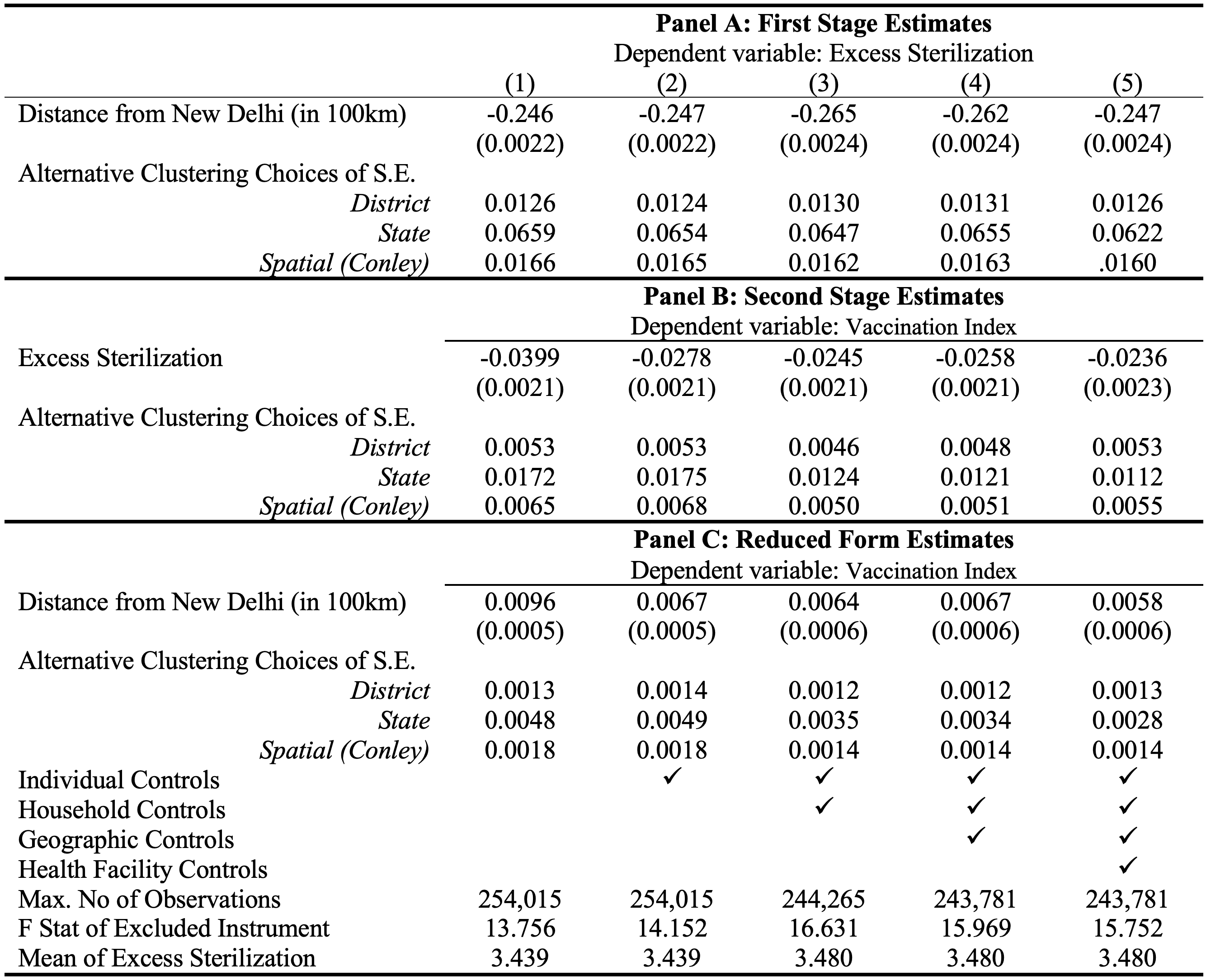}
\end{center}
{\footnotesize Notes: The table presents IV results along with the first stage and reduced form estimates. Data are from India's National Family and Health Survey 2015-16 (NFHS-4). The unit of observation is a child below the age of 5. Vaccination Index is an index that includes BCG, measles, and three doses each of DPT, four doses of polio, and four doses of hepatitis B. Distance from New Delhi measures the distance from New Delhi (the national capital) to state capitals expressed in 100 kilometers. Excess Sterilization measures the number of excess sterilizations performed in 1976-77 (compared with 1975-76 numbers) normalized by the sterilization performed in 1975-76 at the state level. Individual controls are for a gender indicator variable of the child, month by year of birth fixed effects, an indicator for whether the child is twin, and the birth order of the child. Household controls include age and sex of the household head, household size, number of household members below the age of 5, seven religion fixed effects, four caste fixed effects, 20 education of the mother fixed effects, four household wealth index fixed effects, and an indicator for whether any household member is covered by health insurance. Geographic controls include the altitude of the cluster in meters, altitude squared, state-level population density per square kilometers (in log), and an indicator of whether the place of residence is urban. Health facility controls include hospitals per 1000 population and doctors per 1000 population at the state level. Below each coefficient, four standard errors are reported. The first, reported in parentheses, are standard errors adjusted for clustering at the NFHS-4 cluster (PSU) level. The Second—District—is standard errors adjusted for clustering at the current district level. The third—State—is standard errors adjusted for clustering at the current state level. The fourth—Spatial (Conley)—is standard errors adjusted for spatial correction proposed by Conley (1999). The reported F Statistics of Excluded Instrument is the most conservative estimate based on adjusting standard errors for clustering at the state level.
}
\end{table}


\clearpage
\begin{table}[htbp]
\begin{center}
\caption{\label{figure:Table3}\textbf{Alternative Measures of Forced Sterilization: INC's Vote Share }}
\includegraphics[width=\textwidth]{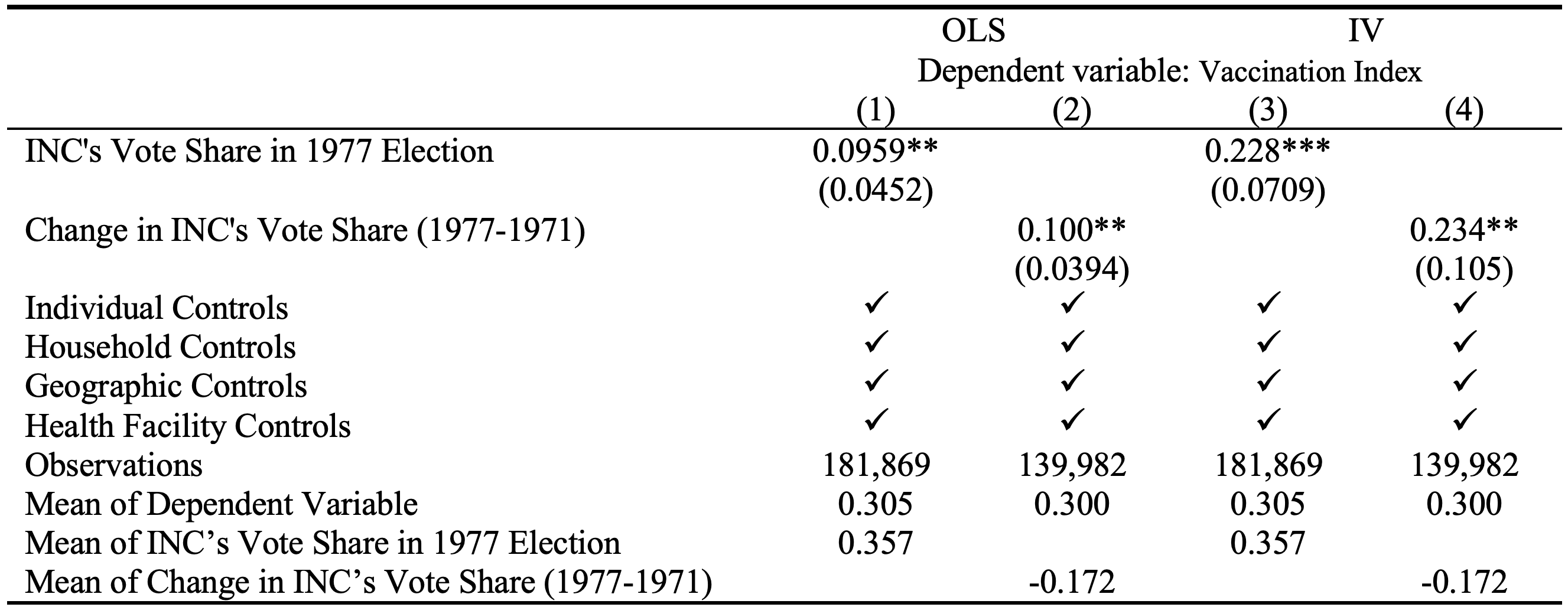}
\end{center}
{\footnotesize Notes: The table presents OLS and IV estimates. Data are from India’s National Family and Health Survey 2015-16 (NFHS-4). The unit of observation is a child below the age of 5. Vaccination Index is an index that includes BCG, measles, and three doses each of DPT, four doses of polio, and four doses of hepatitis B. INC’s Vote Share in 1977 Election measures the constituency level variation in INC candidates’ vote share in the 1977 election (a proxy measure of excess sterilization). Change in INC’s Vote Share (1977-1971) measures the constituency-level change in INC candidates’ vote share in 1977 compared with the 1971 election (as a second proxy measure of excess sterilization). Individual controls are for a gender indicator variable of the child, month by year of birth fixed effects, an indicator for whether the child is twin, and the birth order of the child. Household controls include age and sex of the household head, household size, number of household members below the age of 5, seven religion fixed effects, four caste fixed effects, 20 education of the mother fixed effects, four household wealth index fixed effects, and an indicator for whether any household member is covered by health insurance. Geographic controls include the altitude of the cluster in meters, altitude squared, state-level population density per square kilometers (in log), and an indicator of whether the place of residence is urban. Health facility controls include hospitals per 1000 population and doctors per 1000 population at the state level. Robust standard errors in parentheses clustered at the parliament constituency level. *** p$<0.01$, ** p$<0.05$, * p$<0.1$
}
\end{table}


\clearpage
\begin{table}[htbp]
\begin{center}
\caption{\label{figure:Table4}\textbf{Mechanism – Non-institutional Delivery}}
\includegraphics[width=\textwidth]{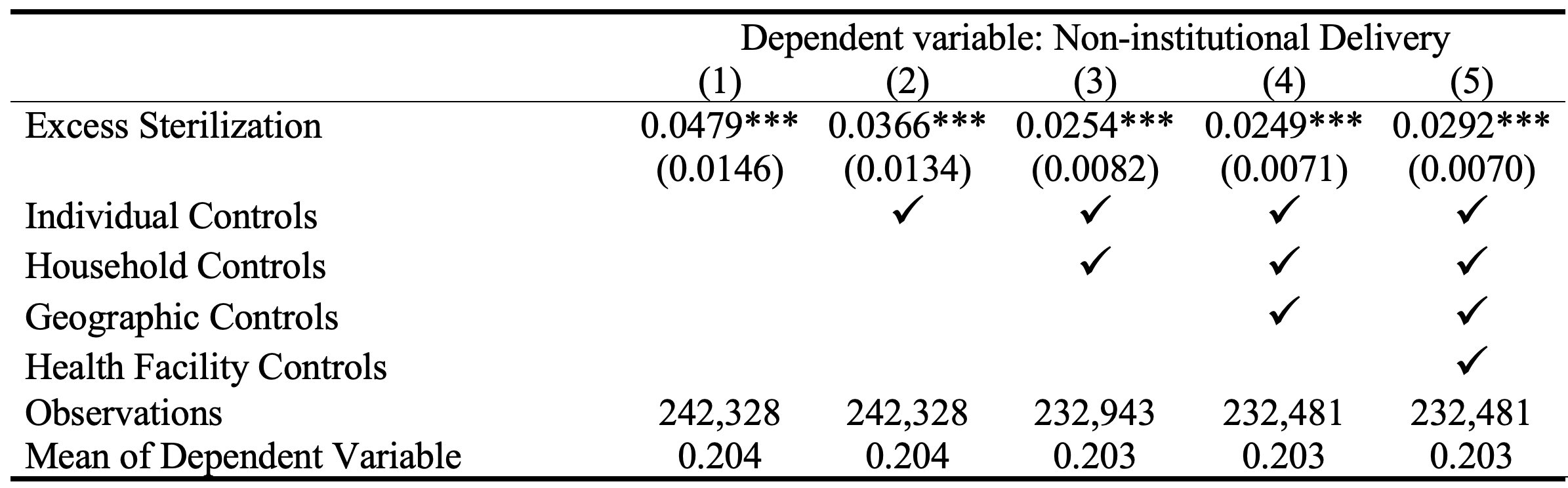}
\end{center}
{\footnotesize Notes: The table presents IV estimates. Data are from India’s National Family and Health Survey 2015-16 (NFHS-4). The unit of observation is a child below the age of 5. Non-institutional Delivery is an indicator variable for a child born at home. Excess Sterilization measures the number of excess sterilizations performed in 1976-77 (compared with 1975-76 numbers) normalized by the sterilization performed in 1975-76 at the state level. Individual controls are for a gender indicator variable of the child, month by year of birth fixed effects, an indicator for whether the child is twin, and the birth order of the child. Household controls include age and sex of the household head, household size, number of household members below the age of 5, seven religion fixed effects, four caste fixed effects, 20 education of the mother fixed effects, four household wealth index fixed effects, and an indicator for whether any household member is covered by health insurance. Geographic controls include the altitude of the cluster in meters, altitude squared, state-level population density per square kilometers (in log), and an indicator of whether the place of residence is urban. Health facility controls include hospitals per 1000 population and doctors per 1000 population at the state level. Robust standard errors in parentheses clustered at the state level. *** p$<0.01$, ** p$<0.05$, * p$<0.1$
}
\end{table}


\clearpage
\begin{table}[htbp]
\begin{center}
\caption{\label{figure:Table5}\textbf{Reasons for Non-institutional Delivery (Indexing reasons)}}
\includegraphics[width=10cm]{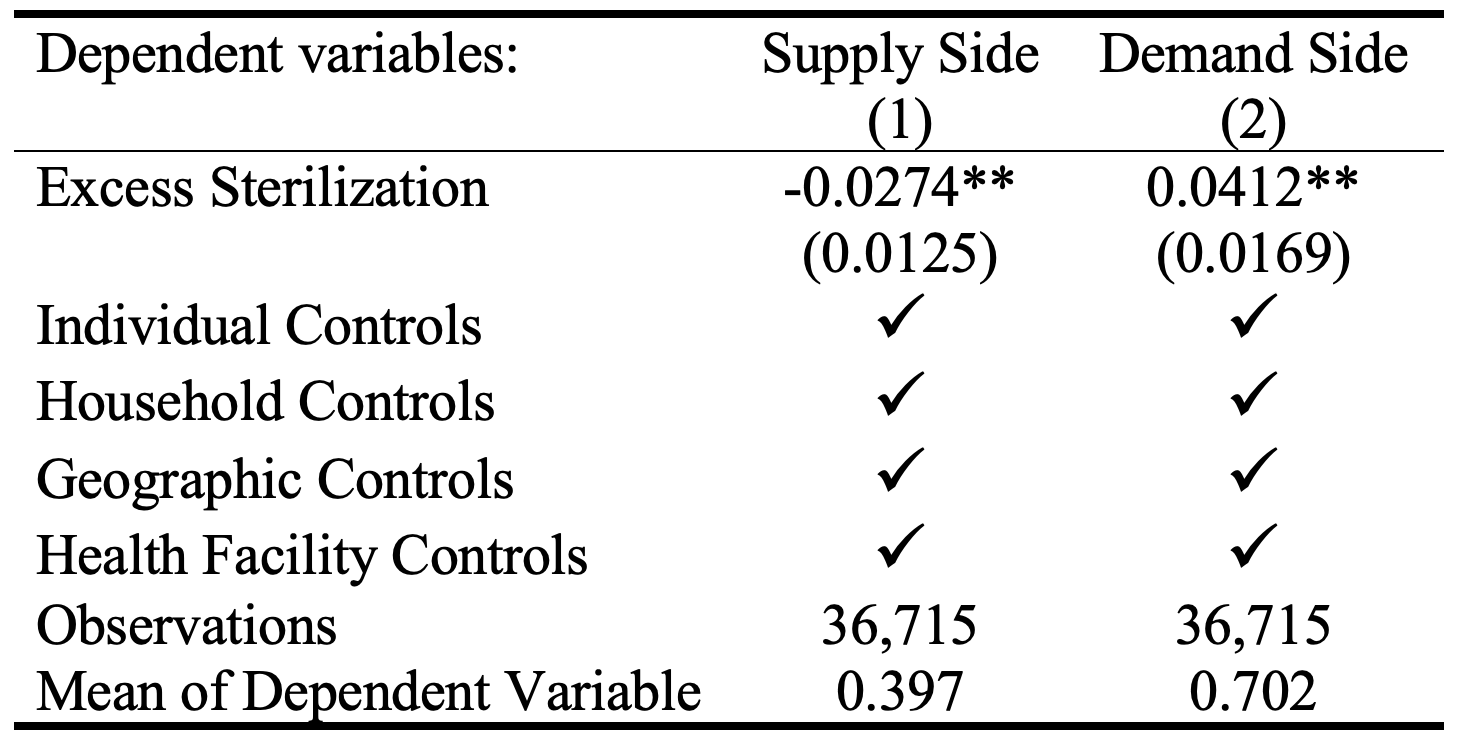}
\end{center}
{\footnotesize Notes: The table presents IV estimates. Data are from India’s National Family and Health Survey 2015-16 (NFHS-4). The unit of observation is a child below the age of 5 who is not born at a healthcare facility. The Supply Side is an index that includes Cost too much, Facility not open, Too far/ no transport, and No female provider. The Demand Side is an index that includes Do not trust facility/ poor service, Husband/family did not allow, Not necessary, Not customary, and Other. The mean of dependent variables (in percentages) does not add to 100 because multiple responses were permitted. Excess Sterilization measures the number of excess sterilizations performed in 1976-77 (compared with 1975-76 numbers) normalized by the sterilization performed in 1975-76 at the state level. Individual controls are for a gender indicator variable of the child, month by year of birth fixed effects, an indicator for whether the child is twin, and the birth order of the child. Household controls include age and sex of the household head, household size, number of household members below the age of 5, seven religion fixed effects, four caste fixed effects, 20 education of the mother fixed effects, four household wealth index fixed effects, and an indicator for whether any household member is covered by health insurance. Geographic controls include the altitude of the cluster in meters, altitude squared, state-level population density per square kilometers (in log), and an indicator of whether the place of residence is urban. Health facility controls include hospitals per 1000 population and doctors per 1000 population at the state level. Robust standard errors in parentheses clustered at the state level. *** p$<0.01$, ** p$<0.05$, * p$<0.1$
}
\end{table}


\clearpage
\begin{table}[htbp]
\begin{center}
\caption{\label{figure:Table6}\textbf{Mechanism – Antenatal Care (ANC)}}
\includegraphics[width=10cm]{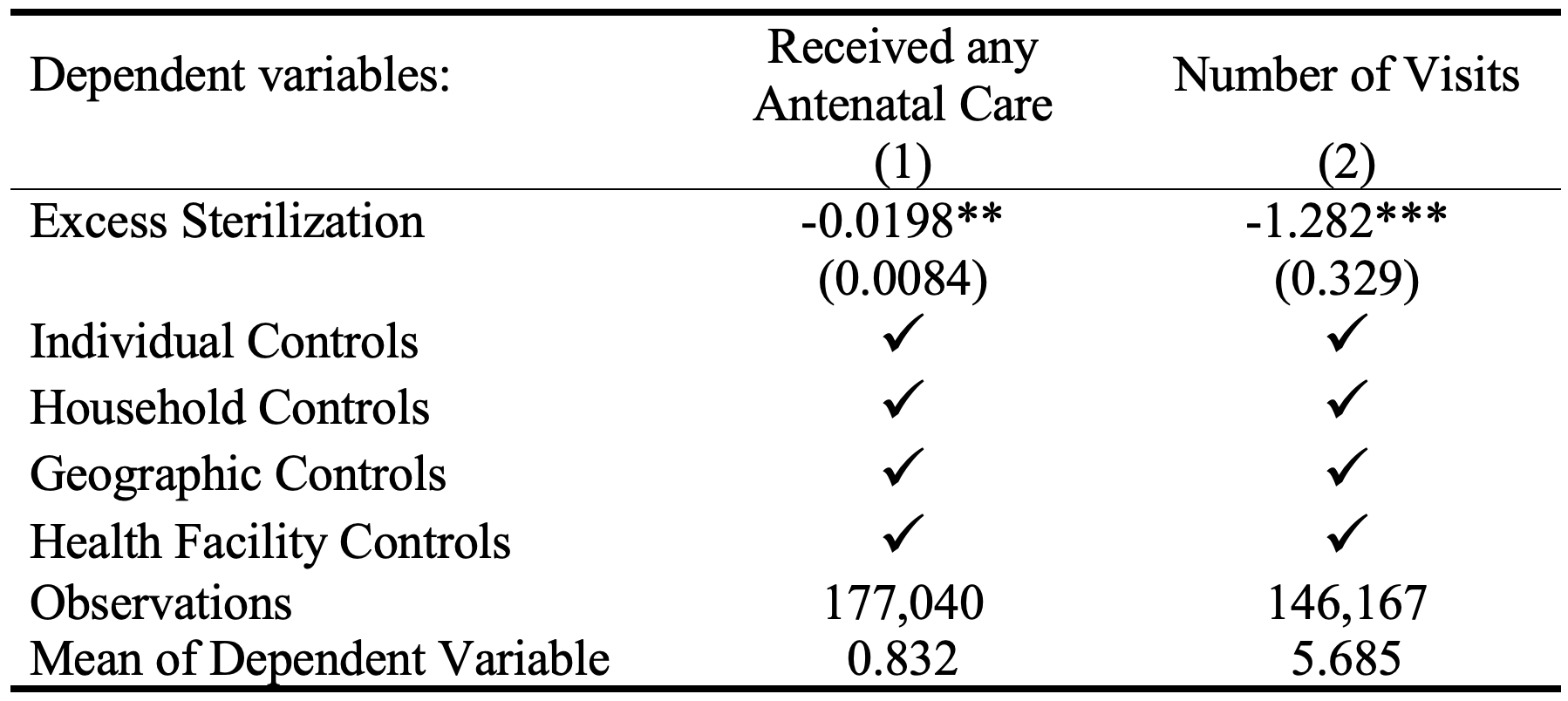}
\end{center}
{\footnotesize Notes: The table presents IV estimates. Data are from India’s National Family and Health Survey 2015-16 (NFHS-4). The unit of observation is the most recent child of a mother below the age of 5.  Received any Antenatal Care is an indicator variable for mothers who received antenatal care in the last pregnancy in the NFHS-4 data. The Number of Visits measures the number of times the mother received antenatal care conditional on receiving any antenatal care in the last pregnancy. Excess Sterilization measures the number of excess sterilizations performed in 1976-77 (compared with 1975-76 numbers) normalized by the sterilization performed in 1975-76 at the state level. Individual controls are for a gender indicator variable of the child, month by year of birth fixed effects, an indicator for whether the child is twin, and the birth order of the child. Household controls include age and sex of the household head, household size, number of household members below the age of 5, seven religion fixed effects, four caste fixed effects, 20 education of the mother fixed effects, four household wealth index fixed effects, and an indicator for whether any household member is covered by health insurance. Geographic controls include the altitude of the cluster in meters, altitude squared, state-level population density per square kilometers (in log), and an indicator of whether the place of residence is urban. Health facility controls include hospitals per 1000 population and doctors per 1000 population at the state level. Robust standard errors in parentheses clustered at the state level.  *** p$<0.01$, ** p$<0.05$, * p$<0.1$
}
\end{table}


\clearpage
\begin{table}[htbp]
\begin{center}
\caption{\label{figure:Table7}\textbf{Consequence – Child Mortality}}
\includegraphics[width=\textwidth]{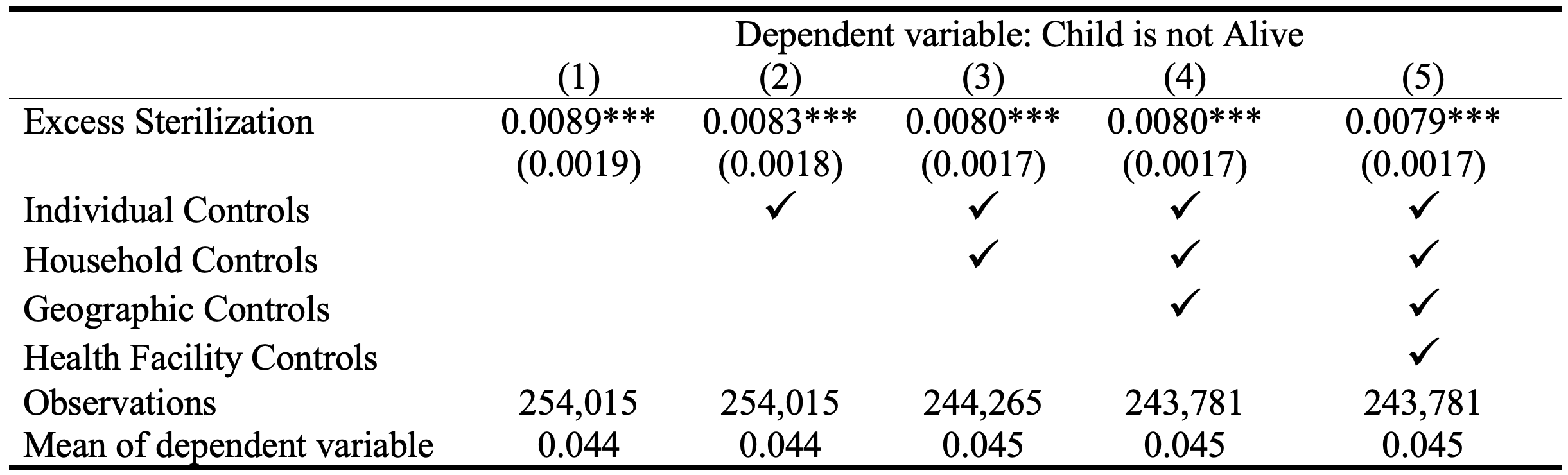}
\end{center}
{\footnotesize Notes: The table presents IV estimates. Data are from India’s National Family and Health Survey 2015-16 (NFHS-4). The unit of observation is a child below the age of 5, including those who are not alive. Excess Sterilization measures the number of excess sterilizations performed in 1976-77 (compared with 1975-76 numbers) normalized by the sterilization performed in 1975-76 at the state level. Individual controls are for a gender indicator variable of the child, month by year of birth fixed effects, an indicator for whether the child is twin, and the birth order of the child. Household controls include age and sex of the household head, household size, number of household members below the age of 5, seven religion fixed effects, four caste fixed effects, 20 education of the mother fixed effects, four household wealth index fixed effects, and an indicator for whether any household member is covered by health insurance. Geographic controls include the altitude of the cluster in meters, altitude squared, state-level population density per square kilometers (in log), and an indicator of whether the place of residence is urban. Health facility controls include hospitals per 1000 population and doctors per 1000 population at the state level. Robust standard errors in parentheses clustered at the state level.  *** p$<0.01$, ** p$<0.05$, * p$<0.1$
}
\end{table}

\appendix
\renewcommand\thefigure{\thesection.\arabic{figure}}
\renewcommand\thetable{\thesection.\arabic{table}}
\setcounter{figure}{0}
\setcounter{table}{0}

\clearpage

\title{\textbf{ \Large {Online Appendix \\
\vskip 1cm
 Understanding Vaccine Hesitancy: Empirical Evidence from India
}}}

\clearpage
\section{Figures}


\begin{figure}[htbp]

\begin{center}
\caption{\label{figure:FigureA1}\textbf{Vaccination Completion Rate by Child’s Background Characteristics (aged 12–23 months)}}

\includegraphics[width=\textwidth]{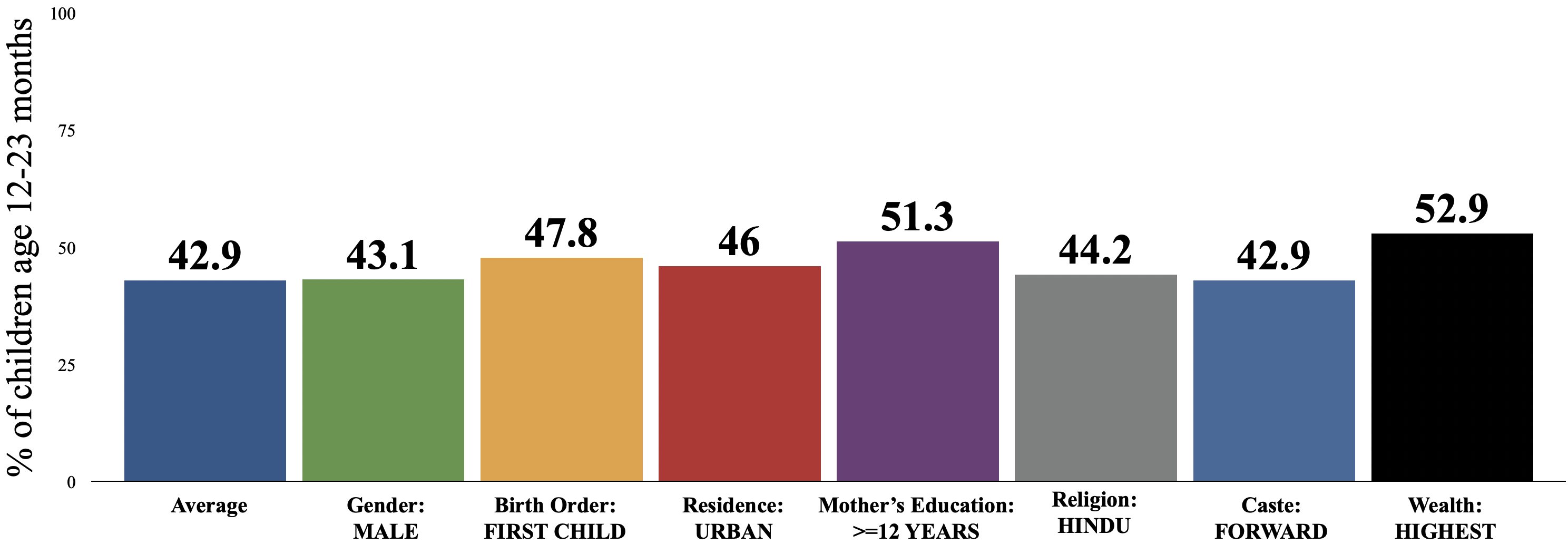}
\end{center}
{\footnotesize Notes: Author’s compilation using data from National Family Health Survey (NFHS-4), 2015-16: India. These estimates are based on 12 reported vaccines (excluding Polio given at birth) for children between 12–23 months of age. See \url{https://dhsprogram.com/pubs/pdf/FR339/FR339.pdf} for details. }
\end{figure}

\clearpage
\begin{figure}[htbp]
\begin{center}
\caption{\label{figure:FigureA2}\textbf{Examples of Archival Records  }}
\subcaption{Panel A: Examples of Archival Data – Yearly Sterilization Figures}
\includegraphics[height=10cm]{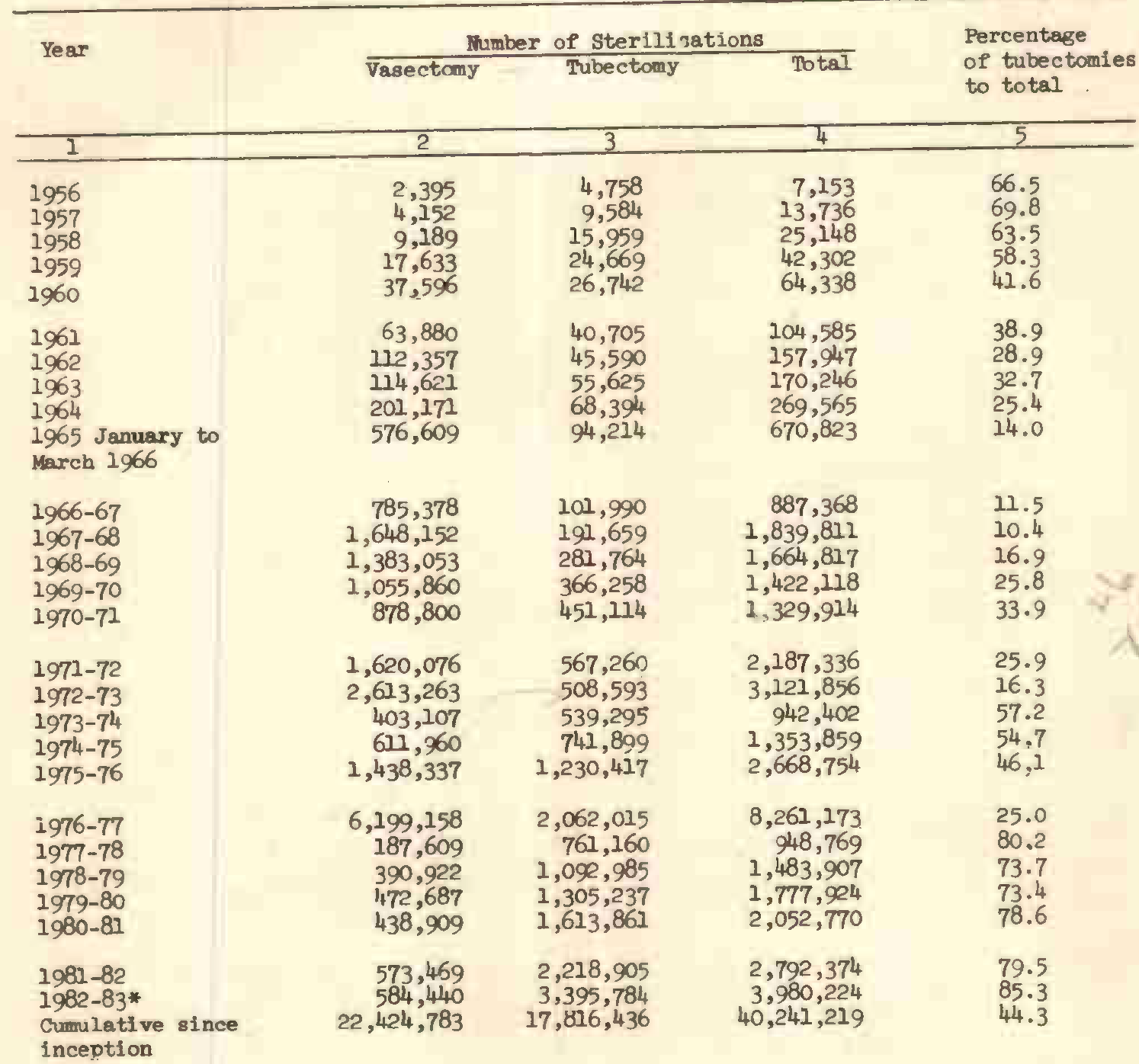}
\subcaption{Panel B: Examples of Archival Data – State Level Sterilization Performance}
\includegraphics[height=10cm]{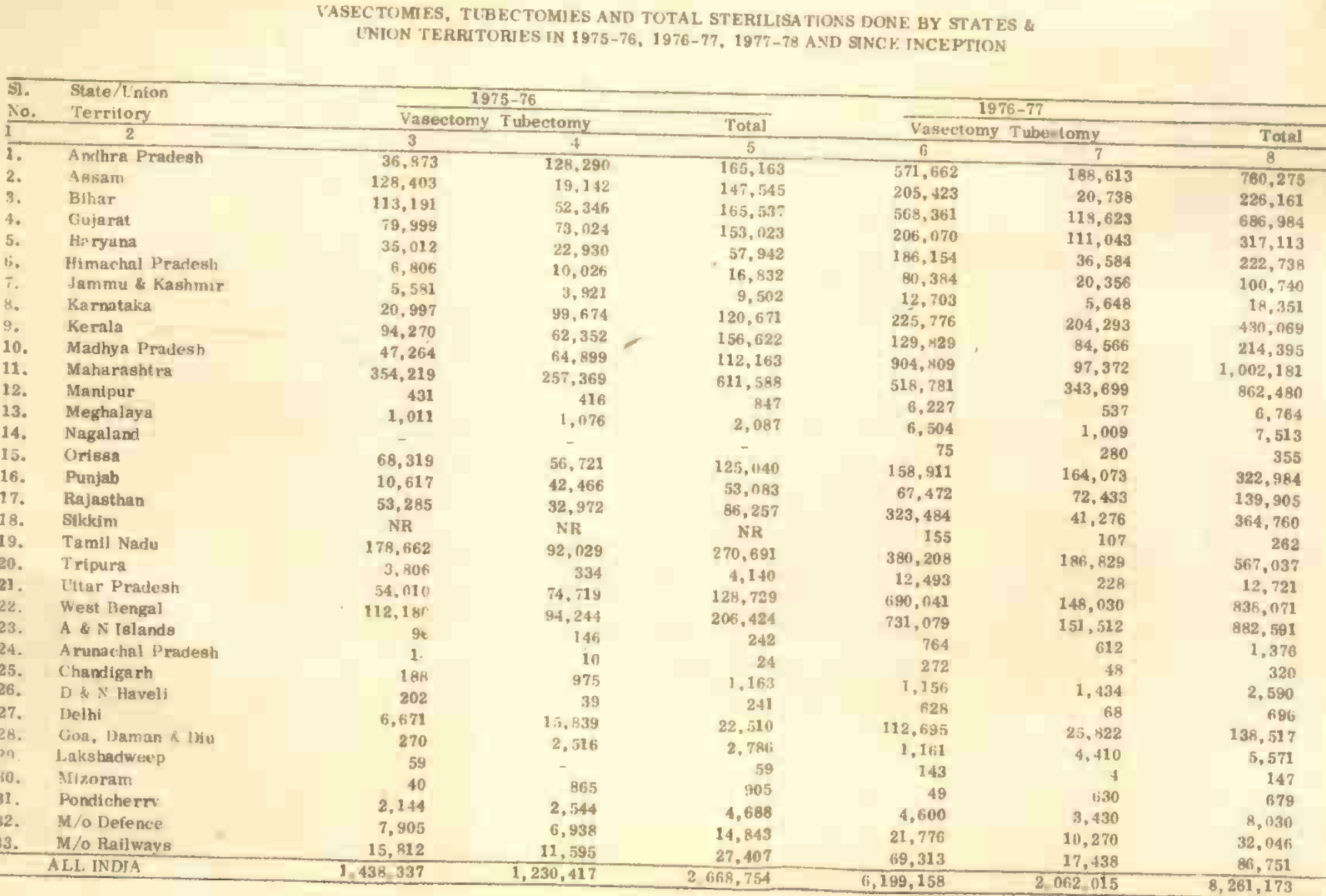}
\end{center}
{\footnotesize Notes: The figure presents some examples of archival data used in this paper. Panel A presents the yearly sterilization figures published in the 1982-83 yearbook. Panel B presents the state-level sterilization performance in 1975-76 published in the 1977-78 yearbook.}
\end{figure}


\clearpage
\begin{figure}[htbp]
\begin{center}
\caption{\label{figure:FigureA3}\textbf{Total Number of Sterilizations Performed in 1975-76 (\textit{Previous year})}}
\includegraphics[height=10cm]{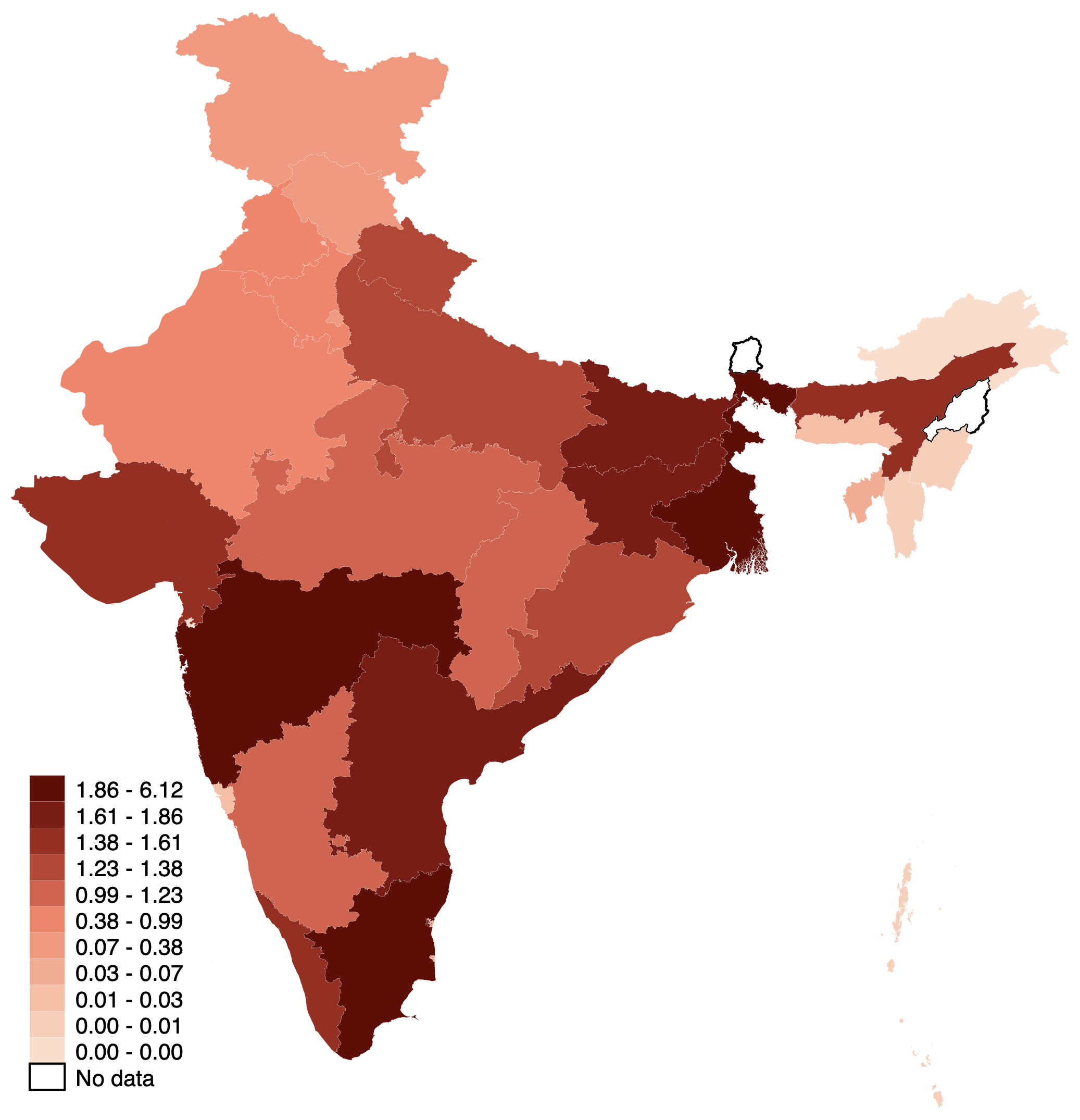}
\end{center}
{\footnotesize Notes: The figure presents the state-level variation in the number of sterilizations performed in 1975–76 (in 100,000). Darker shades denote a greater number of sterilizations performed.}
\end{figure}


\clearpage
\begin{figure}[htbp]
\begin{center}
\caption{\label{figure:FigureA4}\textbf{Percentage of Children Who are Fully Vaccinated (Vaccination Index) }}
\includegraphics[height=10cm]{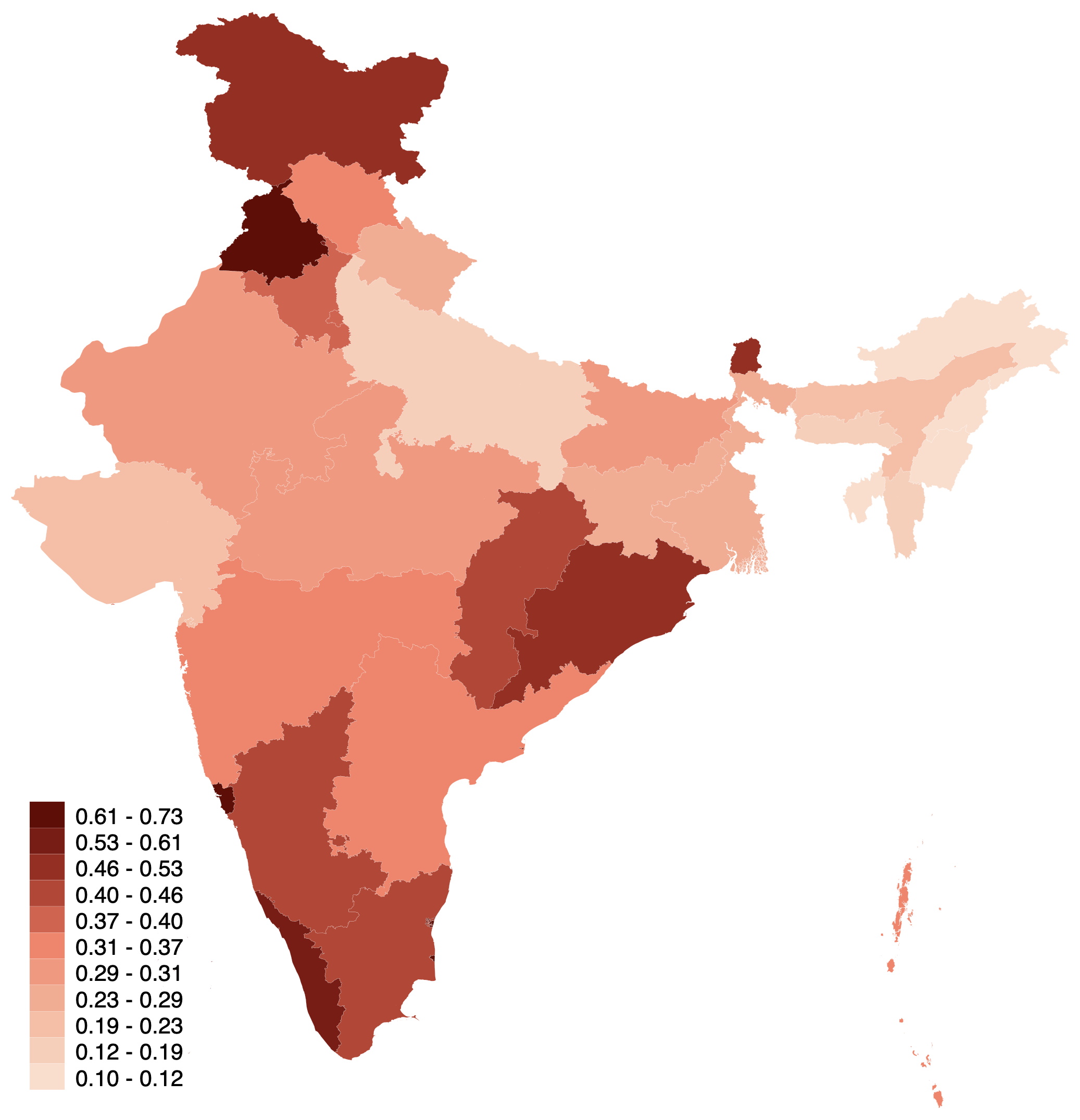}
\end{center}
{\footnotesize Notes: The figure presents the state-level variation in the percentage of children under the age of 5 who received all vaccines from NFHS-4 dataset. Darker shades denote a higher percentage of vaccination.}
\end{figure}


\clearpage
\begin{figure}[htbp]
\begin{center}
\caption{\label{figure:FigureA5}\textbf{A Descriptive Comparison of Madhya Pradesh and Odisha  }}
\includegraphics[width=\textwidth]{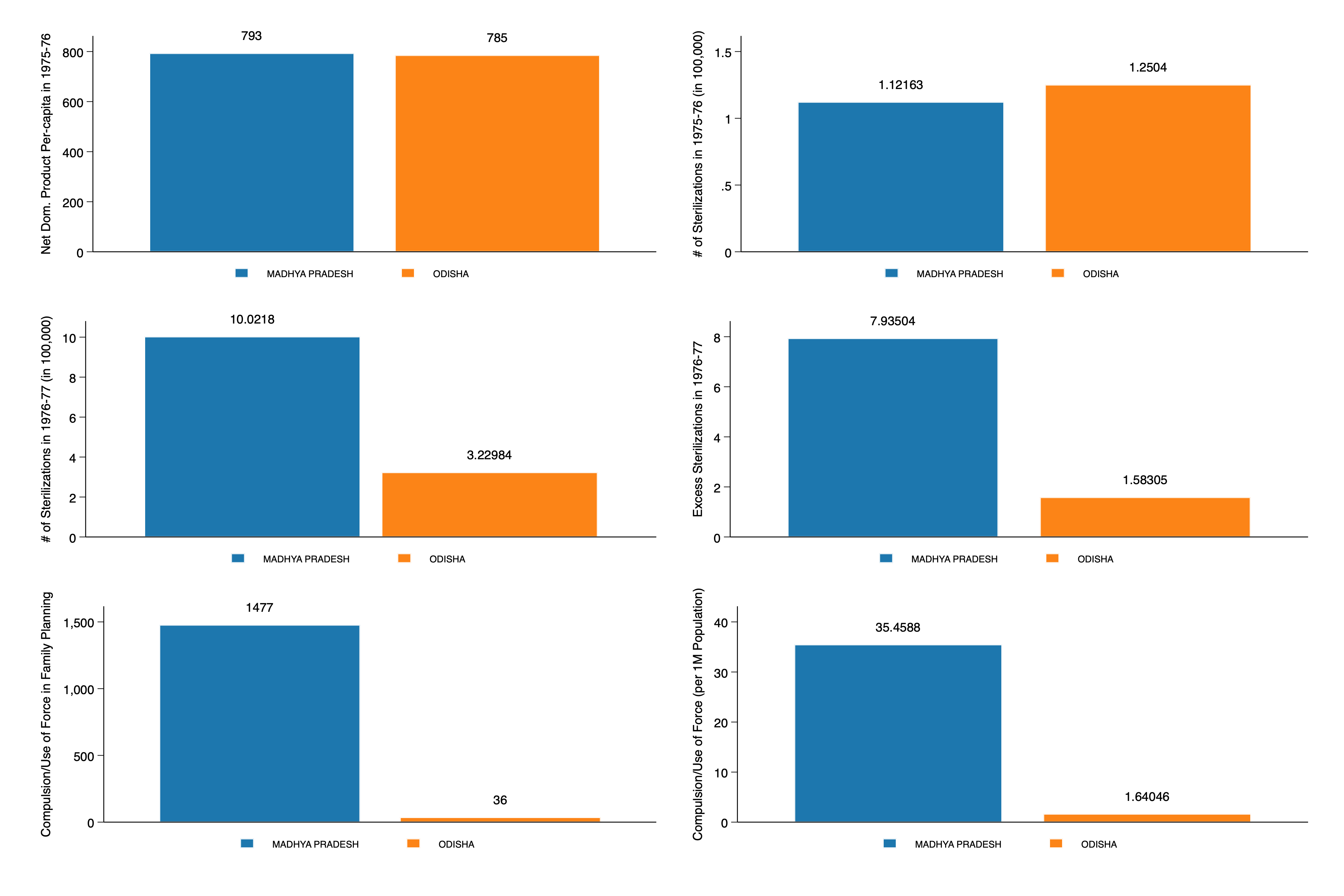}
\end{center}
{\footnotesize Notes: The figure presents a descriptive comparison of the state of  Madhya Pradesh and Odisha. The top figure presents the differences in the Net GDP per capita (in Rupees) and the number of sterilizations performed (in 100,000) in 1975-76, respectively. The middle figure presents the differences in the number of sterilizations (in 100,000) and excess sterilizations performed in 1976-77, respectively. The bottom figure presents the total number of complaints and the number of complaints per one million population received in relation to compulsion or the use of force in family planning during the force sterilization period.}
\end{figure}


\clearpage
\begin{figure}[htbp]
\begin{center}
\caption{\label{figure:FigureA6}\textbf{Density Test of the Forcing Variable}}
\includegraphics[width=\textwidth]{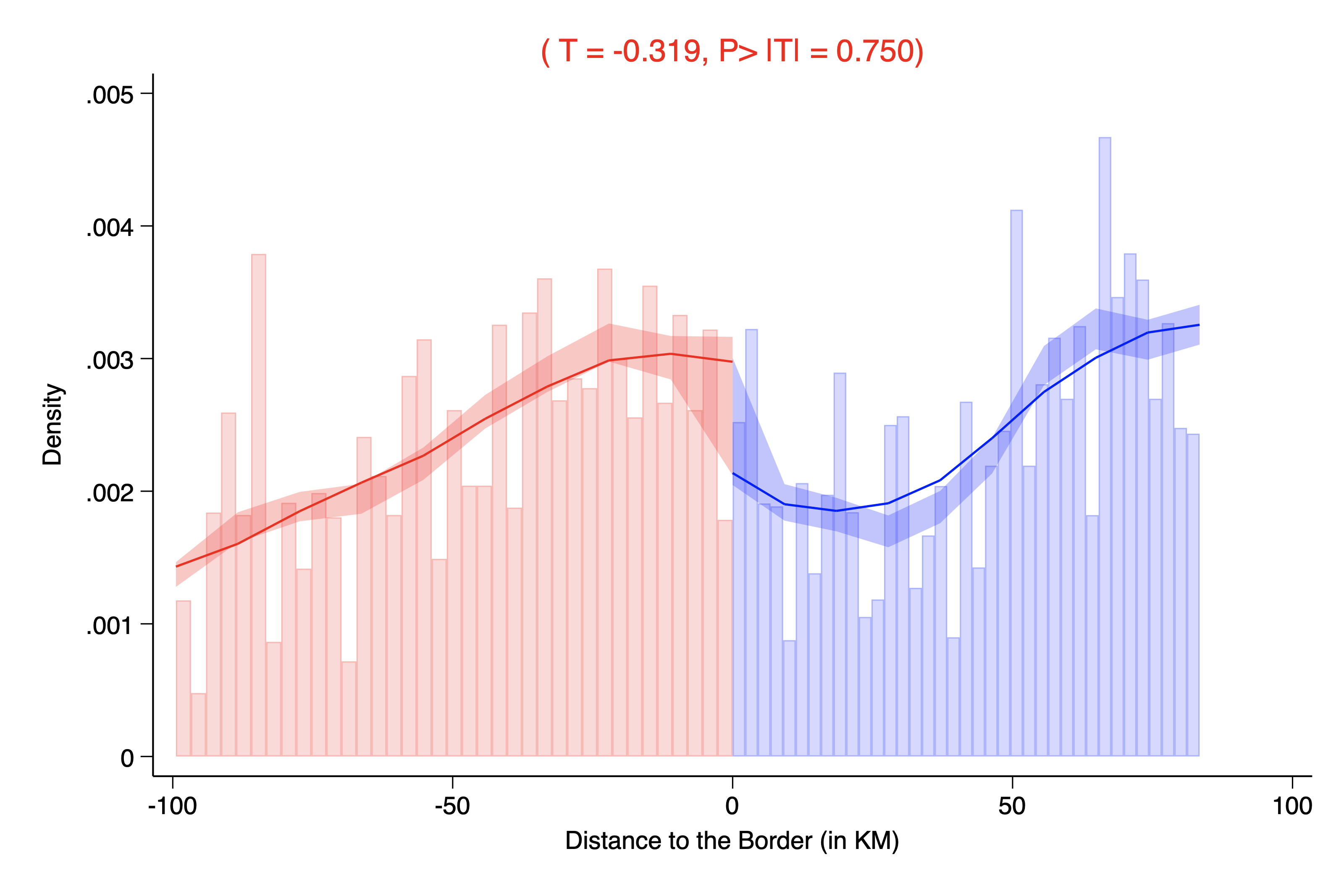}
\end{center}
{\footnotesize Notes: The figure presents the density test of the forcing variable (distance to the border) between Odisha and Chhattisgarh. The left-hand side of the graph presents the samples of Odisha, and the right-hand side presents the samples of Chhattishgarh.}
\end{figure}

\clearpage
\begin{figure}[htbp]
\begin{center}
\caption{\label{figure:FigureA7}\textbf{Placebo Cutoff Points at Artificial Boundaries}}
\subcaption{Panel A: Odisha Sample Only}
\includegraphics[height=8cm]{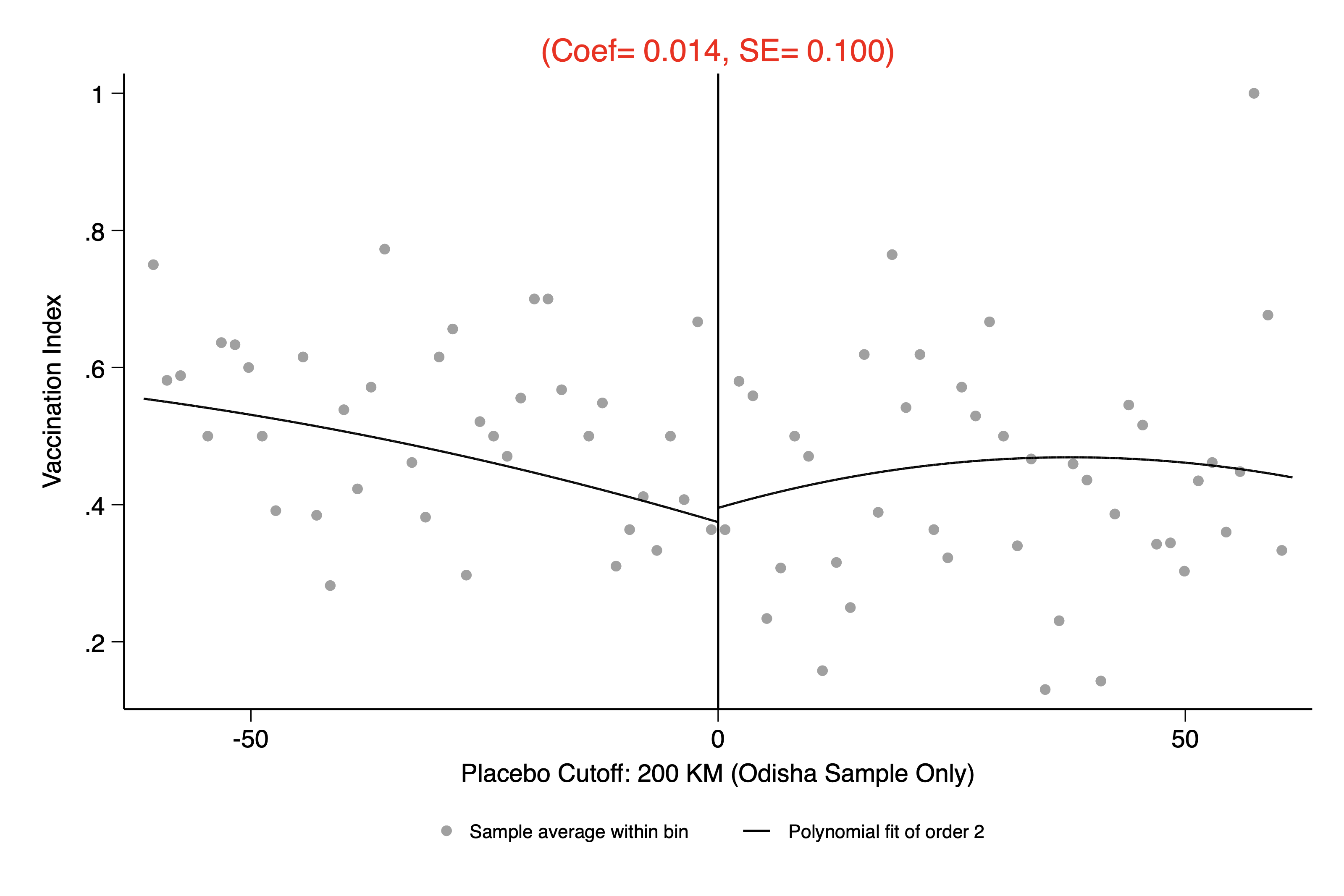}
\subcaption{Panel B: Chhattisgarh Sample Only}
\includegraphics[height=8cm]{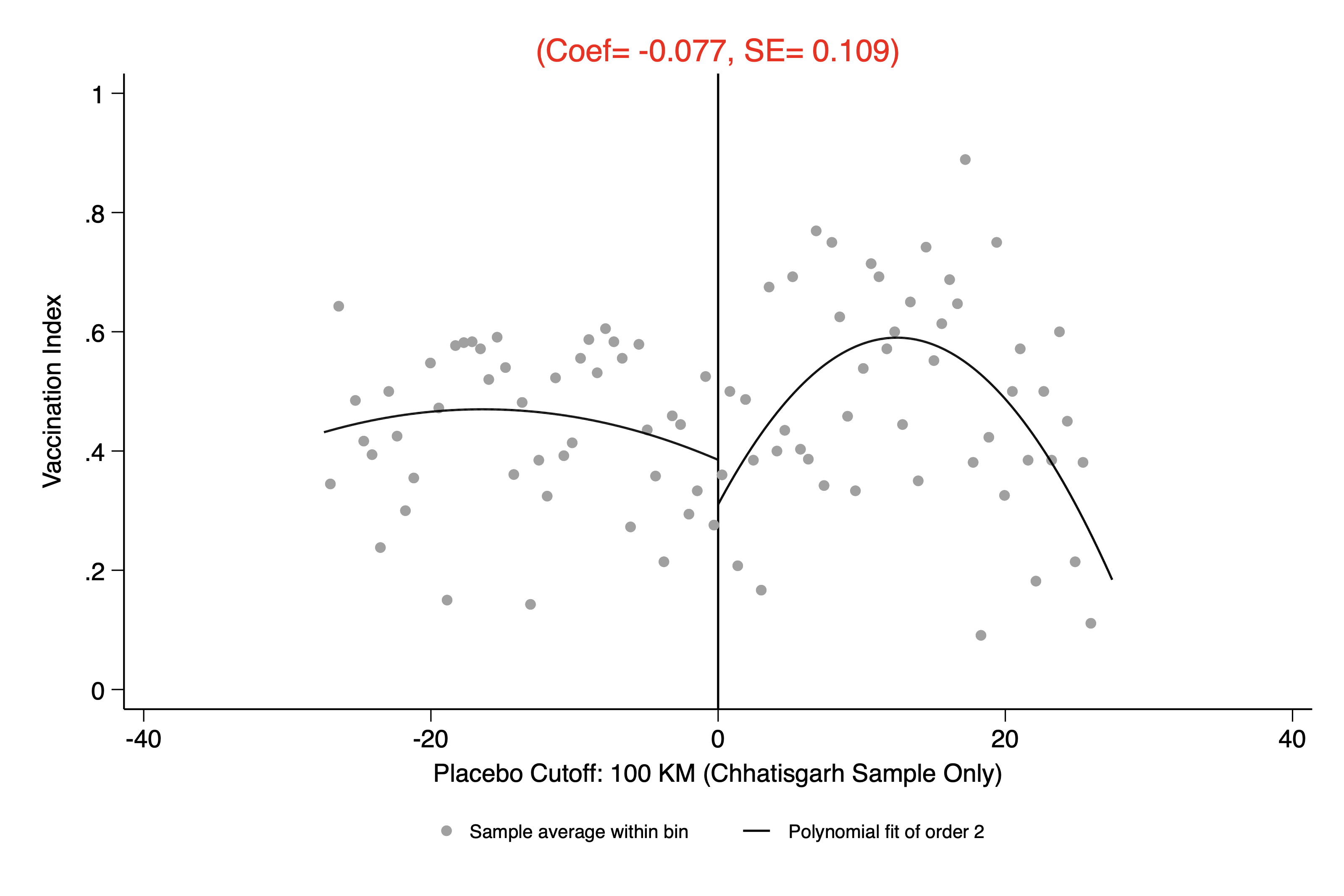}
\end{center}
{\footnotesize Notes: The figure presents the RDD specification by relocating the cutoff points to artificial boundaries on either side of the border at the median points (or close to the median). Panel A presents the artificial cutoff points of the Odisha sample at 200 kilometers. Panel B presents the artificial cutoff points of the Chhattisgarh sample at 100 kilometers. All the figures use optimal bandwidths based on the data-driven method proposed by Calonico et al. (2015). We set the local polynomial of order 2, which has a minimum asymptotic mean squared error (MSE) of the RD point estimator in our sample, as proposed by Pei et al. (2021). We use a uniform kernel, as suggested by Imbens and Lemieux (2008) and Lee and Lemieux (2010), which is simple, transparent, and easy to interpret. (Use of triangular kernel also produces similar results.)}
\end{figure}

\clearpage
\begin{figure}[htbp]
\begin{center}
\caption{\label{figure:FigureA8}\textbf{Placebo Exercise Between Madhya Pradesh and Chhattisgarh}}
\subcaption{Panel A: Placebo Exercise Between Madhya Pradesh and Chhattisgarh (General RDD Approach)}
\includegraphics[height=8cm]{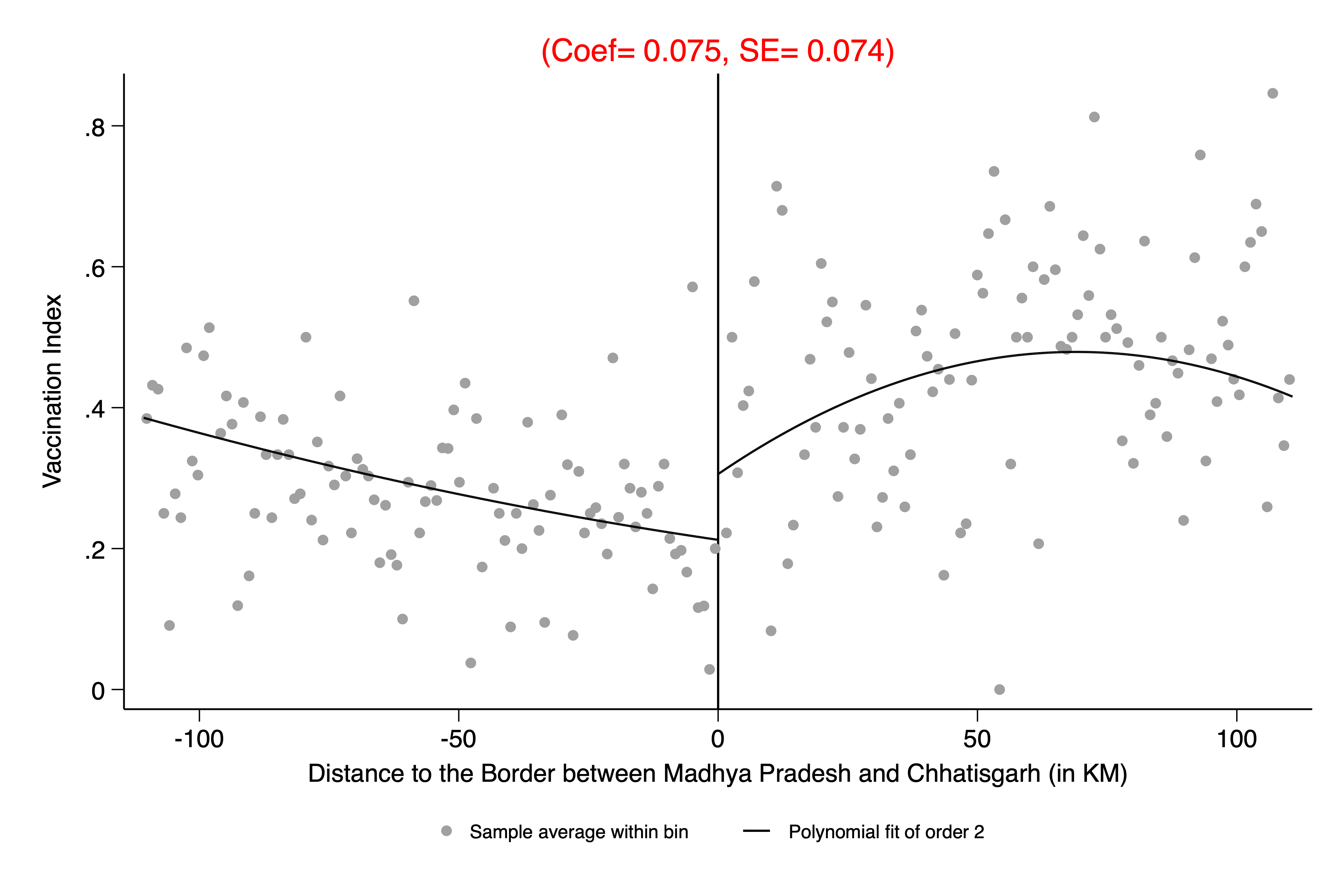}
\subcaption{Panel B: Placebo Exercise Between Madhya Pradesh and Chhattisgarh (Using Donut Hole RD Approach: Excluding the observations within 5km from the border)}
\includegraphics[height=8cm]{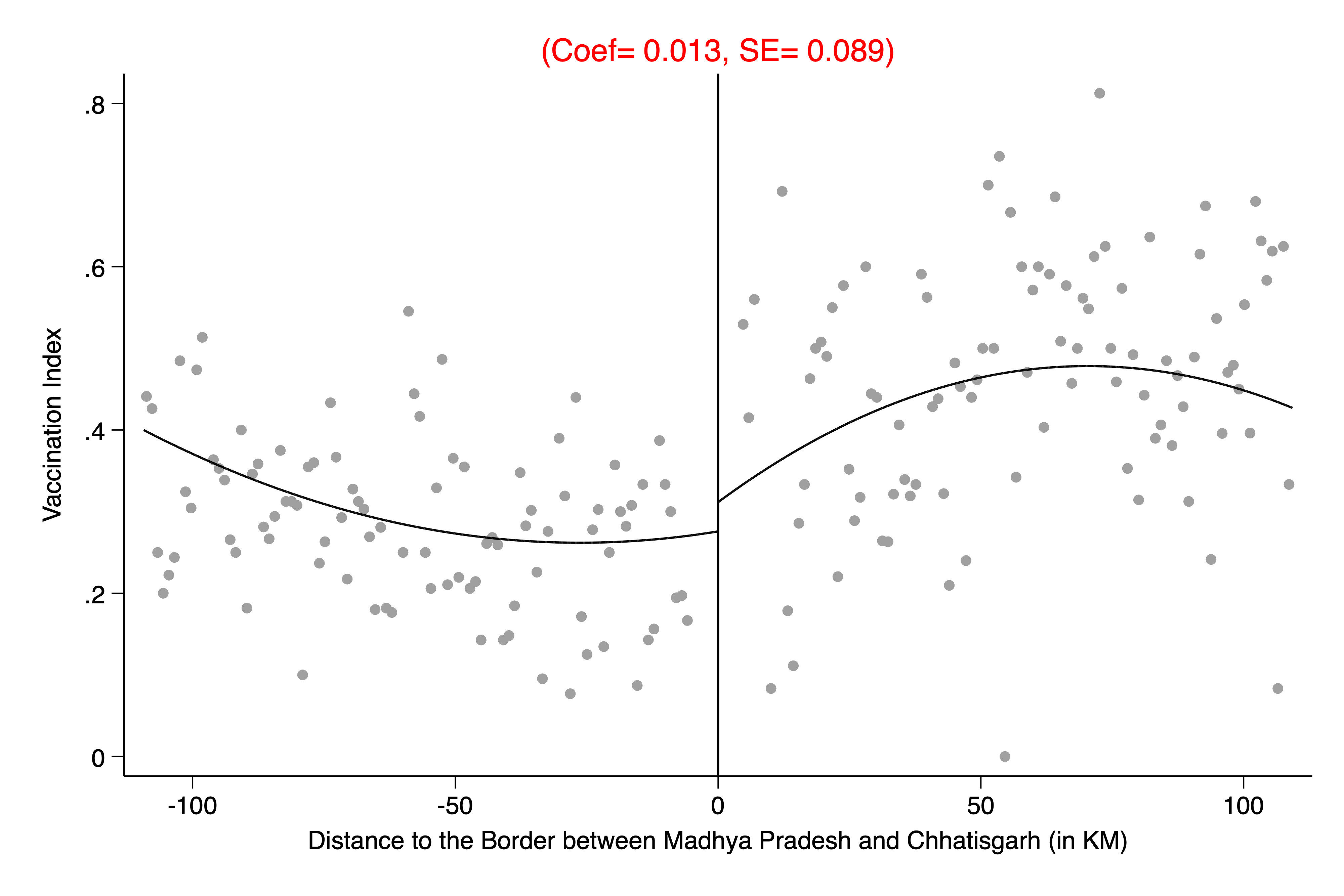}
\end{center}
{\footnotesize Notes: The figure presents the placebo RDD between Madhya Pradesh and Chhattisgarh, which belonged to the same state during the forced sterilization period. Panel A presents the result using the general RDD approach. Panel B presents the result using the donut hole RDD approach, excluding the observations within 5 km of the border due to the random allocation of the GPS coordinates up to 5 km in the NFHS-4 survey. The left-hand side of each graph presents the samples of Madhya Pradesh, and the right-hand side presents the samples of Chhattisgarh. All the figures use optimal bandwidths based on the data-driven method proposed by Calonico et al. (2015). We set the local polynomial of order 2, which has a minimum asymptotic mean squared error (MSE) of the RD point estimator in our sample, as proposed by Pei et al. (2021). We use a uniform kernel, as suggested by Imbens and Lemieux (2008) and Lee and Lemieux (2010), which is simple, transparent, and easy to interpret. (Use of triangular kernel also produces similar results.)}
\end{figure}

\clearpage
\begin{figure}[htbp]
\begin{center}
\caption{\label{figure:FigureA9}\textbf{RDD Approach with Covariates (Including the Urban Variable in the Regression)}}
\includegraphics[height=10cm]{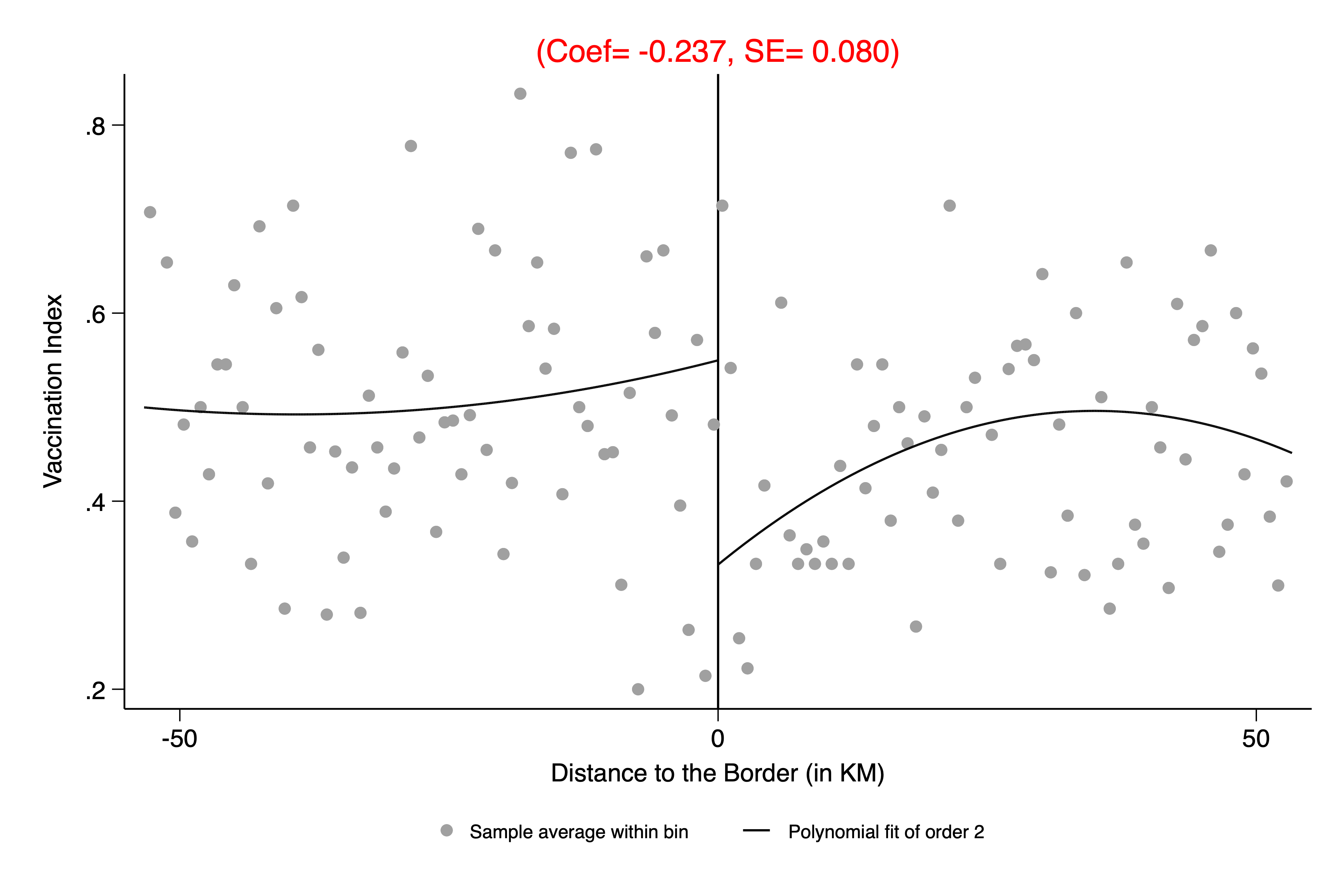}
\end{center}
{\footnotesize Notes: The figure presents the RDD approach with covariates. In particular, it produces the estimates, including the urban variable in the regression. The left-hand side of the graph presents the samples of Odisha, and the right-hand side presents the samples of Chhattisgarh. The figure uses optimal bandwidths based on the data-driven method proposed by Calonico et al. (2015). We set the local polynomial of order 2, which has a minimum asymptotic mean squared error (MSE) of the RD point estimator in our sample, as proposed by Pei et al. (2021). We use a uniform kernel, as suggested by Imbens and Lemieux (2008) and Lee and Lemieux (2010), which is simple, transparent, and easy to interpret. (Use of triangular kernel also produces similar results.)}
\end{figure}

\clearpage
\begin{figure}[htbp]
\begin{center}
\caption{\label{figure:FigureA10}\textbf{RDD Results with Alternative Bandwidth Selection}}
\includegraphics[width=\textwidth]{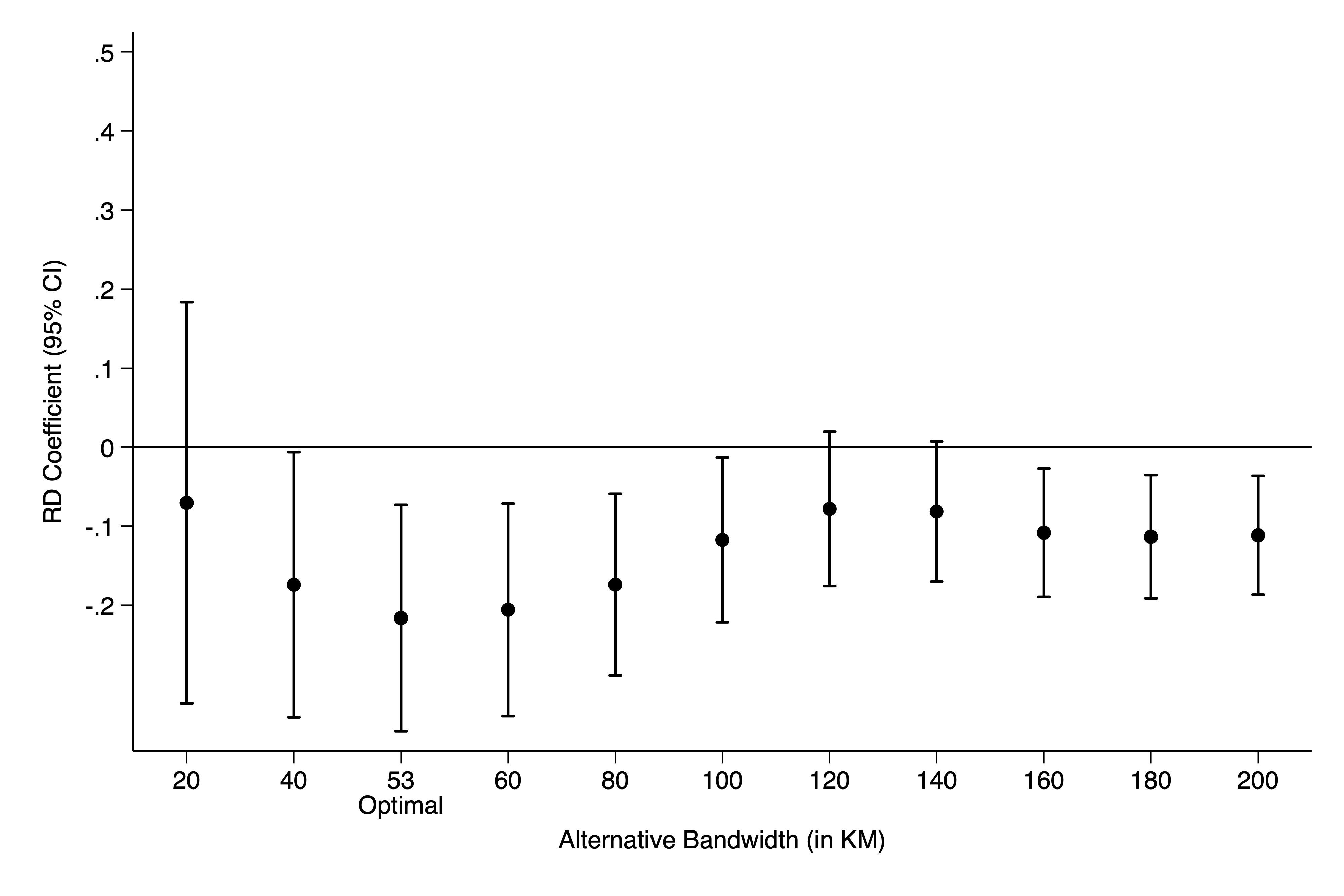}
\end{center}
{\footnotesize Notes: The figure presents the RDD approach with different bandwidth choices between 20 and 200 kilometers from the border between Odisha and Chhattisgarh. The dots are the estimated coefficients, and the vertical lines represent the 95\% percent confidence intervals. We set the local polynomial of order 2, which has a minimum asymptotic mean squared error (MSE) of the RD point estimator in our sample, as proposed by Pei et al. (2021). We use a uniform kernel, as suggested by Imbens and Lemieux (2008) and Lee and Lemieux (2010), which is simple, transparent, and easy to interpret. (Use of triangular kernel also produces similar results.)}
\end{figure}

\clearpage
\begin{table}[htbp]
\begin{center}
\section{Robustness to OLS Estimates – Different Measures of Sterilization}
\end{center}

This section presents the robustness results of OLS estimates reported in Table 1. Tables B1-B5 present results, including each set of controls sequentially. Table B6 presents results for the cohort of children between 12-23 months. Table B7 presents results considering Excess Vasectomy as an alternative measure of forced sterilization policy.
\vskip 1cm
\begin{center}
\caption{\label{figure:TableB1}\textbf{Total Sterilizations Performed in 1976-77 (in 100,000)}}
\includegraphics[width=\textwidth]{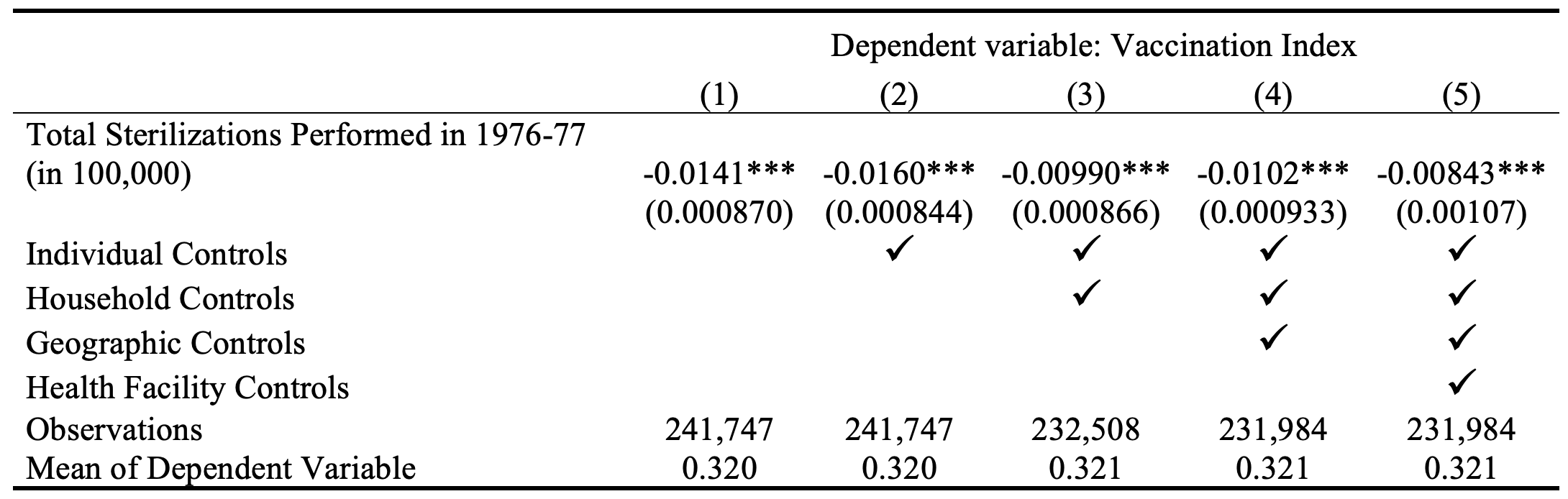}
\end{center}
{\footnotesize Notes: Data are from India’s National Family and Health Survey 2015-16 (NFHS-4). The unit of observation is a child below the age of 5. Vaccination Index is an index that includes BCG, measles, and three doses each of DPT, four doses of polio, and four doses of hepatitis B. Total Sterilizations Performed in 1976-77 (in 100,000) measures the total number of sterilizations performed in a state in 1976-77 (expressed in 100,000 individuals).  Individual controls are for a gender indicator variable of the child, month by year of birth fixed effects, an indicator for whether the child is twin, and the birth order of the child. Household controls include age and sex of the household head, household size, number of household members below the age of 5, seven religion fixed effects, four caste fixed effects, 20 education of the mother fixed effects, four household wealth index fixed effects, and an indicator for whether any household member is covered by health insurance. Geographic controls include the altitude of the cluster in meters, altitude squared, state-level population density per square kilometers (in log), and an indicator of whether the place of residence is urban. Health facility controls include hospitals per 1000 population and doctors per 1000 population at the state level. Robust standard errors in parentheses clustered at the NFHS-4 cluster (PSU) level. *** p$<0.01$, ** p$<0.05$, * p$<0.1$
}
\end{table}

\clearpage
\begin{table}[htbp]
\begin{center}
\caption{\label{figure:TableB2}\textbf{Total Sterilizations Performed in 1976-77 (in log)}}
\includegraphics[width=\textwidth]{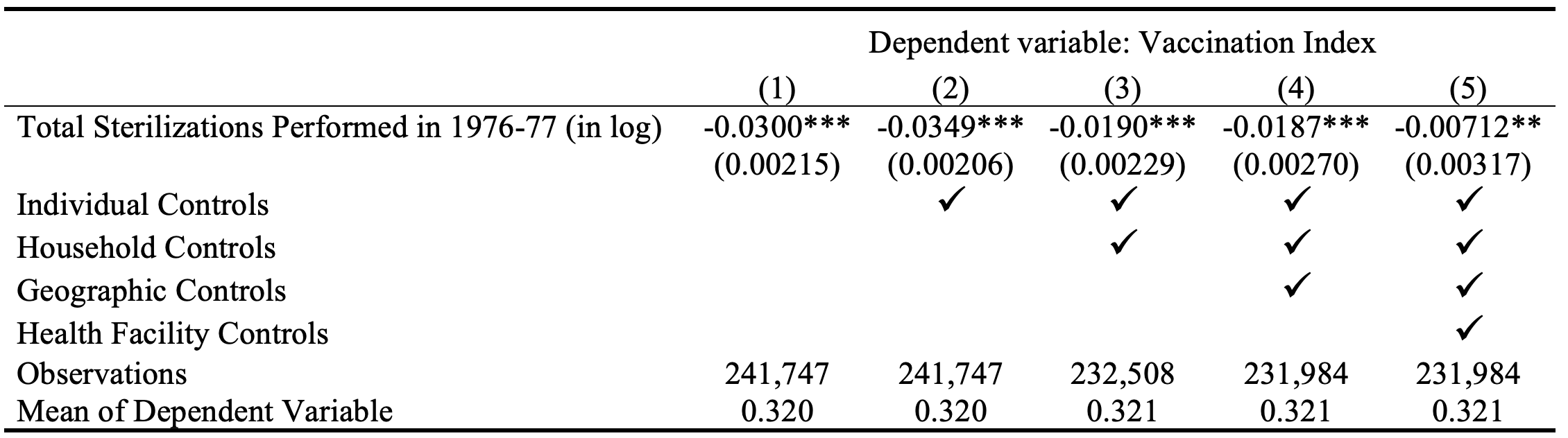}
\end{center}
{\footnotesize Notes: Data are from India’s National Family and Health Survey 2015-16 (NFHS-4). The unit of observation is a child below the age of 5. Vaccination Index is an index that includes BCG, measles, and three doses each of DPT, four doses of polio, and four doses of hepatitis B. Total Sterilizations Performed in 1976-77 (in log) measures the natural log of the number of sterilizations performed in 1976-77. Individual controls are for a gender indicator variable of the child, month by year of birth fixed effects, an indicator for whether the child is twin, and the birth order of the child. Household controls include age and sex of the household head, household size, number of household members below the age of 5, seven religion fixed effects, four caste fixed effects, 20 education of the mother fixed effects, four household wealth index fixed effects, and an indicator for whether any household member is covered by health insurance. Geographic controls include the altitude of the cluster in meters, altitude squared, state-level population density per square kilometers (in log), and an indicator of whether the place of residence is urban. Health facility controls include hospitals per 1000 population and doctors per 1000 population at the state level. Robust standard errors in parentheses clustered at the NFHS-4 cluster (PSU) level.  *** p$<0.01$, ** p$<0.05$, * p$<0.1$
}
\end{table}

\clearpage
\begin{table}[htbp]
\begin{center}
\caption{\label{figure:TableB3}\textbf{Excess Sterilization Performed in 1976-77 (in 100,000)}}
\includegraphics[width=\textwidth]{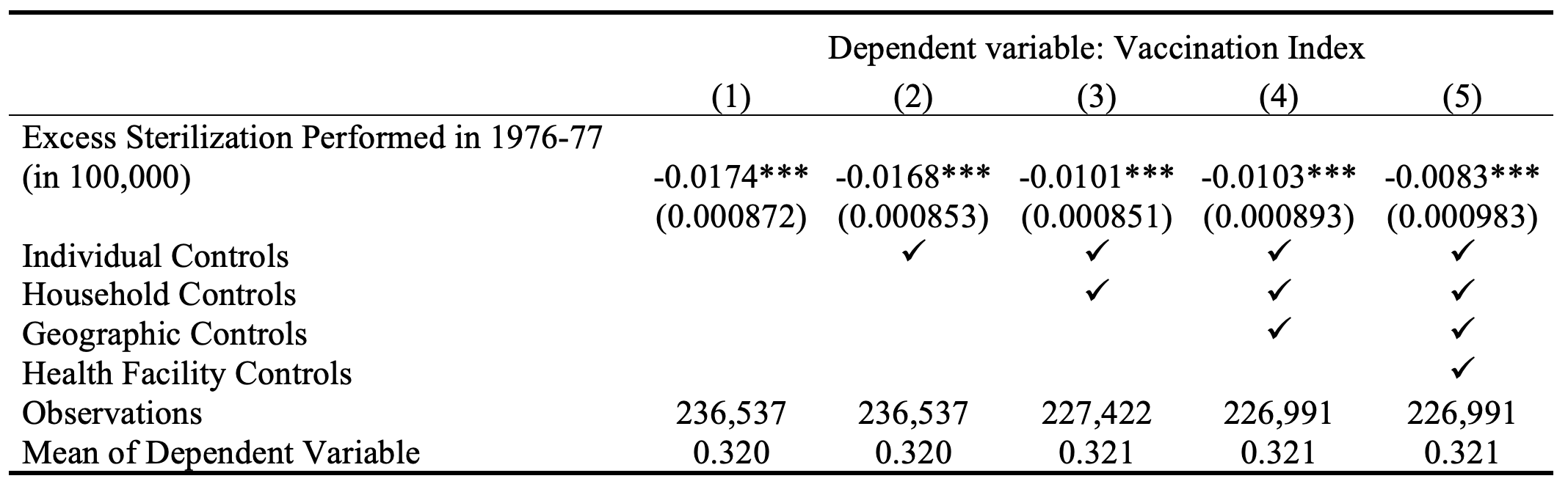}
\end{center}
{\footnotesize Notes: Data are from India’s National Family and Health Survey 2015-16 (NFHS-4). The unit of observation is a child below the age of 5. Vaccination Index is an index that includes BCG, measles, and three doses each of DPT, four doses of polio, and four doses of hepatitis B. Excess Sterilization Performed in 1976-77 (in 100,000) measures the number of excess sterilizations performed in 1976-77 over and above the 1975-76 numbers (expressed in 100,000 individuals). Individual controls are for a gender indicator variable of the child, month by year of birth fixed effects, an indicator for whether the child is twin, and the birth order of the child. Household controls include age and sex of the household head, household size, number of household members below the age of 5, seven religion fixed effects, four caste fixed effects, 20 education of the mother fixed effects, four household wealth index fixed effects, and an indicator for whether any household member is covered by health insurance. Geographic controls include the altitude of the cluster in meters, altitude squared, state-level population density per square kilometers (in log), and an indicator of whether the place of residence is urban. Health facility controls include hospitals per 1000 population and doctors per 1000 population at the state level. Robust standard errors in parentheses clustered at the NFHS-4 cluster (PSU) level.  *** p$<0.01$, ** p$<0.05$, * p$<0.1$
}
\end{table}

\clearpage
\begin{table}[htbp]
\begin{center}
\caption{\label{figure:TableB4}\textbf{Excess Sterilization Performed in 1976-77 (in log)}}
\includegraphics[width=\textwidth]{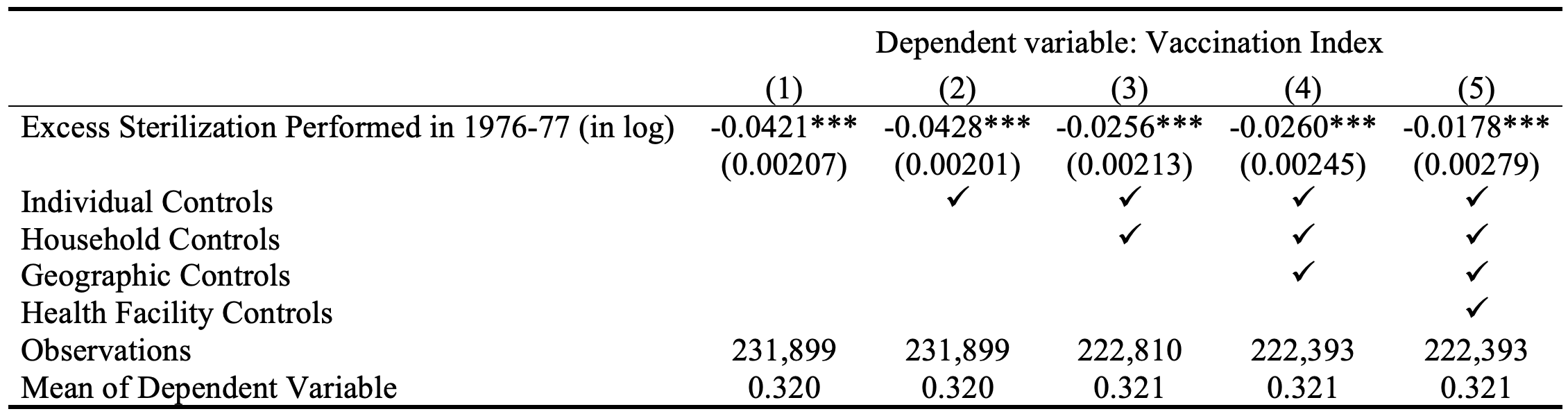}
\end{center}
{\footnotesize Notes: Data are from India’s National Family and Health Survey 2015-16 (NFHS-4). The unit of observation is a child below the age of 5. Vaccination Index is an index that includes BCG, measles, and three doses each of DPT, four doses of polio, and four doses of hepatitis B. Excess Sterilization Performed in 1976-77 (in log) measures the natural log of the excess number of sterilizations performed in 1976-77 over and above the 1975-76 numbers. Individual controls are for a gender indicator variable of the child, month by year of birth fixed effects, an indicator for whether the child is twin, and the birth order of the child. Household controls include age and sex of the household head, household size, number of household members below the age of 5, seven religion fixed effects, four caste fixed effects, 20 education of the mother fixed effects, four household wealth index fixed effects, and an indicator for whether any household member is covered by health insurance. Geographic controls include the altitude of the cluster in meters, altitude squared, state-level population density per square kilometers (in log), and an indicator of whether the place of residence is urban. Health facility controls include hospitals per 1000 population and doctors per 1000 population at the state level. Robust standard errors in parentheses clustered at the NFHS-4 cluster (PSU) level.   *** p$<0.01$, ** p$<0.05$, * p$<0.1$
}
\end{table}

\clearpage
\begin{table}[htbp]
\begin{center}
\caption{\label{figure:TableB5}\textbf{Excess Sterilization }}
\includegraphics[width=\textwidth]{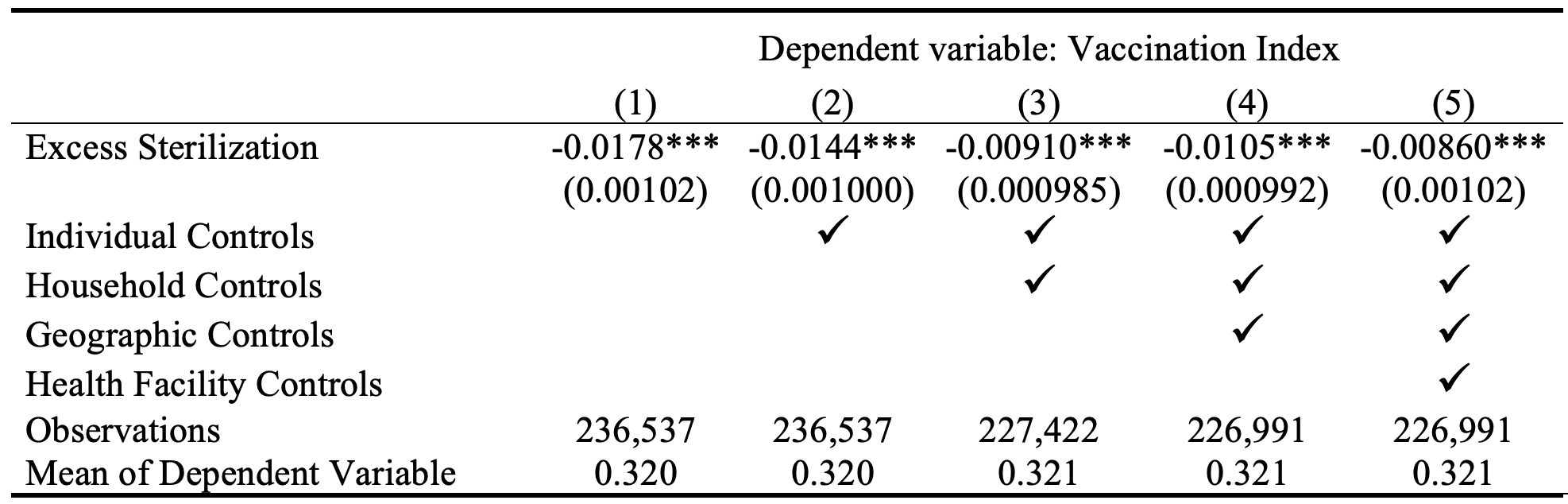}
\end{center}
{\footnotesize Notes: Data are from India’s National Family and Health Survey 2015-16 (NFHS-4). The unit of observation is a child below the age of 5. Vaccination Index is an index that includes BCG, measles, and three doses each of DPT, four doses of polio, and four doses of hepatitis B. Excess Sterilization measures the number of excess sterilizations performed in 1976-77 (compared with 1975-76 numbers) normalized by the sterilization performed in 1975-76 at the state level. Individual controls are for a gender indicator variable of the child, month by year of birth fixed effects, an indicator for whether the child is twin, and the birth order of the child. Household controls include age and sex of the household head, household size, number of household members below the age of 5, seven religion fixed effects, four caste fixed effects, 20 education of the mother fixed effects, four household wealth index fixed effects, and an indicator for whether any household member is covered by health insurance. Geographic controls include the altitude of the cluster in meters, altitude squared, state-level population density per square kilometers (in log), and an indicator of whether the place of residence is urban. Health facility controls include hospitals per 1000 population and doctors per 1000 population at the state level. Robust standard errors in parentheses clustered at the NFHS-4 cluster (PSU) level.   *** p$<0.01$, ** p$<0.05$, * p$<0.1$
}
\end{table}

\clearpage
\begin{table}[htbp]
\begin{center}
\caption{\label{figure:TableB6}\textbf{Children Between 12-23 Months }}
\includegraphics[width=\textwidth]{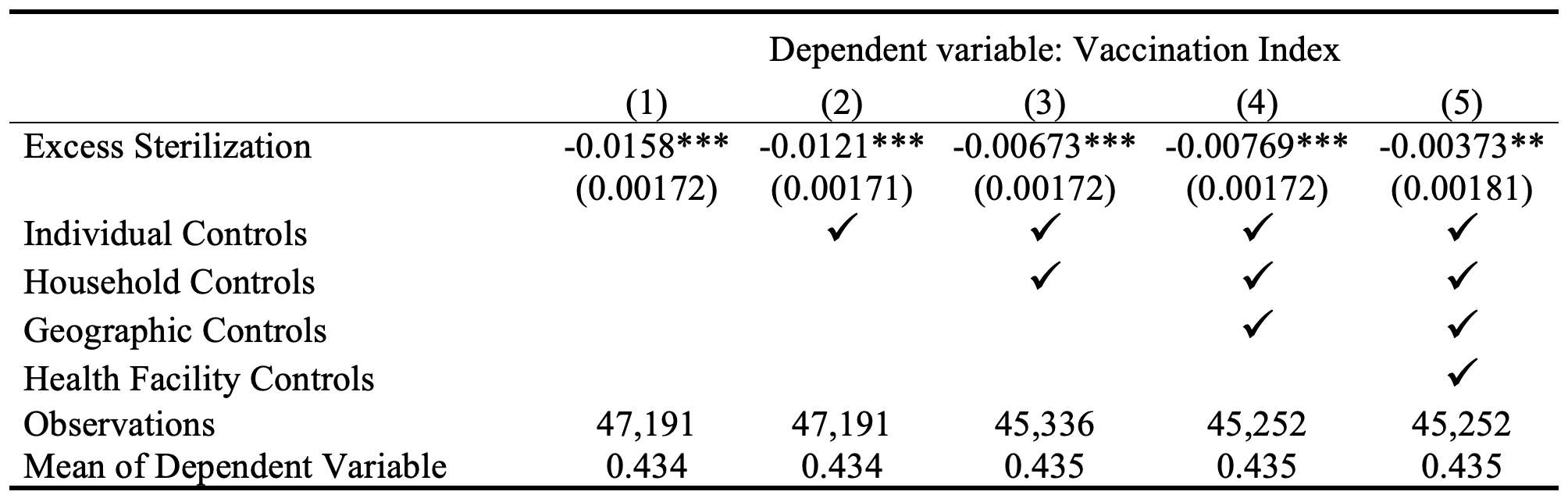}
\end{center}
{\footnotesize Notes: Data are from India’s National Family and Health Survey 2015-16 (NFHS-4). The unit of observation is a child between 12-23 months. Vaccination Index is an index that includes BCG, measles, and three doses each of DPT, four doses of polio, and four doses of hepatitis B. Excess Sterilization measures the number of excess sterilizations performed in 1976-77 (compared with 1975-76 numbers) normalized by the sterilization performed in 1975-76 at the state level. Individual controls are for a gender indicator variable of the child, month by year of birth fixed effects, an indicator for whether the child is twin, and the birth order of the child. Household controls include age and sex of the household head, household size, number of household members below the age of 5, seven religion fixed effects, four caste fixed effects, 20 education of the mother fixed effects, four household wealth index fixed effects, and an indicator for whether any household member is covered by health insurance. Geographic controls include the altitude of the cluster in meters, altitude squared, state-level population density per square kilometers (in log), and an indicator of whether the place of residence is urban. Health facility controls include hospitals per 1000 population and doctors per 1000 population at the state level. Robust standard errors in parentheses clustered at the NFHS-4 cluster (PSU) level. *** p$<0.01$, ** p$<0.05$, * p$<0.1$
}
\end{table}

\clearpage
\begin{table}[htbp]
\begin{center}
\caption{\label{figure:TableB7}\textbf{Alternative Measures of Force Sterilization Policy – Male Sterilization }}
\includegraphics[width=\textwidth]{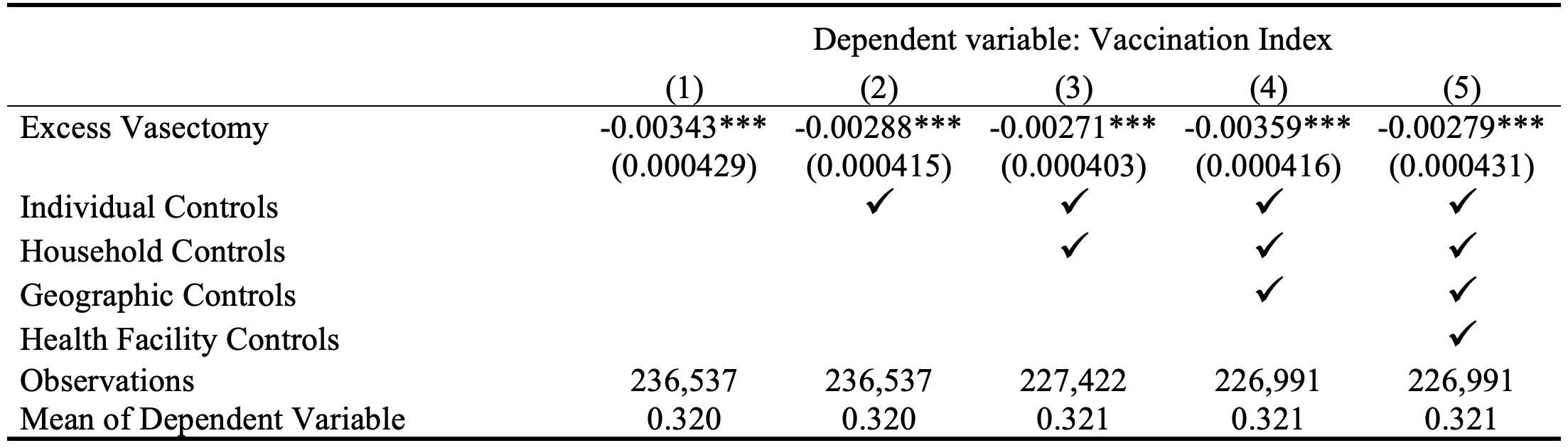}
\end{center}
{\footnotesize Notes: Data are from India’s National Family and Health Survey 2015-16 (NFHS-4). The unit of observation is a child below the age of 5. Vaccination Index is an index that includes BCG, measles, and three doses each of DPT, four doses of polio, and four doses of hepatitis B. Excess Vasectomy measures the number of excess vasectomies performed in 1976-77 (compared with 1975-76 numbers) normalized by the vasectomies performed in 1975-76 at the state level. Individual controls are for a gender indicator variable of the child, month by year of birth fixed effects, an indicator for whether the child is twin, and the birth order of the child. Household controls include age and sex of the household head, household size, number of household members below the age of 5, seven religion fixed effects, four caste fixed effects, 20 education of the mother fixed effects, four household wealth index fixed effects, and an indicator for whether any household member is covered by health insurance. Geographic controls include the altitude of the cluster in meters, altitude squared, state-level population density per square kilometers (in log), and an indicator of whether the place of residence is urban. Health facility controls include hospitals per 1000 population and doctors per 1000 population at the state level. Robust standard errors in parentheses clustered at the NFHS-4 cluster (PSU) level.  *** p$<0.01$, ** p$<0.05$, * p$<0.1$
}
\end{table}

\clearpage
\begin{table}[htbp]
\setcounter{table}{0}
\begin{center}
\section{Robustness to IV Estimates}
\end{center}
This section presents the robustness results of IV estimates reported in Table 2. Table C1 presents results for the cohort of children between 12-23 months. Table C2 presents results considering Excess Vasectomy as an alternative measure of forced sterilization policy.

\begin{center}
\caption{\label{figure:TableC1}\textbf{Children Between 12-23 Months}}
\includegraphics[height=11cm]{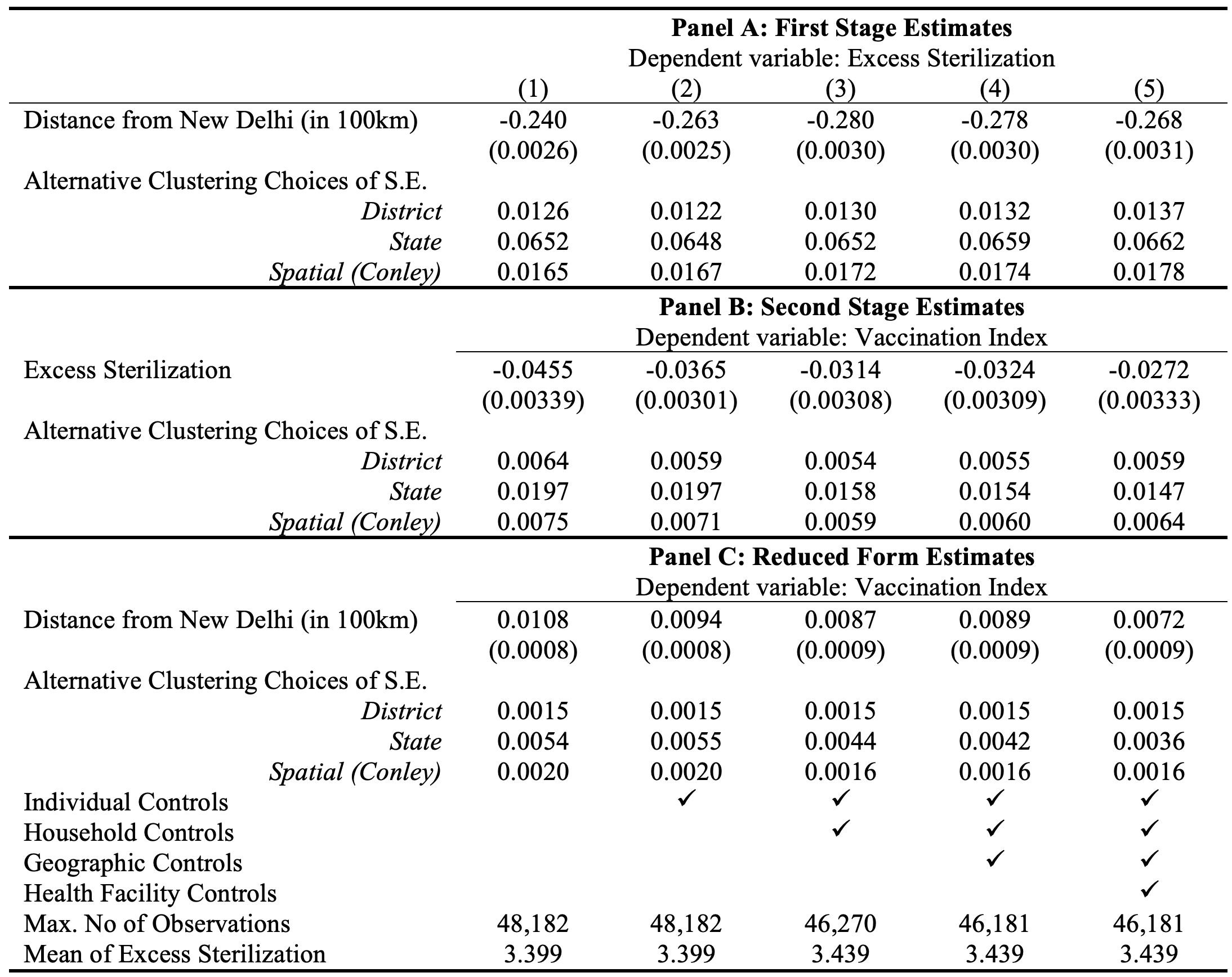}
\end{center}
{\footnotesize Notes: The table presents IV results along with the first stage and reduced form estimates. Data are from India's National Family and Health Survey 2015-16 (NFHS-4). The unit of observation is a child between 12-23 months. Vaccination Index is an index that includes BCG, measles, and three doses each of DPT, four doses of polio, and four doses of hepatitis B. Distance from New Delhi measures the distance from New Delhi (the national capital) to state capitals expressed in 100 kilometers. Excess Sterilization measures the number of excess sterilizations performed in 1976-77 (compared with 1975-76 numbers) normalized by the sterilization performed in 1975-76 at the state level. Individual controls are for a gender indicator variable of the child, month by year of birth fixed effects, an indicator for whether the child is twin, and the birth order of the child. Household controls include age and sex of the household head, household size, number of household members below the age of 5, seven religion fixed effects, four caste fixed effects, 20 education of the mother fixed effects, four household wealth index fixed effects, and an indicator for whether any household member is covered by health insurance. Geographic controls include the altitude of the cluster in meters, altitude squared, state-level population density per square kilometers (in log), and an indicator of whether the place of residence is urban. Health facility controls include hospitals per 1000 population and doctors per 1000 population at the state level. Below each coefficient, four standard errors are reported. The first, reported in parentheses, are standard errors adjusted for clustering at the NFHS-4 cluster (PSU) level. The Second—District—is standard errors adjusted for clustering at the current district level. The third—State—is standard errors adjusted for clustering at the current state level. The fourth—Spatial (Conley)—is standard errors adjusted for spatial correction proposed by Conley (1999).
}
\end{table}

\clearpage
\begin{table}[htbp]
\begin{center}
\caption{\label{figure:TableC2}\textbf{Alternative Measures of Force Sterilization Policy – Male Sterilization}}
\includegraphics[width=\textwidth]{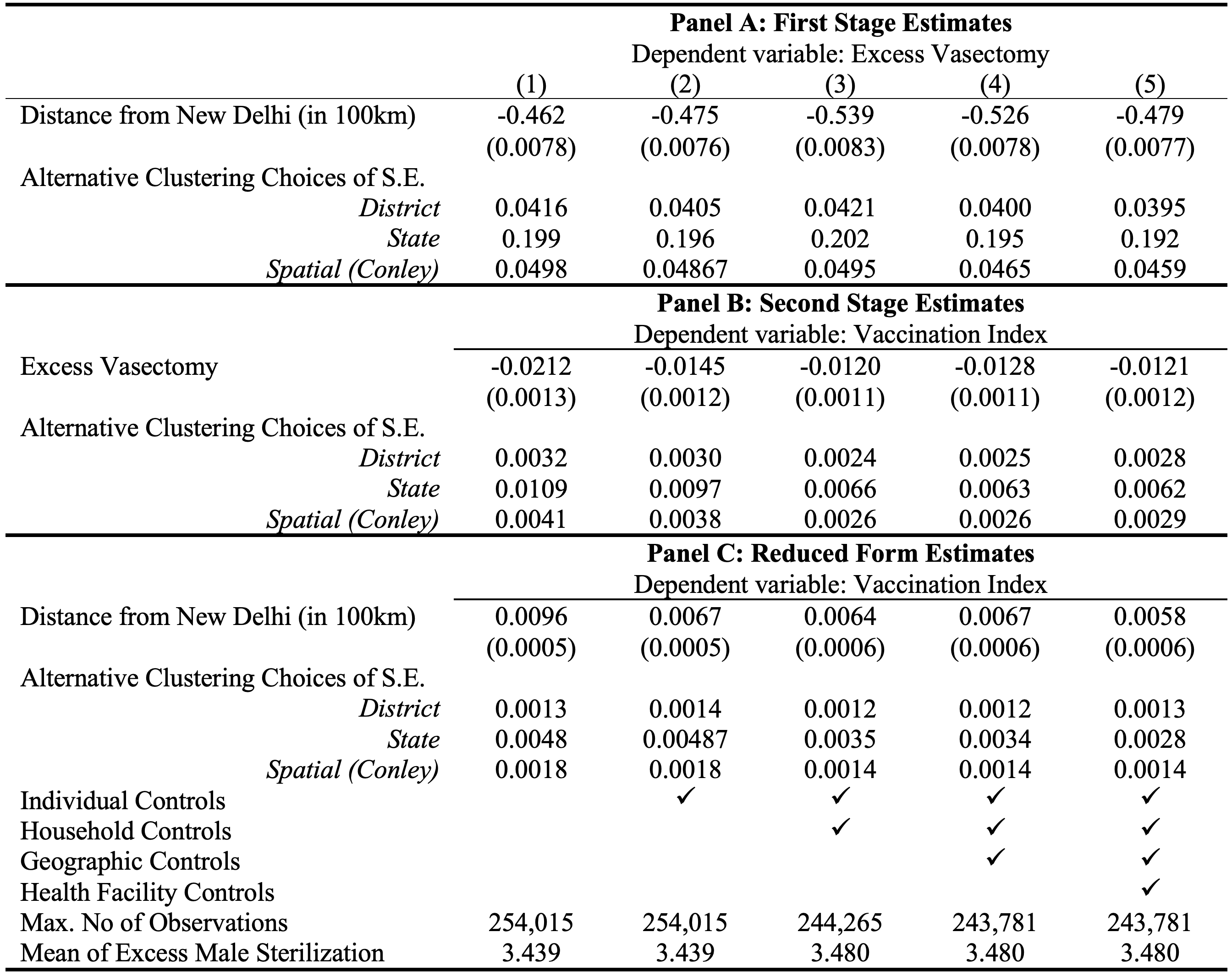}
\end{center}
{\footnotesize Notes: The table presents IV results along with first stage and reduced form estimates. Data are from India's National Family and Health Survey 2015-16 (NFHS-4). The unit of observation is a child below the age of 5. Vaccination Index is an index that includes BCG, measles, and three doses each of DPT, four doses of polio, and four doses of hepatitis B. Distance from New Delhi measures the distance from New Delhi (the national capital) to state capitals expressed in 100 kilometers. Excess Vasectomy measures the number of excess vasectomies performed in 1976-77 (compared with 1975-76 numbers) normalized by the vasectomy performed in 1975-76 at the state level. Individual controls are for a gender indicator variable of the child, month by year of birth fixed effects, an indicator for whether the child is twin, and the birth order of the child. Household controls include age and sex of the household head, household size, number of household members below the age of 5, seven religion fixed effects, four caste fixed effects, 20 education of the mother fixed effects, four household wealth index fixed effects, and an indicator for whether any household member is covered by health insurance. Geographic controls include the altitude of the cluster in meters, altitude squared, state-level population density per square kilometers (in log), and an indicator of whether the place of residence is urban. Health facility controls include hospitals per 1000 population and doctors per 1000 population at the state level. Below each coefficient, four standard errors are reported. The first, reported in parentheses, are standard errors adjusted for clustering at the NFHS-4 cluster (PSU) level. The Second—District—is standard errors adjusted for clustering at the current district level. The third—State—is standard errors adjusted for clustering at the current state level. The fourth—Spatial (Conley)—is standard errors adjusted for spatial correction proposed by Conley (1999).
}
\end{table}

\clearpage
\begin{sidewaystable}[htbp]
\setcounter{table}{0}
\begin{center}
\section{Heterogenous Effects and Robustness}
\end{center}
This section presents the additional results of heterogenous effects reported in Figure 9 and Figure 10. Table D1 presents the results of Figure 9 in tabular format. Table D2 presents results for the cohort of children between 12-23 months. Table D3 presents results considering Excess Vasectomy as an alternative measure of forced sterilization policy. Table D4-D6 presents the results of Figure 10 in tabular Format.

\begin{center}
\caption{\label{figure:TableD1}\textbf{Heterogenous Effects NFHS-4 (2015-16)}}
\includegraphics[width=\textwidth]{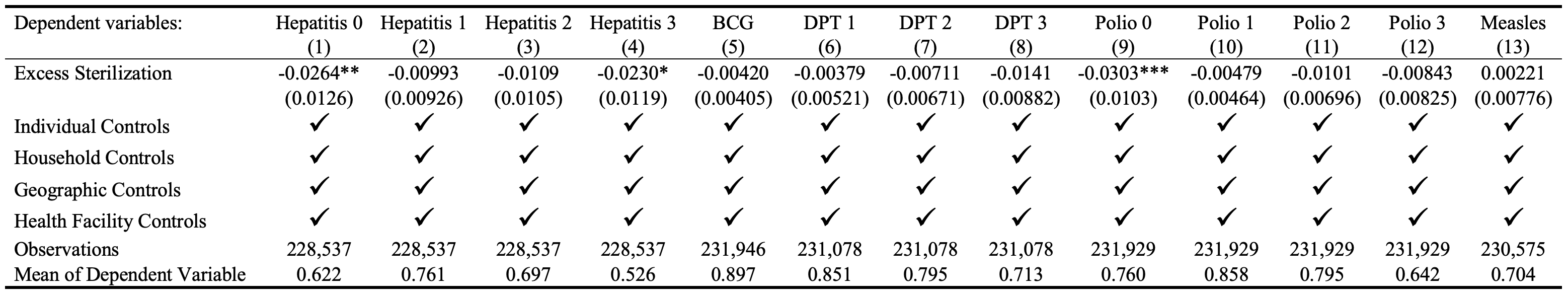}
\end{center}
{\footnotesize Notes: The table presents IV estimates. Data are from India's National Family and Health Survey 2015-16 (NFHS-4). The unit of observation is a child below the age of 5. Excess Sterilization measures the number of excess sterilizations performed in 1976-77 (compared with 1975-76 numbers) normalized by the sterilization performed in 1975-76 at the state level. Individual controls are for a gender indicator variable of the child, month by year of birth fixed effects, an indicator for whether the child is twin, and the birth order of the child. Household controls include age and sex of the household head, household size, number of household members below the age of 5, seven religion fixed effects, four caste fixed effects, 20 education of the mother fixed effects, four household wealth index fixed effects, and an indicator for whether any household member is covered by health insurance. Geographic controls include the altitude of the cluster in meters, altitude squared, state-level population density per square kilometers (in log), and an indicator of whether the place of residence is urban. Health facility controls include hospitals per 1000 population and doctors per 1000 population at the state level. Robust standard errors in parentheses clustered at the state level. *** p$<0.01$, ** p$<0.05$, * p$<0.1$
}
\end{sidewaystable}

\clearpage
\begin{sidewaystable}[htbp]
\begin{center}
\caption{\label{figure:TableD2}\textbf{Children Between 12-23 Months (NFHS-4)}}
\includegraphics[width=\textwidth]{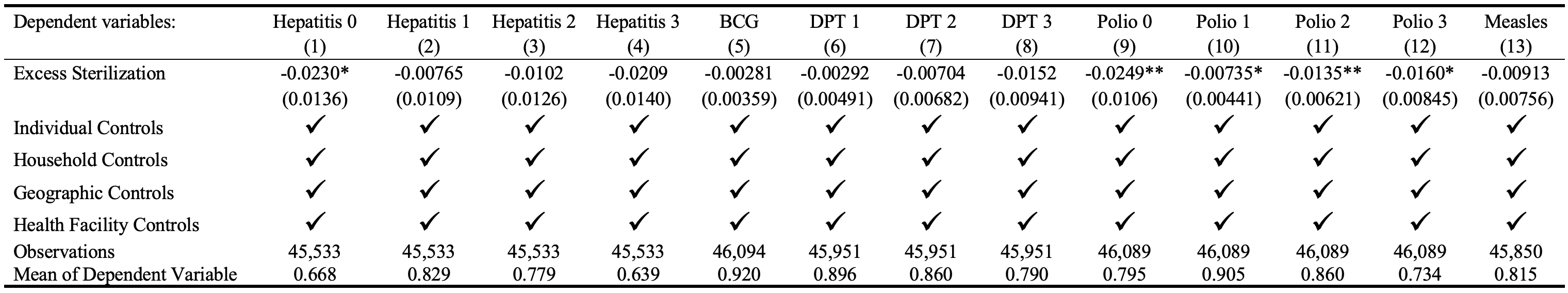}
\end{center}
{\footnotesize Notes: The table presents IV estimates. Data are from India's National Family and Health Survey 2015-16 (NFHS-4). The unit of observation is a child between 12-23 months. Excess Sterilization measures the number of excess sterilizations performed in 1976-77 (compared with 1975-76 numbers) normalized by the sterilization performed in 1975-76 at the state level. Individual controls are for a gender indicator variable of the child, month by year of birth fixed effects, an indicator for whether the child is twin, and the birth order of the child. Household controls include age and sex of the household head, household size, number of household members below the age of 5, seven religion fixed effects, four caste fixed effects, 20 education of the mother fixed effects, four household wealth index fixed effects, and an indicator for whether any household member is covered by health insurance. Geographic controls include the altitude of the cluster in meters, altitude squared, state-level population density per square kilometers (in log), and an indicator of whether the place of residence is urban. Health facility controls include hospitals per 1000 population and doctors per 1000 population at the state level. Robust standard errors in parentheses clustered at the state level.  *** p$<0.01$, ** p$<0.05$, * p$<0.1$
}
\end{sidewaystable}

\clearpage
\begin{sidewaystable}[htbp]
\begin{center}
\caption{\label{figure:TableD3}\textbf{Alternative Measures of Force Sterilization Policy (NFHS-4)- Male Sterilization}}
\includegraphics[width=\textwidth]{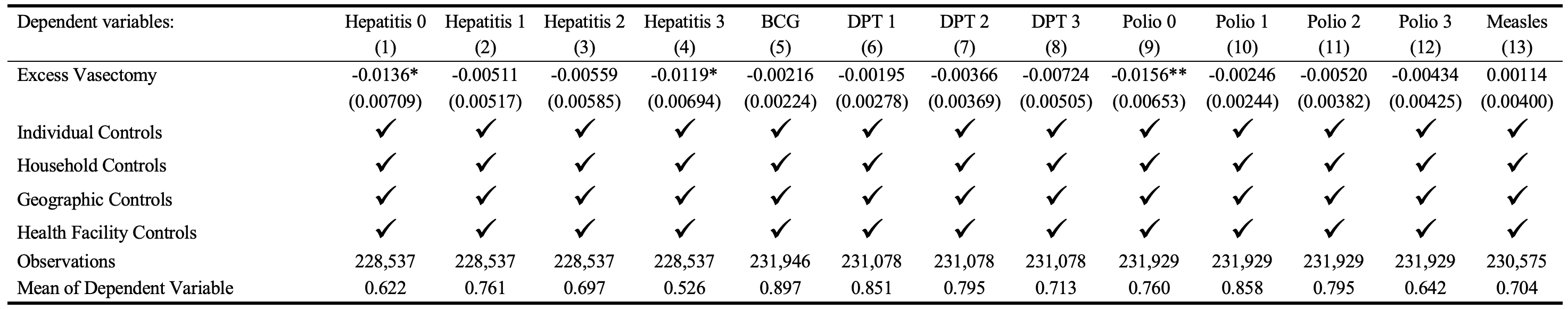}
\end{center}
{\footnotesize Notes: The table presents IV estimates. Data are from India's National Family and Health Survey 2015-16 (NFHS-4). The unit of observation is a child below the age of 5. Excess Vasectomy measures the number of excess vasectomies performed in 1976-77 (compared with 1975-76 numbers) normalized by the vasectomy performed in 1975-76 at the state level.  Individual controls are for a gender indicator variable of the child, month by year of birth fixed effects, an indicator for whether the child is twin, and the birth order of the child. Household controls include age and sex of the household head, household size, number of household members below the age of 5, seven religion fixed effects, four caste fixed effects, 20 education of the mother fixed effects, four household wealth index fixed effects, and an indicator for whether any household member is covered by health insurance. Geographic controls include the altitude of the cluster in meters, altitude squared, state-level population density per square kilometers (in log), and an indicator of whether the place of residence is urban. Health facility controls include hospitals per 1000 population and doctors per 1000 population at the state level. Robust standard errors in parentheses clustered at the state level.  *** p$<0.01$, ** p$<0.05$, * p$<0.1$
}
\end{sidewaystable}

\clearpage
\begin{sidewaystable}[htbp]
\begin{center}
\caption{\label{figure:TableD4}\textbf{Heterogenous Effects NFHS-1 (1992-93)}}
\includegraphics[width=\textwidth]{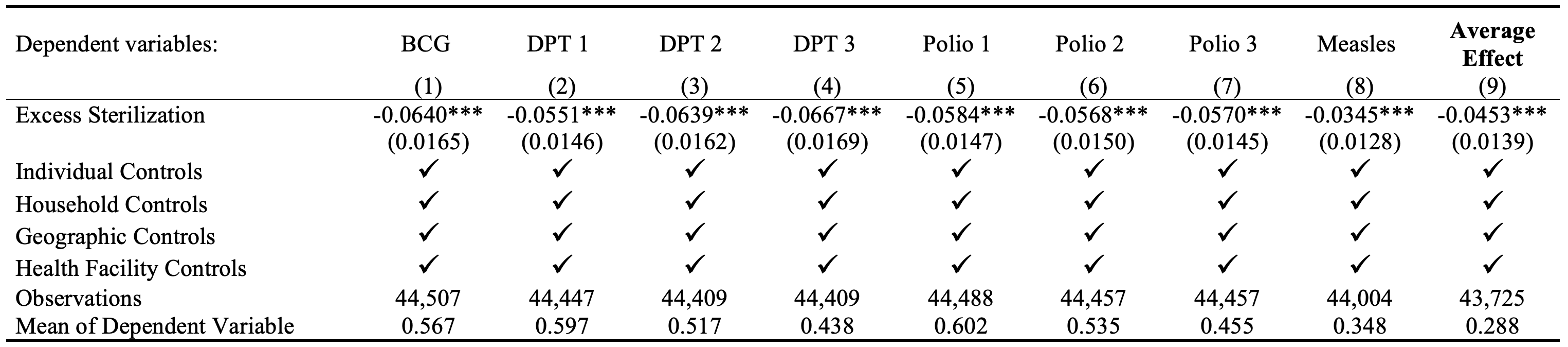}
\end{center}
{\footnotesize Notes: The table presents IV estimates. Data are from India's National Family and Health Survey 1992-93 (NFHS-1). The unit of observation is a child below the age of 5. Excess Sterilization measures the number of excess sterilizations performed in 1976-77 (compared with 1975-76 numbers) normalized by the sterilization performed in 1975-76 at the state level. Individual controls are for a gender indicator variable of the child, month by year of birth fixed effects, an indicator for whether the child is twin, and the birth order of the child. Household controls include age and sex of the household head, household size, number of household members below the age of 5, religion fixed effects, caste fixed effects, and education of the mother fixed effects. Geographic controls include state-level population density per square kilometers (in log) in 1991 and an indicator of whether the place of residence is urban. Health facility controls include hospitals per 1000 population and doctors per 1000 population at the state level. Robust standard errors in parentheses clustered at the state level. *** p$<0.01$, ** p$<0.05$, * p$<0.1$
}
\end{sidewaystable}

\clearpage
\begin{sidewaystable}[htbp]
\begin{center}
\caption{\label{figure:TableD5}\textbf{Heterogenous Effects NFHS-2 (1998-99)}}
\includegraphics[width=\textwidth]{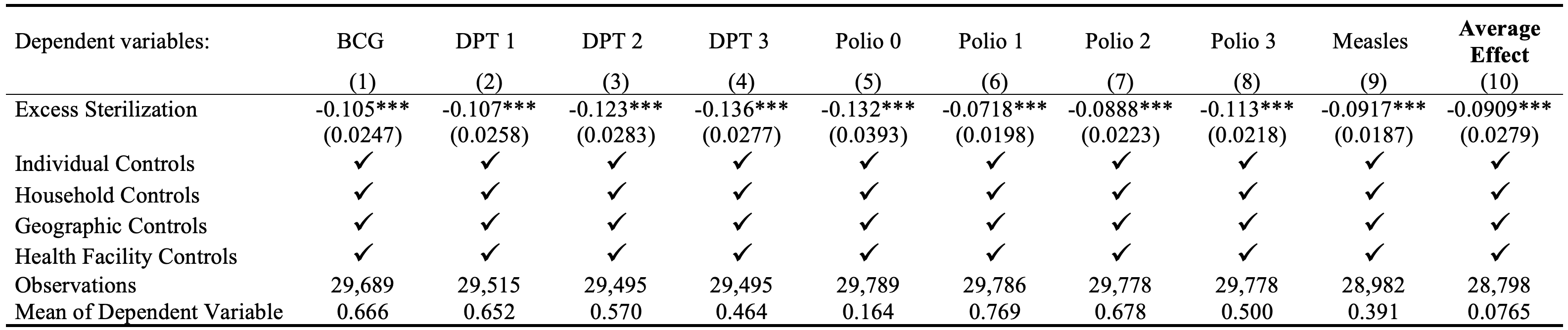}
\end{center}
{\footnotesize Notes: The table presents IV estimates. Data are from India's National Family and Health Survey 1998-99 (NFHS-2). The unit of observation is a child below the age of 5. Excess Sterilization measures the number of excess sterilizations performed in 1976-77 (compared with 1975-76 numbers) normalized by the sterilization performed in 1975-76 at the state level. Individual controls are for a gender indicator variable of the child, month by year of birth fixed effects, an indicator for whether the child is twin, and the birth order of the child. Household controls include age and sex of the household head, household size, number of household members below the age of 5, religion fixed effects, caste fixed effects, and education of the mother fixed effects. Geographic controls include the altitude of the cluster in meters, altitude squared, state-level population density per square kilometers (in log) in 1991, and an indicator of whether the place of residence is urban. Health facility controls include hospitals per 1000 population and doctors per 1000 population at the state level. Robust standard errors in parentheses clustered at the state level.  *** p$<0.01$, ** p$<0.05$, * p$<0.1$
}
\end{sidewaystable}

\clearpage
\begin{sidewaystable}[htbp]
\begin{center}
\caption{\label{figure:TableD6}\textbf{Heterogenous Effects NFHS-3 (2005-06)}}
\includegraphics[width=\textwidth]{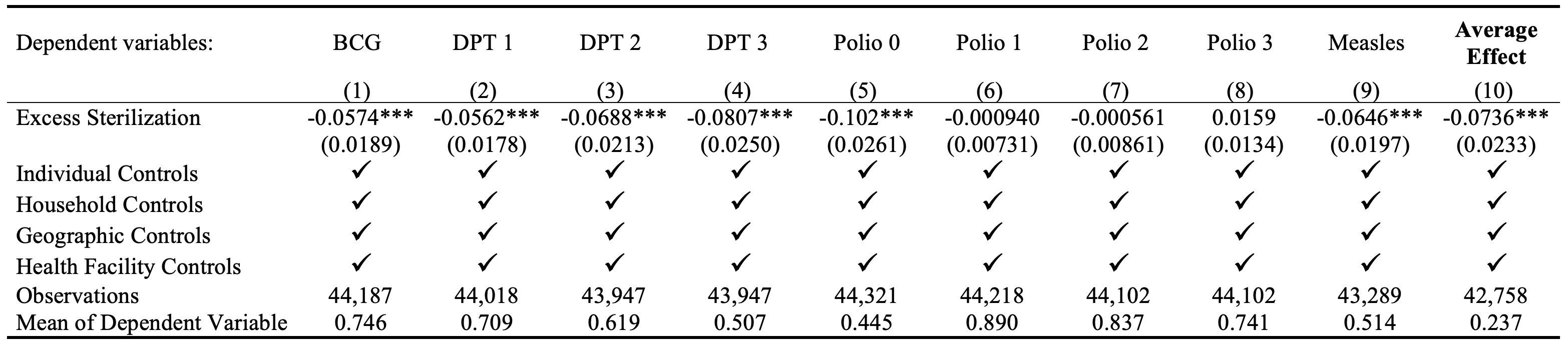}
\end{center}
{\footnotesize Notes: The table presents IV estimates. Data are from India's National Family and Health Survey 2005-06 (NFHS-3). The unit of observation is a child below the age of 5. Excess Sterilization measures the number of excess sterilizations performed in 1976-77 (compared with 1975-76 numbers) normalized by the sterilization performed in 1975-76 at the state level. Individual controls are for a gender indicator variable of the child, month by year of birth fixed effects, an indicator for whether the child is twin, and the birth order of the child. Household controls include age and sex of the household head, household size, number of household members below the age of 5, religion fixed effects, caste fixed effects, education of the mother fixed effects, household wealth index fixed effects, and an indicator for whether any household member is covered by health insurance. Geographic controls include the altitude of the cluster in meters, altitude squared, state-level population density per square kilometers (in log) in 2001, and an indicator of whether the place of residence is urban. Health facility controls include hospitals per 1000 population and doctors per 1000 population at the state level. Robust standard errors in parentheses clustered at the state level.   *** p$<0.01$, ** p$<0.05$, * p$<0.1$
}
\end{sidewaystable}

\clearpage
\begin{table}[htbp]
\setcounter{table}{0}
\begin{center}
\section{Mechanisms and Robustness}
\end{center}
This section presents additional results to the mechanism reported in Section 6. Table E1 presents the results of Table 4 considering Excess Vasectomy as an alternative measure of forced sterilization policy. Table E2 presents the results of Figure 12 in tabular format. Table E3 presents the results of Table 5 considering Excess Vasectomy as an alternative measure of forced sterilization policy.

\begin{center}
\caption{\label{figure:TableE1}\textbf{Robustness to Non-institutional Delivery Using Alternative Measures of Force Sterilization Policy - Male Sterilization}}
\includegraphics[width=\textwidth]{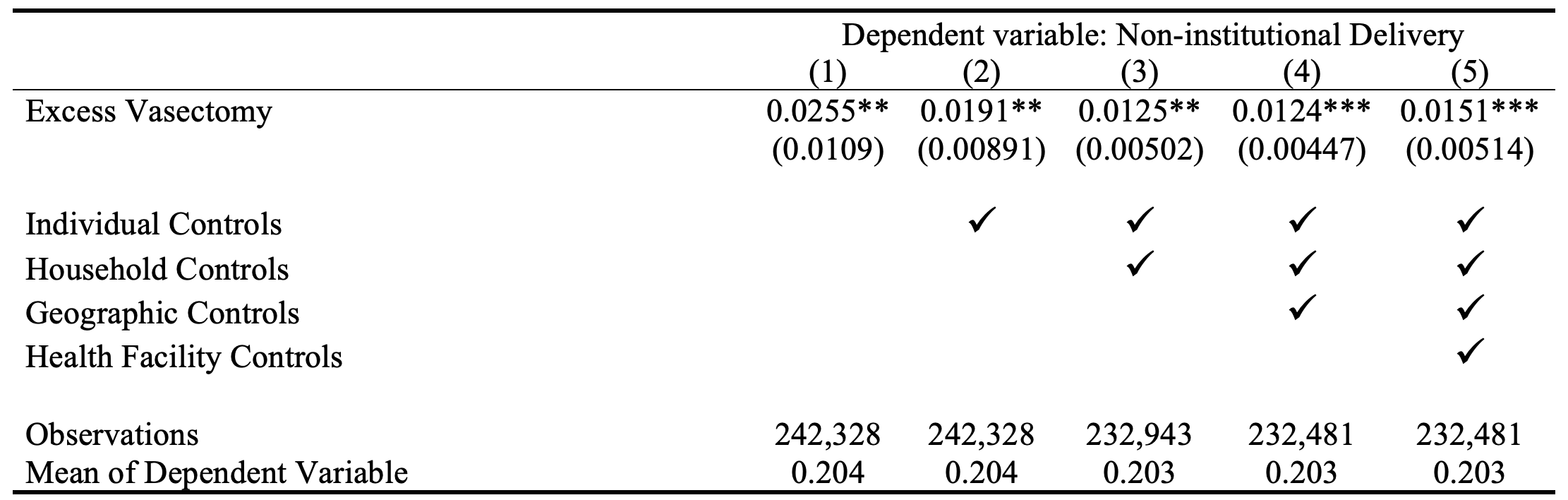}
\end{center}
{\footnotesize Notes: The table presents IV estimates. Data are from India's National Family and Health Survey 2015-16 (NFHS-4). The unit of observation is a child below the age of 5. Non-institutional Delivery is an indicator variable for a child born at home in the NFHS-4 data. Excess Vasectomy measures the number of excess vasectomies performed in 1976-77 (compared with 1975-76 numbers) normalized by the vasectomy performed in 1975-76 at the state level.  Individual controls are for a gender indicator variable of the child, month by year of birth fixed effects, an indicator for whether the child is twin, and the birth order of the child. Household controls include age and sex of the household head, household size, number of household members below the age of 5, seven religion fixed effects, four caste fixed effects, 20 education of the mother fixed effects, four household wealth index fixed effects, and an indicator for whether any household member is covered by health insurance. Geographic controls include the altitude of the cluster in meters, altitude squared, state-level population density per square kilometers (in log), and an indicator of whether the place of residence is urban. Health facility controls include hospitals per 1000 population and doctors per 1000 population at the state level. Robust standard errors in parentheses clustered at the state level. *** p$<0.01$, ** p$<0.05$, * p$<0.1$
}
\end{table}

\clearpage
\begin{sidewaystable}[htbp]
\begin{center}
\caption{\label{figure:TableE2}\textbf{Mechanism - Reasons for Non-Institutional Delivery (Tabular format)}}
\includegraphics[width=\textwidth]{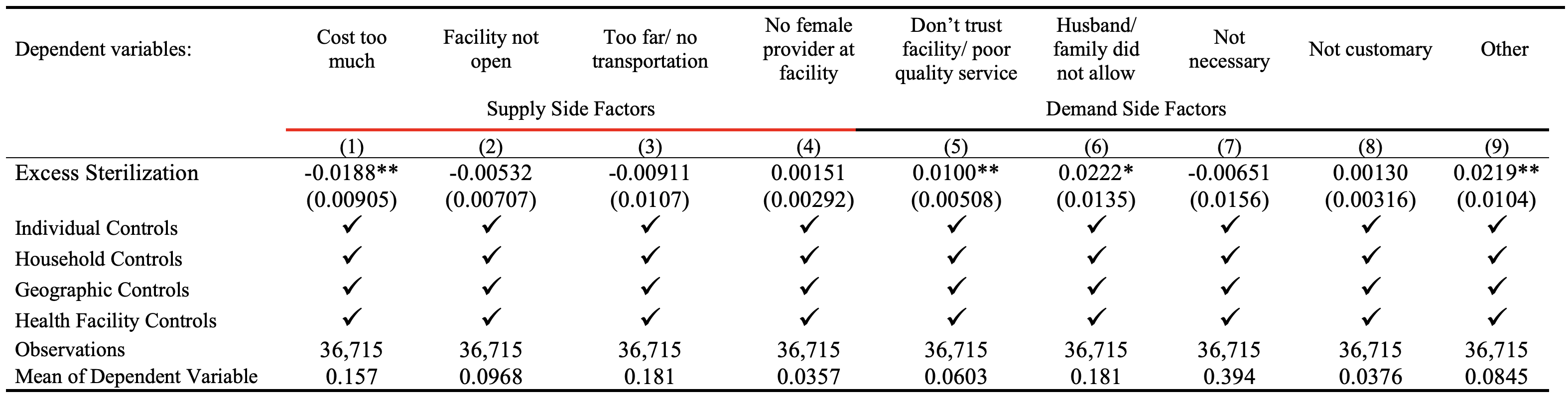}
\end{center}
{\footnotesize Notes: The table presents IV estimates. Data are from India's National Family and Health Survey 2015-16 (NFHS-4). The unit of observation is a mother’s last child below the age of 5 who is not born at a health care facility. The mean of dependent variables (in percentages) does not add to 100 because multiple responses were permitted. Excess Sterilization measures the number of excess sterilizations performed in 1976-77 (compared with 1975-76 numbers) normalized by the sterilization performed in 1975-76 at the state level. Individual controls are for a gender indicator variable of the child, month by year of birth fixed effects, an indicator for whether the child is twin, and the birth order of the child. Household controls include age and sex of the household head, household size, number of household members below the age of 5, seven religion fixed effects, four caste fixed effects, 20 education of the mother fixed effects, four household wealth index fixed effects, and an indicator for whether any household member is covered by health insurance. Geographic controls include the altitude of the cluster in meters, altitude squared, state-level population density per square kilometers (in log), and an indicator of whether the place of residence is urban. Health facility controls include hospitals per 1000 population and doctors per 1000 population at the state level. Robust standard errors in parentheses clustered at the state level.  *** p$<0.01$, ** p$<0.05$, * p$<0.1$
}
\end{sidewaystable}

\clearpage
\begin{table}[htbp]
\begin{center}
\caption{\label{figure:TableE3}\textbf{Robustness to Information Provision through Antenatal Care (ANC) Using Alternative Measures of Force Sterilization Policy - Male Sterilization}}
\includegraphics[width=10cm]{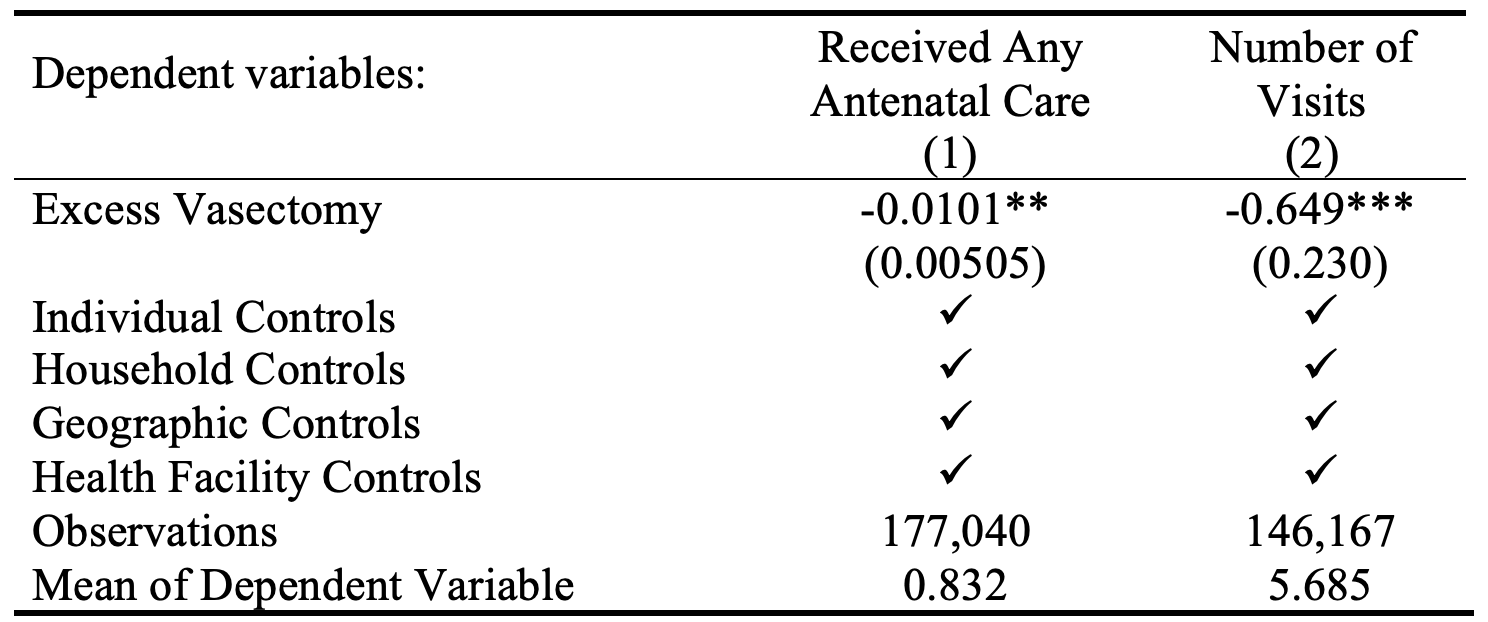}
\end{center}
{\footnotesize Notes: The table presents IV estimates. Data are from India's National Family and Health Survey 2015-16 (NFHS-4). The unit of observation is a mother’s last child below the age of 5.  Received Antenatal Care is an indicator variable for mothers who received antenatal care in the last pregnancy in the NFHS-4 data. The number of Visits measures the number of times the mother received antenatal care conditional on receiving any antenatal care in the last pregnancy. Excess Vasectomy measures the number of excess vasectomies performed in 1976-77 (compared with 1975-76 numbers) normalized by the vasectomy performed in 1975-76 at the state level.  Individual controls are for a gender indicator variable of the child, month by year of birth fixed effects, an indicator for whether the child is twin, and the birth order of the child. Household controls include age and sex of the household head, household size, number of household members below the age of 5, seven religion fixed effects, four caste fixed effects, 20 education of the mother fixed effects, four household wealth index fixed effects, and an indicator for whether any household member is covered by health insurance. Geographic controls include the altitude of the cluster in meters, altitude squared, state-level population density per square kilometers (in log), and an indicator of whether the place of residence is urban. Health facility controls include hospitals per 1000 population and doctors per 1000 population at the state level. Robust standard errors in parentheses clustered at the state level.  *** p$<0.01$, ** p$<0.05$, * p$<0.1$
}
\end{table}

\clearpage
\begin{table}[htbp]
\setcounter{table}{0}
\begin{center}
\section{Robustness to Consequence}
\end{center}
This section presents the robustness results of the consequences reported in Section 7. Table F1  presents the robustness results of Table 7, considering Excess Vasectomy as an alternative measure of forced sterilization policy. Table F2 presents the robustness results of Table 7, considering the death of a male child only.  

\begin{center}
\caption{\label{figure:TableF1}\textbf{Robustness to Consequence Using Alternative Measures of Force Sterilization Policy - Male Sterilization}}
\includegraphics[width=\textwidth]{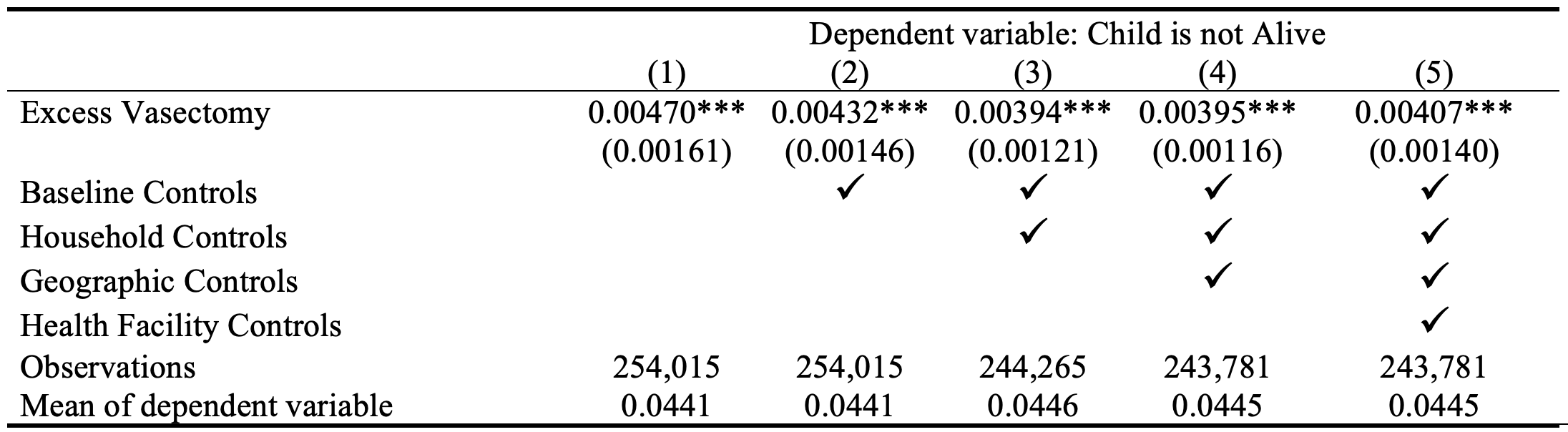}
\end{center}
{\footnotesize Notes: The table presents IV estimates. Data are from India's National Family and Health Survey 2015-16 (NFHS-4). The unit of observation is a child below the age of 5, including who is not alive. Excess Vasectomy measures the number of excess vasectomies performed in 1976-77 (compared with 1975-76 numbers) normalized by the vasectomy performed in 1975-76 at the state level.  Individual controls are for a gender indicator variable of the child, month by year of birth fixed effects, an indicator for whether the child is twin, and the birth order of the child. Household controls include age and sex of the household head, household size, number of household members below the age of 5, seven religion fixed effects, four caste fixed effects, 20 education of the mother fixed effects, four household wealth index fixed effects, and an indicator for whether any household member is covered by health insurance. Geographic controls include the altitude of the cluster in meters, altitude squared, state-level population density per square kilometers (in log), and an indicator of whether the place of residence is urban. Health facility controls include hospitals per 1000 population and doctors per 1000 population at the state level. Robust standard errors in parentheses clustered at the state level.  *** p$<0.01$, ** p$<0.05$, * p$<0.1$
}
\end{table}

\clearpage
\begin{table}[htbp]
\begin{center}
\caption{\label{figure:TableF2}\textbf{Robustness to Consequence Using Death of \textit{Male} Child Only}}
\includegraphics[width=\textwidth]{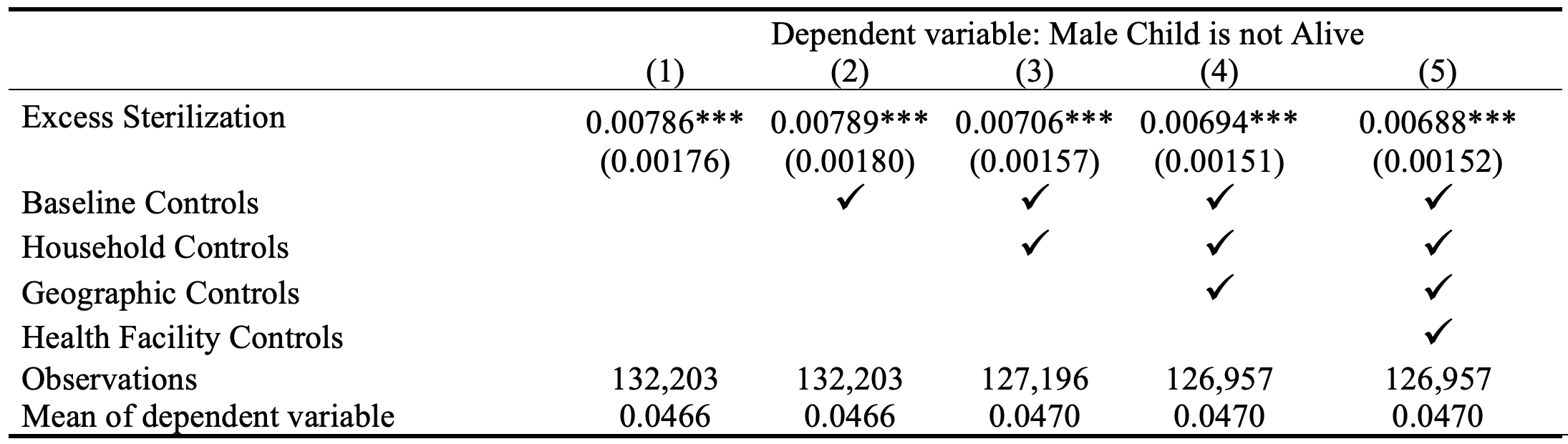}
\end{center}
{\footnotesize Notes: The table presents IV estimates. Data are from India's National Family and Health Survey 2015-16 (NFHS-4). The unit of observation is a male child below the age of 5, including one who is not alive. Excess Sterilization measures the number of excess sterilizations performed in 1976-77 (compared with 1975-76 numbers) normalized by the sterilizations performed in 1975-76 at the state level.  Individual controls are month-by-year of birth fixed effects, an indicator for whether the child is twin, and the birth order of the child. Household controls include age and sex of the household head, household size, number of household members below the age of 5, seven religion fixed effects, four caste fixed effects, 20 education of the mother fixed effects, four household wealth index fixed effects, and an indicator for whether any household member is covered by health insurance. Geographic controls include the altitude of the cluster in meters, altitude squared, state-level population density per square kilometers (in log), and an indicator of whether the place of residence is urban. Health facility controls include hospitals per 1000 population and doctors per 1000 population at the state level. Robust standard errors in parentheses clustered at the state level.   *** p$<0.01$, ** p$<0.05$, * p$<0.1$
}
\end{table}

\end{document}